%% file: ex_article.tex
\documentclass[review,onefignum,onetabnum]{siamart171218}
\nolinenumbers
\input{ex_shared}

\myexternaldocument{ex_supplement}

\begin{document}

\maketitle

\begin{abstract}
  This work develops a model-order reduction framework for the meshless weakly-compressible smoothed-particle hydrodynamics (SPH) method. The proposed framework introduces the concept of modal reference spaces to overcome the challenges of discovering low-dimensional subspaces from unstructured, dynamic, and mixing numerical topology that occurs in SPH simulations. These modal reference spaces enable a low-dimensional representation of the SPH field equations while maintaining their inherent meshless qualities. Modal reference spaces are constructed by projecting SPH snapshot data onto a reference space where low-dimensionality of field quantities can be discovered via traditional modal decomposition techniques (e.g., the proper orthogonal decomposition (POD)). Modal quantities are mapped back to the meshless SPH space via scattered data interpolation during the online predictive stage. The proposed model-order reduction framework is cast into the \emph{meshless} Galerkin POD (GPOD) and the Adjoint Petrov--Galerkin (APG) projection model-order reduction (PMOR) formulation. The PMORs are tested on three numerical experiments: 1) the Taylor--Green vortex; 2) the lid-driven cavity; and 3) the flow past an open cavity. Results show good agreement in reconstructed and predictive velocity fields, which showcase the ability of this framework to evolve the field equations in a low-dimensional subspace that originate from approximations in an unstructured, dynamic, and mixing numerical topology. Results also show that the pressure field is sensitive to the projection error due to the stiff weakly-compressible assumption made in the current SPH framework, but this sensitivity can be alleviated through nonlinear approximations, such as the APG approach. The proposed meshless model-order reduction framework reports a dimensionality compression factor of up to 90,000$\times$ within $10\%$ error in quantities of interest, marking a step toward drastic cost reduction in SPH simulations.
\end{abstract}


 \pagestyle{plain}{ }


\section{Introduction} \label{Section:Introduction}

In the past five decades, mathematical and numerical methods, in conjunction with improved computational resources, have enabled the modeling and simulation of complex multiphysics systems, from magneto-hydrodynamics in astrophysics to chemically-reacting and turbulent flows in combustion. However, even with great progress in the capabilities of numerical solvers and state-of-the-art high-performance computing, obtaining solutions to many complex physical systems remains computationally intractable and time-intensive. This resource and wall-time expense poses a barrier for realistic deployment of high-fidelity modeling and simulation in multi-query (MQ) settings, where many solutions (on the order of thousands or even hundreds of thousands) need to be performed to characterize parametric variations. These MQ settings include uncertainty quantification \cite{peherstorfer2018survey}, optimization \cite{rozza2022advanced}, and control \cite{benner2017model,brunton2022data}. However, recent advances in data-driven scientific computing and scientific machine learning are beginning to enable drastic reduction in the expense of modeling and simulation\cite{benner2015survey,brunton2022data}. Projection-based model-order reduction (PMOR) has been at the forefront of these advancements, and \emph{mesh-based} methods, such as the finite element (FE), finite volume (FV), and finite difference (FD) methods have dominated the progression of PMORs \cite{rozza2022advanced}. Yet, \emph{meshless} numerical frameworks, a cornerstone for modeling extreme events \cite{griffin2020smoothed,becker2019numerical,pearl2022fsisph} , large deforming flows \cite{monaghan2005smoothed,marrone2019extreme}, multiphase and multiphysics phenomena \cite{pearl2022fsisph,fuchs2021sph}, have seen minimal adoption of PMOR to reduce their modeling and simulation expense drastically.

The current work presents a model-order reduction framework for the smoothed-particle hydrodynamics (SPH) method, a numerical method that exemplifies meshless frameworks. This model-reduction framework adopts the ubiquitous architecture of intrusive PMORs where governing ordinary differential equations of dynamical systems are projected onto a low-dimensional subspace to realize cost savings in both computational resources and time. The motivation behind the present work is to construct a meshless PMOR (mPMOR) framework that is founded on the infrastructure of traditional PMORs. Thereby, the suite of rigorous Galerkin or Petrov--Galerkin PMORs \cite{parish2020adjoint, carlberg2011efficient, carlberg2017galerkin} that exist in the literature can also be applied to the proposed method while retaining the benefits and flexibility of a meshless numerical topology. To the best of the authors' knowledge, there exists no other PMOR framework that reconciles the \emph{mapping} from the unstructured, dynamic, and mixing numerical topology of the SPH method to a low-dimensional embedding, all while retaining a meshless infrastructure. The remainder of this section provides a brief introduction the SPH literature and its current state in Subsection \ref{Section:Introduction_SPHReview}, a review on the state of PMORs in \ref{Section:Introduction_ROMReview}, followed by an overview of the proposed method's contribution in Section \ref{Section:Introduction_Contributions}.

\subsection{Smoothed-particle hydrodynamics}\label{Section:Introduction_SPHReview}
SPH was first developed in \cite{lucy1977numerical, gingold1977smoothed} to perform astrophysics simulations of unbounded domains. Since its inception, SPH has evolved into an exemplary meshless numerical method widely used across a range of industries that include biomedical, naval, civil, mechanical, and aerospace engineering \cite{shadloo2016smoothed, liu2010smoothed}. The wide ranging adoption of SPH can be traced back to the several benefits it has to offer, especially its intrinsic numerical adaptivity. Because SPH is meshless and Lagrangian, its numerical topology is adaptive to the dynamics of the governing system of equations. In other words, the spatial discretization of SPH does not have fixed or pre-specified connectivity, and its numerical integration points are free to move in space, as dictated by the dynamics of the problem of interest. This intrinsic numerical adaptivity makes SPH extremely beneficial for the intrinsic tracking of free-surfaces or interfaces, modeling large deforming flows, multiphase flows, compressible flow, and complex fluid-structure interactions. For more on SPH, its benefits, and applications the reader is referred to the following non-exhaustive list of texts and literature \cite{sigalotti2021mathematics,violeau2012fluid,liu2003smoothed,monaghan1992smoothed}. While SPH has shown promise in its versatility across scientific disciplines, it remains a nascent numerical method with several well-known ``grand challenges" (GC), listed in \cite{vacondio2021grand} by the SPH rEsearch and engineeRing International Community (SPHERIC), hindering its complete adoption in theory and practice. Specifically, the following five grand challenges facing SPH have been highlighted by SPHERIC: GC1) convergence, consistency, and stability; GC2) boundary conditions; GC3) adaptivity (numerical refinement); GC4) coupling to other methods; GC5) applicability to industry. In recent years, however, these grand challenges have advanced considerably, especially GC1-GC3. The reader is referred to the literature for specific details on these improvements, and the following references provide a good starting point \cite{rastelli2022implicit, rastelli2023arbitrarily,english2022modified, antuono2023clone, sun2021accurate, ricci2024multiscale}. 

The current work seeks to address the technical barriers of GC5, namely, to reduce the computational costs of time-critical industrial MQ tasks using SPH. A primary drawback of SPH within the industrial setting stems from GC1, i.e., its limited convergence rate, which makes it computationally expensive to achieve desired levels of accuracy, especially for large problems. Recent work has leveraged the parallelizability of SPH, especially with GPU programming \cite{harada2007smoothed, dominguez2013new, crespo2011gpus, valdez2013towards, tafuni2018versatile, o2021fluid, dominguez2022dualsphysics}, to reduce compute times.  However, GPU implementations of SPH still cannot offer sufficient memory or time savings for time-critical industrial MQ tasks with high-fidelity constraints, e.g., uncertainty quantification, design optimization, inverse problems, control, etc. The goal of the present work is to address the cost limitation posed by convergence rates in SPH by introducing its meshless numerical infrastructure into a projection-based moder-order reduction framework, which to the best of the authors' knowledge has yet to be accomplished. An SPH PMOR would enable a low-dimensional representation of the SPH formalism, allowing the meshless framework to evolve in a subspace that is manageable for high-fidelity time-critical MQ tasks. Unfortunately, to date, PMORs for meshless methods are severely underdeveloped, if developed at all; this is especially true of Galerkin and Petrov--Galerkin methods, which are a cornerstone of model-order reduction. The current work aims to bridge the gap between meshless numerical methods and PMORs in a general sense by way of the SPH method due to its representative meshless numerical formalism. 

\subsection{Projection-based model-order reduction}\label{Section:Introduction_ROMReview}
Projection-based model-or\-der reduction can be cast into non-intrusive or intrusive frameworks. Both of these  approaches seek to evolve the full-order model (FOM) (a high-dimensional semi-discrete system of ordinary differential equations (ODEs)), such as  SPH, FE, FV, FD equations, on a low-dimensional subspace or manifold. Projecting onto these low-dimensional subspaces or manifolds is performed by way of Galerkin or Petrov--Galerkin projections, where the test and trial bases are determined \emph{a posteriori} in a data-driven manner through dimensionality reduction of FOM solution data over a parametric space of interest. The proper-orthogonal decomposition (POD), the principal components analysis (PCA) \cite{taira2017modal}, and deep learning architectures (such as convolutional \cite{lee2020model}, beta-variational \cite{solera2024beta}, graph \cite{magargal2024projection}, or traditional auto-encoders \cite{kim2020efficient}), are a few common dimensionality-reduction techniques used to determine the test and trial bases for non-intrusive and intrusive PMOR. In a way, PMOR can be thought of as a data-driven perspective of the classical separation of variables method to solve partial differential equations, where the eigenfunctions are determined \emph{a posteriori} from the FOM solution data rather than being defined \emph{a priori}.

Non-intrusive PMORs are motivated by the limited or lack of access to the numerical infrastructure of FOMs. For instance, non-intrusive methods are particularly useful for developing PMORs from proprietary commercial codes, where the practitioner has no access to the internal numerical infrastructure, operators, or actions of operators on state vectors. Nevertheless, non-intrusive frameworks leverage knowledge of known governing equations and their structure for the problem at hand, and a low-dimensional representation of the problem can be constructed for the state trajectories of the FOM from an abstract dynamical systems point of view, i.e., data-driven operators acting on the state dictating its trajectory and evolution \cite{ghattas2021learning}. Examples of non-intrusive PMORs can be seen in \cite{peherstorfer2016data, schmid2010dynamic, brunton2016discovering, williams2015data, mezic2013analysis, haller2016nonlinear,geelen2023operator, khodabakhshi2022non,mcquarrie2023nonintrusive}.  

Intrusive PMORs are the focus of the present work. Unlike their non-intrusive counterparts, intrusive PMORs require access to the underlying infrastructure of the numerical method at hand. For intrusive frameworks, the data-driven trial basis is directly embedded into the numerical infrastructure as part of an expansion representing a low-dimensional approximation of the solution state vector. The embedded trial basis produces a system of equations that approximates the FOM. The approximate system of equations is then projected onto the test basis, which gives way for a low-dimensional representation of the FOM. Examples of intrusive PMORs include, Galerkin POD \cite{benner2017model, carlberg2017galerkin,rowley2004model}, balanced POD \cite{willcox2002balanced,rowley2005model}, the reduced-basis method \cite{benner2017model, rozza2022advanced}, and least-squares Petrov--Galerkin projection \cite{carlberg2011efficient,carlberg2013gnat,amsallem2012nonlinear}, among many others. It is important to note, however, that the action of the test basis on the embedded system of equations does not guarantee that the resulting PMOR will be a low-dimensional system of equations due to the potential for persistent nonlinear and parametric dependencies in the numerical infrastructure that scale with the dimensionality of the FOM. Hyper-reduction, also known as empirical interpolation, can be performed to alleviate persistent high-dimensional scaling in the PMOR. In essence, hyper-reduction serves as empirical quadrature, which interpolates the high-dimensional nonlinear and parametric dependencies left over from the test basis projection on a sparse set of control points in the numerical domain. The ``empirical" component of hyper-reduction stems from its approach in leveraging the data-derived test and trial basis functions to achieve accurate approximates of the nonlinear components in the PMOR. Examples of hyper-reduction or empirical interpolation, include the Gauss--Newton method with Approximated Tensors (GNAT) \cite{carlberg2011efficient,carlberg2013gnat}, energy-conserving sampling and weighting (ECSW) \cite{farhat2014dimensional,farhat2015structure}, and the Empirical Interpolation method (EIM) \cite{barrault2004empirical} and its variants; DEIM \cite{chaturantabut2010nonlinear,tiso2013discrete}, Q-DEIM \cite{drmac2016new}, and Empirical Quadrature Procedure (EQP) \cite{yano2019lp}. 

It is important to highlight that most non-intrusive and intrusive PMOR frameworks have been developed with mesh-based numerical methods in mind. In fact, very few published works exist on non-intrusive PMOR methods for meshless numerical methods, and even fewer exist for intrusive approaches. Some recent works on meshless PMOR are as follows. The ``projection-tree" reduced-order modeling (PTROM), introduced in \cite{rodriguez2022projection}, is an intrusive PMOR framework that combined ideas from hierarchical decomposition and PMORs to enable rapid computations of meshless \emph{N-}body problems. However, their approach did not account for significant mixing in the numerical topology. Work presented in \cite{rodriguez2022multiscale} has also incorporated the least-squares Petrov--Galerkin projection equipped with GNAT hyper-reduction into SPH for problems modeling heat-deposition processes. However, the SPH formulation in their study accounted for the SPH framework cast in an Eulerian reference frame, which did not encounter any issues with dimensionality reduction of unstructured, dynamic, and mixing numerical topology. Recently, the work presented in \cite{chen2023model} developed a PMOR approach for the meshless material point method (MPM), where an implicit neural representation approach approximates the continuous deformation map that provides the method with a resolution-agnostic point of view. The resulting deformation map enables the computation of the dynamics in a low-dimensional manifold over a hyper-reduced sampled set of MPM particles. Their method demonstrated good accuracy and an order of magnitude speed-up relative to their FOM. However, their work was restricted to the infrastructure of the MPM method and to elastic problems in solid mechanics. Follow-up work in \cite{chen2022crom, chen2022multiscaling} presented a non-intrusive and continuous reduced-order modeling approach that aims to solve PDEs from discretization-agnostic training data. 

\subsection{Contributions}\label{Section:Introduction_Contributions}

The scope of the current work is concerned with the development of an intrusive PMOR framework for meshless methods. Our work aims to develop a model-order reduction infrastructure that is generalizable across meshless numerical frameworks and maintains the ubiquitous infrastructure of intrusive PMORs. Specifically, the proposed method seeks an approximate solution of the meshless system of equations by embedding a data-driven trial basis into the meshless FOM, followed by projecting the approximate system onto a low-dimensional subspace. The objective of our proposed method is equivalent to executing traditional intrusive PMORs as done for FE, FV, and FD methods. However, the challenge with na{\"i}vely exercising this traditional approach, which is addressed in our method, is that the resulting dimensionality reduction is severely impeded and rendered inefficient by the unstructured, dynamic, and mixing numerical topology, resulting in a slow-decaying Kolmogorov \emph{n}-width. Our work showcases these limitations on three benchmark problems that are widely used in the literature to test the robustness of SPH frameworks due to significant mixing in numerical topology and/or sharp boundaries and corners: 1) the Taylor--Green vortex; 2) the lid-driven cavity flow; and 3) flow past an open cavity. To overcome the aforementioned limitation, the present work introduces a \emph{reference space}, where meshless dynamics are mapped to perform dimensionality reduction, and then mapped back to exercise a meshless PMOR. For the remainder of this paper, a meshless PMOR is implied to be an intrusive approach. 

The contributions of our work include: 
\begin{enumerate}
	\item A meshless PMOR formulation that inherits the structure of traditional intrusive Galerkin and/or Petrov--Galerkin PMOR frameworks while retaining inherent qualities of the underlying meshless numerical method.
	\item A mathematical construct, known as a ``reference space", that enables effective dimensionality reduction of meshless numerical methods with unstructured, dynamic, and mixing numerical topology.
	\item A mapping that evolves the meshless dynamics in a low-dimensional subspace.
	\item Numerical examples of the proposed meshless PMOR method.
\end{enumerate}

It is important to note that, to the best of the authors' knowledge, the current work marks a first step enabling intrusive meshless PMOR. Therefore, we demonstrate this method on canonical numerical examples where solutions and behaviors are well understood. Future work will consider problems where meshless methods provide invaluable capabilities, such as free-surface flows and multiphase flows. Finally, the scope of the present work is to develop an approximation infrastructure via meshless projection, i.e., to construct low-dimensional approximations of SPH solutions. Therefore, hyper-reduction to achieve efficient and rapid cost savings is not included in this work but is currently under development. 

The remainder of this paper is structured as follows. Section \ref{Section:Problem_formulation} introduces the SPH formulation employed, which is the meshless FOM of choice in the current work. Section \ref{Section:PBROM} presents the proposed meshless PMOR, including the construction of reference spaces and the composition operators that evolve meshless dynamics in a low-dimensional subspace. Numerical experiments are showcased in Section \ref{Section:NumericalExperiments}. Finally, conclusions and outlooks are provided in Section \ref{Section:Conclusion}.
\\
\\
Mathematical notation: matrices are bold-italicized uppercase letters (e.g., $\bm{X}$ or $\bm{\Psi}$) and vectors are bold-italicized lowercase letters (e.g., $\bm{x}$ or $\bm{\phi}$). Italicized lowercase and uppercase letters (e.g., $t$, $\rho$, $N$, and $M$) are scalars, except as subscripts or superscripts which correspond to tensor, scalar indexing, or dimensional descriptors. Vector spaces and special operators are denoted by calligraphic letters (e.g., $\mathcal{G}$). Matrix functions and vector functions are denoted with parentheses following their corresponding letters (e.g., $\bm{\Psi}(\bm{x})$ or $\bm{x}(t)$).

\section{Background: SPH full order model} \label{Section:Problem_formulation}

The proposed work employs the meshless SPH method. A brief overview of SPH is provided herein for context and to illustrate the use of notation. The current work adopts and modifies SPH notation of \cite{sigalotti2021mathematics} to fit the present effort. This section provides essential background material that we leverage extensively to build our reduced-order modeling framework. The reader is referred to reference texts \cite{liu2003smoothed} and \cite{violeau2012fluid} for more detail, and an approximation overview is provided by \cite{sigalotti2021mathematics}. 

\subsection{Function approximation}
The SPH approximation relies on the Dirac sifting property, where
\begin{equation}
	f(\bm{x})=\int_{\Omega} f(\bm{x}^{\prime})\delta(\bm{x}-\bm{x}^{\prime})d^n\bm{x}^{\prime}.
	\label{eqn:field_function}
\end{equation}
Here, $\bm{x}$ is an $n$-tuple of spatial dimensions $n$, where normally $n=1, 2,$ or $3$, i.e., $\bm{x}:=\lbrace{x,y,z}\rbrace$ in three-dimensional Cartesian space. Next, $f(\bm{x})\in \mathbb{R}$ is a smooth function in the domain $\Omega\subseteq\mathbb{R}^n$ evaluated at some arbitrary point in space, $\bm{x}\in \Omega$. In SPH, the Dirac distribution is generalized to a smoothing kernel, $W\in \mathcal{W}$, where $\mathcal{W}$ is the space of functions that satisfy the following conditions:

\begin{enumerate}
	\item $\int_{\Omega} W(\lVert \bm{x} - \bm{x}^{\prime} \rVert, h)d^n \bm{x}^{\prime}=1$ 
	\item $\underset{{h\rightarrow0}}{\text{lim}}W(\lVert \bm{x} - \bm{x}^{\prime} \rVert, h)=\delta(\bm{x}-\bm{x}^{\prime})$
	\item $W(\lVert \bm{x} - \bm{x}^{\prime} \rVert, h)\geq 0$
	\item $W(\lVert \bm{x} - \bm{x}^{\prime} \rVert, h)=W(\lVert \bm{x}^{\prime} - \bm{x} \rVert, h)$
	\item If $\lVert \bm{x} - \bm{x}^{\prime} \rVert \leq \lVert \bm{y} - \bm{y}^{\prime} \rVert$ then $ W(\lVert \bm{x} - \bm{x}^{\prime} \rVert, h) \geq W(\lVert \bm{y} - \bm{y}^{\prime} \rVert, h)  $
\end{enumerate}
Here, $\lVert \cdot \rVert= \lVert \cdot \rVert_2$ is the Euclidean norm unless stated otherwise. The smoothing kernel also takes in the argument, $h\in\mathbb{R}_+$, which is a characteristic length scale termed the \emph{smoothing length}. With these definitions the function in Eq.~\ref{eqn:field_function} can be approximated by $\langle f(\bm{x})\rangle \approx f(\bm{x})$, where
\begin{equation}	
	\langle f(\bm{x})\rangle:=\int_{\Omega} f(\bm{x}^{\prime})W(\lVert \bm{x} - \bm{x}^{\prime} \rVert, h)d^n\bm{x}^{\prime}.
	\label{eqn:smothed_function}
\end{equation}
Gradients of the function $f(\bm{x})$ can also be approximated, where $\langle \nabla f(\bm{x})\rangle \approx \nabla f(\bm{x})$, and
\begin{equation}	
	\langle \nabla f(\bm{x})\rangle:=\int_{\Omega} f(\bm{x}^{\prime})\nabla W(\lVert \bm{x} - \bm{x}^{\prime} \rVert, h)d^n\bm{x}^{\prime}.
	\label{eqn:smothed_function_gradient}
\end{equation}
The smoothing kernel to approximate $f(\bm{x})$ must be chosen with care, such that the kernel belongs to $\mathcal{W}$ while satisfying consistency relations in the Taylor expansions of Eq.~\ref{eqn:smothed_function}. Similar consistency relations must be satisfied by Eq.~\ref{eqn:smothed_function_gradient} so that it can approximate higher-order derivatives. Further details on consistency and convergence rates of the SPH approximation can be found in references \cite{sigalotti2021mathematics, violeau2012fluid, liu2003smoothed, quinlan2006truncation}.

\subsection{Discretization}
\begin{figure}[tbh]
	\hspace*{3.125cm}\includegraphics[trim = 4cm 2cm 0 0, scale=0.6]{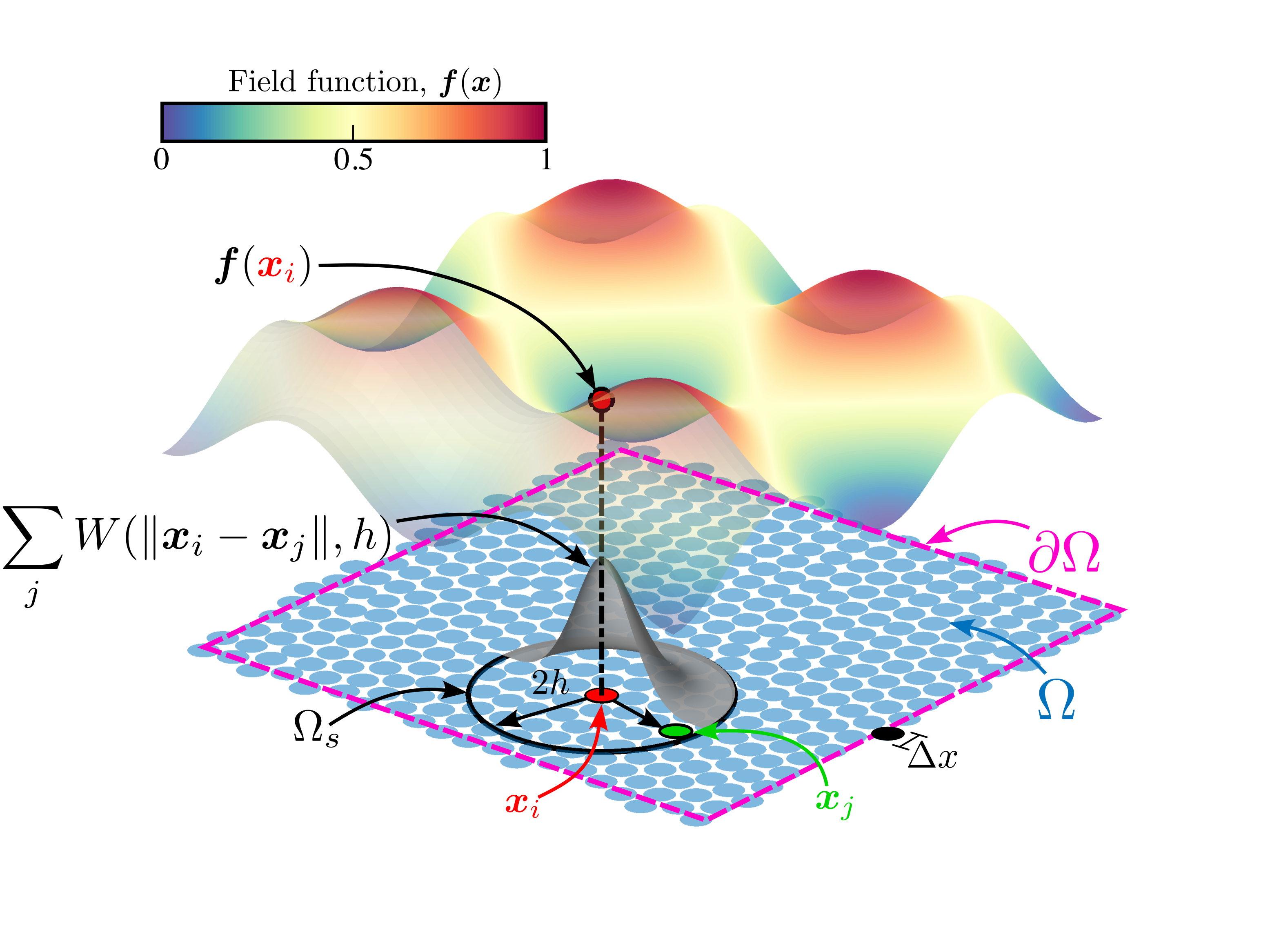}
	\caption{Illustration of the SPH numerical discretization of a periodic field function, $\bm{f}(\bm{x})$, over domain, $\Omega$, with boundary, $\partial \Omega$. The discretization of the field function is projected onto a two-dimensional plane, where each blue particle represents a subdomain, $\Omega_i$, with diameter, $\Delta x$, that has a kernel support, $\Omega_s$. } 
	\centering
	\label{fig:sphInterpolation_figure}
\end{figure}

Approximations of $f(\bm{x})$ and its gradient $\nabla f(\bm{x})$ have so far been cast in a continuum, $\Omega$. The SPH discretization divides the continuum into $N\in\mathbb{N}$ subdomains, of infinitesimal diameter $\Delta x=V_i^{1/n}$ with infinitesimal volume (considered a numerical weight) $V_i\in\mathbb{R}_+$. Here, $\Omega_i$, corresponds to the $i$\textsuperscript{th} subdomain, which contains an integration point referred to as a \emph{particle} in SPH, where the particle is located at $\bm{x}_i\in\Omega_i$ and $i=1, \ldots, N$. The set of all particles is denoted as $\mathcal{N}$. Each particle and its subdomain is free to move in space. The subdomain has a boundary, $\partial \Omega_i$. Every particle has an isotropic kernel support, $\Omega_s$, with a radius of $r_{\Omega_s}:=\kappa h$, where $\kappa\in\mathbb{R}_+$. The kernel support follows that $W(\lVert \bm{x} - \bm{x}^{\prime} \rVert, h)=0$ when $\lVert \bm{x} - \bm{x}^{\prime} \rVert>r_{\Omega_s}$. Every particle, $j\in\mathcal{N}$ that falls within the kernel support of the $i$\textsuperscript{th} particle is considered a neighboring particle and belongs to the set $\mathcal{N}_{\Omega_s}^i$. Therefore, the discrete form of Eq.~\ref{eqn:smothed_function} for a point $\bm{x}_i$ in space can be numerically approximated by 
\begin{equation}
	f_i:=	\langle f(\bm{x}_i)\rangle=\int_{\Omega} f(\bm{x}^{\prime})W(\lVert \bm{x}_i - \bm{x}^{\prime} \rVert, h)d^n\bm{x}_j\approx \sum_{j}^{\mathcal{N}_{\Omega_s}^i} f(\bm{x}_j)W_{ij}(\lVert \bm{x}_i - \bm{x}_j\rVert, h)V_j.
	\label{eqn:smoothing_function_interpolation}
\end{equation}
Here, the kernel takes in discrete arguments, hence the subscript notational convention $W_{ij}(\lVert \bm{x}_i - \bm{x}_j\rVert, h)$ is adopted. The gradient of a function can similarly be approximated by 
\begin{equation}
	\nabla f_i:=	\langle \nabla f(\bm{x}_i)\rangle=\int_{\Omega} f(\bm{x}^{\prime})\nabla W(\lVert \bm{x}_i - \bm{x}^{\prime}\rVert, h)d^n\bm{x}^{\prime}\approx \sum_{j}^{\mathcal{N}_{\Omega_s}^i} f(\bm{x}_j)\nabla_i W_{ij}(\lVert \bm{x}_i - \bm{x}_j \rVert, h)V_j,
	\label{eqn:smoothing_function_gradient_interpolation}
\end{equation}
where the $n-$dimensional Cartesian gradient operator, $\nabla_i:=\{\frac{\partial}{\partial x_i^n}\}$, acts on the $i$\textsuperscript{th} particle position vector. It is important to highlight that Eq.~\ref{eqn:smoothing_function_interpolation} is a summation interpolation as is Eq.~\ref{eqn:smoothing_function_gradient_interpolation}, where the product of $f(\bm{x}_j)V_j$ weighs the interpolation function, $W_{ij}(\lVert \bm{x}_i - \bm{x}_j \rVert, h)$. This interpolatory property is a cornerstone in the development of reference spaces in Section \ref{Section:PBROM}. Figure \ref{fig:sphInterpolation_figure} illustrates the SPH approximation and interpolation scheme. Finally, for notational convenience the upper limit ${\mathcal{N}_{\Omega_s}^i}$ in the summation interpolation and the arguments in the smoothing kernel are implied and omitted for the remainder of this paper. 

\subsection{The $\delta^{+}-$SPH formulation for the weakly compressible Navier--Stokes equations}
The SPH approximation and discretization presented in the previous subsection can be employed to solve a variety of partial differential equations numerically, including the model equation of elasticity, magnetohydrodynamics, and fluid dynamics, among many others \cite{price2012smoothed,liu2003smoothed,violeau2012fluid,dominguez2022dualsphysics}. The current work focuses on fluid dynamics problems, where the weakly compressible Navier--Stokes  (WCNS) equations are discretized using the $\delta^{+}-$SPH formulation presented in \cite{sun2019consistent}, due to its wide application in the SPH literature. The derivation of the adopted SPH formulation is beyond the scope of this work, but a brief description is presented. More detail can be found in \cite{sun2019consistent, antuono2021smoothed}. The governing system of equations employed in this work read as follows:

\vspace*{0.5cm}
\noindent \textbf{Continuity}
\begin{equation}
	\begin{split}
		\frac{\text{d}\rho_i}{\text{d}t} =&- \rho_i \sum\limits_{j} [(\bm{u}_j+\delta\bm{u}_j)-(\bm{u}_i+\delta\bm{u}_i)]\cdot\nabla_iW_{ij}V_j\\&+\sum_j(\rho_j\delta\bm{u}_j+\rho_i\delta\bm{u}_i)\cdot\nabla_iW_{ij}V_j+\delta h c_0\mathcal{D}_i,
	\end{split}
	\label{eqn:sph_wcns_continuity}
\end{equation}
\noindent \textbf{Momentum}
\begin{equation}
	\begin{split}
		\frac{\text{d}\bm{u}_i}{\text{d}t} =&-\frac{1}{\rho_i}\sum_{j} \left(p_i+p_j\right) \nabla W_{ij}V_j + \frac{\rho_0}{\rho_i}\sum_{j}\pi_{ij}\nabla W_{ij}V_j \\
		&+\frac{\rho_0}{\rho_i}\sum_{j}(\bm{u}_j\boldsymbol{\otimes}\delta\bm{u}_j+\bm{u}_i\boldsymbol{\otimes}\delta\bm{u}_i)\cdot\nabla_iW_{ij}V_j
		\\&-\frac{\rho_0}{\rho_i}\bm{u}_i\sum_{j}(\delta\bm{u}_j-\delta\bm{u}_i)\cdot\nabla_iW_{ij}V_j+\bm{b}_i,
	\end{split}
	\label{eqn:sph_wcns_momentum}
\end{equation}
\noindent \textbf{Particle advection, equation of state, and particle volume evolution}
\begin{equation}
	\begin{split}
		\frac{\text{d}\bm{x}_i}{\text{d}t} =& \bm{u}_i+\delta\bm{u}_i, \hspace*{0.5cm} p_i = c_0^2 \left(\rho_i-\rho_0\right), \hspace*{0.5cm} 	V_i = \frac{m_i}{\rho_i}.
	\end{split}
	\label{eqn:sph_wcns_advection_eos_volume}
\end{equation}

In the present work, we employ the $C^2$ quintic Wendland kernel \cite{wendland1995piecewise, violeau2012fluid} and a cell-linked list algorithm is used to enable an efficient neighbor search, presented later in Section \ref{Section:PBROM}. Here, $\lVert \bm{x}_{ij}\rVert:=\lVert \bm{x}_{i}-\bm{x}_{j}\rVert$ and $t\in\left[0, T_f \right]$ denotes time with the final time $T_f\in\mathbb{R}_+$. For particle $i$, the density of a particle is defined by $\rho_i\in\mathbb{R}_{+}$, its velocity is defined by $\bm{u}_i\in\mathbb{R}^n$, and its position in Cartesian space is $\bm{x}_i\in\mathbb{R}^n$. The velocity perturbation, $\delta\bm{u}_i\in\mathbb{R}^n$, is defined as a regularization technique known as particle shifting, which enables regular particle distribution during simulations to avoid tensile instabilities and improve fidelity \cite{sun2019consistent, sun2017deltaplus}. Here, $\bm{b}_i\in\mathbb{R}^n$, defines a body force. The current framework imbues the SPH material derivative with this velocity perturbation such that for an arbitrary field variable, $f$, 
\begin{equation}
	\frac{\text{d} f}{\text{d} t}:=\frac{\partial f}{\partial t}+\nabla f \cdot \left(\bm{u}+\delta \bm{u}\right)=\frac{D f}{D t}+\nabla f \cdot \delta \bm{u}.
\end{equation}
The particle shifting formulation will be defined momentarily. Next, we adopt a linear equation of state to define particle pressure, $p_i\in\mathbb{R}$ in a weakly compressible sense. Here,  $c_0\in\mathbb{R}_+$ is the speed of sound (assumed constant), and is chosen so that density variations stay within a weakly compressible range, i.e., below a 1\% density variation from its reference density $\rho_0\in\mathbb{R}_+$. The speed of sound is defined by $c_0:=\frac{U_{\text{max}}}{M_a}$, where $U_{\text{max}}$ is the maximum expected velocity in a simulation and $M_a\in\mathbb{R}_+$ is the mach number taken to be $M_a=0.1$ for the adopted weakly compressible assumption, unless stated otherwise. The mass of all particles is assumed fixed and determined by the initialized volume, $V_0\in\mathbb{R}_+$, and reference density, such that $m_i=\rho_0 V_0$. The initialized volume is determined by the subdomain volume where, $V_0:=\Delta x^2$ for two-dimensional problems or $V_0:=\Delta x^3$ in three-dimensions. The particle volume $V_i$ is then updated according to Eq. \ref{eqn:sph_wcns_advection_eos_volume}. 

Equation \ref{eqn:sph_wcns_continuity} contains a numerically diffusive component, $\delta h c_0 \mathcal{D}_i$, to reduce spurious numerical artifacts from the weak-compressibility assumption. Here, $\delta\in\mathbb{R}_+$ is a tuning parameter conventionally taken to be 0.1, and a diffusive smoothing operator, $\mathcal{D}_i$, defined by
\begin{equation}
	\mathcal{D}_i:=2 \sum_j \left[(\rho_j-\rho_i)-\frac{1}{2}\left(\langle\nabla \rho\rangle_j^L+\langle\nabla \rho\rangle_i^L\right)\cdot(\bm{x}_{j}-\bm{x}_{i})\right]\frac{(\bm{x}_{j}-\bm{x}_{i})}{\lVert \bm{x}_{ij} \rVert^2}\cdot\nabla_i W_{ij}V_j,\\
\end{equation}
where
\begin{equation}
	\langle\nabla \rho\rangle_i^L:=\sum_j(\rho_j-\rho_i)\bm{L}_i\nabla_iW_{ij}V_j, \:\:\text{and}\:\: \bm{L}_i:=\left[\sum_j(\bm{x}_{j}-\bm{x}_{i})\boldsymbol{\otimes}\nabla_iW_{ij}V_j\right] ^{-1}.
\end{equation}
Equation \ref{eqn:sph_wcns_momentum} contains the viscous operator, 
\begin{equation}
	\pi_{ij}:=K\frac{(\bm{u}_{j}-\bm{u}_{i})\cdot(\bm{x}_{j}-\bm{x}_{i})}{\lVert \bm{x}_{ij} \rVert^2},\:\: \text{where} \:\: K:=2(n+2) \frac{\mu}{\rho_0} . 
\end{equation}
The coefficient, $\mu\in\mathbb{R}_+$, is the dynamic viscosity, where $\nu=\frac{\mu}{\rho_0}$ is known as the kinematic viscosity. The presented work only considers laminar viscosity, but recent works on large eddy simulation (LES) for the current framework can be referenced in \cite{antuono2021smoothed}. Future work will consider LES-SPH modeling. 

Finally, the particle shifting velocity, $\delta\bm{u}_i$, is defined by 
\begin{equation}
	\delta\bm{u}_i:=\text{min}\left(\lVert \delta\breve{\bm{u}}_i \rVert, \frac{U_{\text{max}}}{2} \right)\frac{\delta\breve{\bm{u}}_i}{\lVert \delta\breve{\bm{u}_i} \rVert},
    \label{eq:shifting}
\end{equation}
where, 
\begin{equation}
	\delta\breve{\bm{u}}_i :=-2h c_0 M_a \sum_j\left[1+\chi\left(\frac{W_{ij}}{W(\Delta x)}\right)\right]^{\xi}\nabla_i W_{ij}V_j.
\end{equation}
The constants $\chi$ and $\xi$ are set to 0.2 and 4, respectively. The kernel that takes in the subdomain length is defined similarly to Eqs.~\ref{eqn:smothed_function} and \ref{eqn:smothed_function_gradient} except $W(\Delta x)\equiv W(\Delta x,h)$ for all particles. Careful consideration must be taken and corrections must be implemented for particle shifting at and near domain boundaries including free-surfaces, where details on these corrections can be found in \cite{sun2019consistent}. 
%
\subsection{Boundary conditions}
For wall-bounded domains, we adopt the formulation presented by \cite{adami2012generalized}. The present work does not consider dynamic boundaries. In the adopted approach, walls or boundaries are discretized by ``ghost" particles with the same initial spacing of the subdomain for each particle, $\Delta x$. Each ghost particle stays fixed in space but inherits fluid properties of the internal flow by means of interpolation over particles within their kernel support.  Specifically, let the set of ghost particles be defined by $\mathcal{N}_{g}$ where $\mathcal{N}_{g}\cap\mathcal{N}=\emptyset$, then for $i\in\mathcal{N}_g$ and $j\in\mathcal{N}$

\begin{equation}	
	\bm{u}_i=\frac{\sum\limits_j \bm{u}_jW_{ij}}{\sum\limits_j W_{ij}}, \:\:\: p_i=\frac{\sum\limits_j p_jW_{ij}+\bm{b}\cdot\sum\limits_j \rho_j(\bm{x}_i-\bm{x}_j)W_{ij}}{\sum\limits_j W_{ij}},
\end{equation}
where the density of the ghost particles can be back-tracked by the equation of state, such that
\begin{equation}	
	\rho_i=\frac{p_i}{c_0^2}+\rho_0.
\end{equation}
The velocity of the boundary particles can be split into tangent and normal components, $\bm{u}_i:=\bm{u}_i^{\tau} + \bm{u}_i ^{n} $. For slip boundary conditions, the directions of both tangent and normal components remains unchanged. For no-slip boundary conditions, the tangent component is negated, such that $\bm{u}_i=-\bm{u}_i^{\tau} + \bm{u}_i ^{n}  $.

\subsection{Dynamical system abstraction}
The current work rewrites Eqs.~\ref{eqn:sph_wcns_continuity}-\ref{eqn:sph_wcns_advection_eos_volume} in vector form, as it becomes a primary construct in presenting the proposed meshless PMOR method later.  Here, the vector $\bm{w}_i:=\left\{\rho_i, \bm{u}_i^T, \bm{x}_i^T\right\}^T$, contains $d-$states or $d-$degrees-of-freedom of each particle, where $d=2n+1$ in the current work. Then the FOM state vector reads as $\bm{w}:=\left\{\bm{w}_1^T,\ldots, \bm{w}_N^T\right\}^T\in\mathbb{R}^{N_d}$, where $N_d=dN$. Using this notation, Eqs.~\ref{eqn:sph_wcns_continuity}-\ref{eqn:sph_wcns_advection_eos_volume}, can be rewritten as
\begin{equation}
	\frac{\text{d}\bm{w}}{\text{d}t}=\bm{f}(\bm{w}, t; \bm{\mu}),\:\:\: \bm{w}(0,\bm{\mu})=\bm{w}^0(\bm{\mu}),
	\label{eqn:sph_dynamical_system}
\end{equation}
where $\bm{w}:[0, T_f] \times \mathcal{P} \rightarrow \mathbb{R}^{N_d}$ is the time-dependent, parameterized state that is implicitly defined as the solution to Eq.~\ref{eqn:sph_dynamical_system}, with parameters $\bm{\mu}\in\mathcal{P}$. The parametric space of $n_{\mu}$ parameters is denoted by $\mathcal{P}\subseteq \mathbb{R}^{n_{\mu}}$ and $\bm{w}^0:\mathcal{P}\rightarrow\mathbb{R}^{N_d}$ is the parametrized initial condition. Finally, $\bm{f}: \mathbb{R}^{N}\times[0, T_f]\times\mathcal{P}\rightarrow\mathbb{R}^{N_d}$ denotes the semi-discrete SPH functional, i.e., the right-hand side of Eqs.~\ref{eqn:sph_wcns_continuity}-\ref{eqn:sph_wcns_advection_eos_volume} in vector form.

\subsection{Time integration}
The present work adopts a fourth-order Runge--Kutta (RK4) time integration scheme. For notational convenience, the temporal and parametric arguments of the SPH functional are suppresed. The time-discrete form of Eq.~\ref{eqn:sph_dynamical_system} reads as

\begin{equation}
	\begin{split}
		&\bm{w}^{\tilde{n},0}=\bm{w}^{\tilde{n}},\\
		&\bm{w}^{\tilde{n},1}=\bm{w}^{\tilde{n},0}+\frac{\Delta t}{2}\bm{f}(\bm{w}^{\tilde{n},0}),\\
		&\bm{w}^{\tilde{n},2}=\bm{w}^{\tilde{n},0}+\frac{\Delta t}{2}\bm{f}(\bm{w}^{\tilde{n},1}),\\
		&\bm{w}^{\tilde{n},3}=\bm{w}^{\tilde{n},0}+{\Delta t}\bm{f}(\bm{w}^{\tilde{n},2}),\\
		&\bm{w}^{\tilde{n}+1}=\bm{w}^{\tilde{n},0}+\frac{\Delta t}{6}\left[\bm{f}(\bm{w}^{\tilde{n},0}) + 2\bm{f}(\bm{w}^{\tilde{n},1}) + 2\bm{f}(\bm{w}^{\tilde{n},2}) + \bm{f}(\bm{w}^{\tilde{n},3})\right],\\		
	\end{split}
	\label{eqn:sph_rungekutta}
\end{equation}
where the first superscript in $\bm{w}^{\tilde{n},k}$ denotes the $\tilde{n}$\textsuperscript{th} time-step, where $\tilde{n}\in\mathbb{N}(N_t)$, $N_t\in\mathbb{N}$ denotes the final number of time-steps taken, and $\mathbb{N}(N_t):=\{1,\ldots,N_t\}$. The second superscript, $k\in\mathbb{N}$, in the state vector denotes the $k$\textsuperscript{th} Runge-Kutta sub-step. The absence of a second superscript denotes a whole time step. Finally, the current work adopts uniform time-steps and a conservative CFL condition

\begin{equation}
	\Delta t\leq CFL \frac{h}{c_0},
\end{equation}
where $CFL=1.5$. Future work will investigate employing adaptive time-stepping in PMORs, as it is traditionally done for SPH. 
%
\section{Projection-based reduced order modeling} \label{Section:PBROM}
The objective of the proposed work is to transform the $N_d$-dimensional dynamical system presented in Eq.~\ref{eqn:sph_dynamical_system} into a $M$-dimensional dynamical system, where $M\ll N_d$. However, a main challenge with constructing a transformation on a low-dimensional subspace is the unstructured, dynamic, and mixing numerical topology, namely the particle positions, $\bm{x}:=\{\bm{x}_1^T, \ldots, \bm{x}_N^T\}$, where $\bm{x}:[0, T_f] \times\mathcal{P}\rightarrow \mathbb{R}^n$. As particles mix, their trajectories may follow a chaotic path, where a perturbation to the velocity or pressure fields can lead to significant trajectory deviations between initial neighbors. Thereby, initial neighboring particles may encounter vastly different flow characteristics as time evolves, which severely impacts the subspace dimensionality in which the dynamical system evolves; this will be quantitatively shown in Section \ref{Section:NumericalExperiments}. To combat the unstructured, dynamic, and mixing particle behavior, the proposed framework treats particles as topological field probes, separate from the low-dimensional subspace where the density and velocity field evolve. Thereby, the low-dimensional manifold that is to be discovered \emph{excludes} the dynamic contributions of the particles themselves. 

The reformulated dynamical system is now presented. However, before proceeding, it is important to note that while the presented SPH equations are in Lagrangian form, the model reduction framework projects the low-dimensional dynamics onto an Eulerian reference frame, which requires advecting modal information through the reference space. In the reformulated dynamical system, let the vector $\bm{\omega}_i:=\{\rho_i,\bm{u}_i^T\}^T$ contain $\bar{d}$-states for particle $i$, where $\bar{d}=n+1$ in the current work for two dimensions. Then the restructured state vector reads as $\bm{\omega}:=\{\bm{\omega}_1^T, \ldots, \bm{\omega}_N^T\}^T\in\mathbb{R}^{N_{\bar{d}}}$, where ${N_{\bar{d}}}=\bar{d}N$, and the resulting dynamical system reads:
\begin{equation}
	\frac{\text{d}\bm{\omega}}{\text{d}t}=\bar{\bm{f}}(\bm{x}, \bm{\omega},t;\bm{\mu}),\:\:\: \bm{\bm{\omega}}(\bm{x}^0,0,\bm{\mu})=\bm{\bm{\omega}}^0(\bm{x}^0,\bm{\mu}),
	\label{eqn:reference_dynamical_system}
\end{equation}
where the particle advection is updated by
\begin{equation}
	\frac{\text{d} \bm{x}_i}{\text{d} t}=\bm{u}_i+\delta \bm{u}_i, \:\:\: \bm{x}(0)=\bm{x}^0.
\end{equation}
Here, $\bm{\omega}: [0, T_f]\times \mathcal{P} \rightarrow \mathbb{R}^{N_{\bar{d}}}$ is the restructured state and $\bar{\bm{f}}: \mathbb{R}^{{nN}}\times \mathbb{R}^{\bar{d}} \times [0, T_f] \times \mathcal{P} \rightarrow \mathbb{R}^{N_{\bar{d}}}$ denotes  Eqs.~\ref{eqn:sph_wcns_continuity}-\ref{eqn:sph_wcns_advection_eos_volume} \emph{excluding} particle advection but \emph{includes field variable advection}. Finally, $\bm{\omega}^0:\mathcal{P}\rightarrow\mathbb{R}^{N_{\bar{d}}}$ is the parametrized initial condition of the reference state, and $\bm{x}^0 \in \mathbb{R}^{nN}$ is the initial particle configuration.

The projection of the particle dynamics onto a low-dimensional Eulerian reference space requires the modal advection to be subtracted from the SPH functional form, as defined by the material derivative. For an arbitrary field variable in the consistent $\delta^{+}$-SPH method \cite{sun2019consistent}, the quasi-Lagrangian material derivative is expressed as 
\begin{equation}
	\frac{\text{d} f}{ \text{d} t}=\frac{\partial f}{\partial t}+\nabla f \cdot \left(\bm{u}+\delta \bm{u}\right).
\end{equation}
Therefore, to solve the low-dimensional equations in reference space, the advection components of the quasi-Lagrangian material derivative must be subtracted from the SPH equations. In other words,

\begin{equation}
\underbrace{\frac{\partial f}{\partial t}}_{\text{reference space dynamics}}=\underbrace{\frac{\text{d} f}{ \text{d} t}}_{\delta^+\text{SPH}}-\underbrace{\nabla f \cdot \left(\bm{u}-\delta \bm{u}\right)}_{\text{field advection terms}}.
\end{equation}
However, particle shifting advection is maintained to enforce regular particle distribution. The following advection equations are subtracted from Eqs.~\ref{eqn:sph_wcns_continuity} and \ref{eqn:sph_wcns_momentum}, respectively, 
\begin{equation}
	\begin{split}
		\nabla \rho_i \cdot \bm{u}_i=\sum_{j}\left(\rho_j\bm{u}_j+\rho_i\bm{u}_i\right)\cdot \nabla W_{ij}V_j - \rho_i\sum_{j}\left(\bm{u}_j-\bm{u}_i\right)\cdot \nabla W_{ij}V_j 
	\end{split}
	\label{eqn:continuity advection}
\end{equation}
and
\begin{equation}
	\begin{split}
			\nabla \bm{u}_i \cdot \bm{u}_i=\sum_{j}\left(\bm{u}_j\otimes\bm{u}_j+\bm{u}_i\otimes\bm{u}_i\right)\cdot \nabla W_{ij}V_j -\bm{u}_i\sum_{j}\left(\bm{u}_j-\bm{u}_i\right)\cdot \nabla W_{ij}V_j, 
	\end{split}
	\label{eqn:momentum advection}
\end{equation}
which are known as the continuity advection and momentum advection equations. 
Subtracting Eqs.~\ref{eqn:continuity advection} and \ref{eqn:momentum advection} from \ref{eqn:sph_wcns_continuity} and \ref{eqn:sph_wcns_momentum}, respectively, defines the meshless reference space dynamics functional, $\bar{\bm{f}}$.

Next, the proposed PMOR seeks an approximate solution to the restructured dynamical system in Eq.~\ref{eqn:reference_dynamical_system}. Namely, $\tilde{\bm{\omega}}\approx\bm{\omega}$, where the approximate solution can be represented by a linear combination between a trial basis, $\bm{\Phi}\in\mathbb{R}^{N_{\bar{d}}\times M}$, and generalized coordinates, $\hat{\bm{\omega}}\in\mathbb{R}^{M}$,
\begin{equation}
	\tilde{\bm{\omega}}=\bm{\Phi}\hat{\bm{\omega}}.
	\label{eqn:affine_approximation}
\end{equation}	
The trial basis exists on the Stiefel manifold, i.e., for a full-column-rank matrix, $\bm{\mathcal{A}}\in\mathbb{R}^{q\times p}$, the Stiefel manifold is defined by $\mathcal{V}_p(\mathbb{R}^q)\equiv\{\bm{\mathcal{A}}\in\mathbb{R}^{q\times p}\vert \bm{\mathcal{A}}^T\bm{\mathcal{A}}=\bm{I}\}$, where ${\bm{\Phi}}\in\mathcal{V}_M(\mathbb{R}^{N_{\bar{d}}})$ and is comprised of $M$ orthonormal basis vectors,
\begin{equation}
	{\bm{\Phi}}\equiv\left[\bm{\phi}_1,\ldots, \bm{\phi}_M\right],
	\label{eqn:trial_basis}
\end{equation}
where $\bm{\phi}_i\in\mathbb{R}^{N_{\bar{d}}}$. Next, a test basis, $\bm{\Psi}\in{\mathbb{R}^{N_{\bar{d}}\times M}}$, is introduced and is assumed to also exist in the Stiefel manifold, ${\bm{\Psi}}\in\mathcal{V}_M(\mathbb{R}^{N_{\bar{d}}})$. The test basis is also composed of $M$ orthonormal basis vectors,
\begin{equation}
	{\bm{\Psi}}\equiv\left[\bm{\psi}_1,\ldots, \bm{\psi}_M\right],
\end{equation}
where $\bm{\psi}_i\in\mathbb{R}^{N_{\bar{d}}}$. Substituting Eq.~\ref{eqn:affine_approximation} in Eq.~\ref{eqn:reference_dynamical_system} and taking the $L^2$ inner product with $\bm{\Psi}$ yields 
\begin{equation}
	\bm{\Psi}^T\bm{\Phi}\frac{\partial \hat{\bm{\omega}}}{\partial t}=\bm{\Psi}^T\bar{\bm{f}}(\bm{\Phi}\hat{\bm{\omega}},t;\bm{\mu}).
	\label{eqn:L2_projected_RFOM}
\end{equation}
After some algebraic manipulation of Eq.~\ref{eqn:L2_projected_RFOM}, a low-dimensional dynamical system of equations is obtained, 
\begin{equation}
	\frac{\partial \hat{\bm{\omega}}}{\partial t}=	\hat{\bm{f}}(\bm{\Psi},\bm{\Phi}\hat{\bm{\omega}},t;\bm{\mu}),
	\label{eqn:mPMOR}
\end{equation}
where $\hat{\bm{f}}:=\lbrack\bm{\Psi}^T\bm{\Phi}\rbrack^{-1} \bm{\Psi}^T\bar{\bm{f}}(\bm{\Phi}\hat{\bm{\omega}},t;\bm{\mu})$. Finally, the particle advection is enforced by the velocity vector, $\tilde{\bm{u}}_i$, from the approximate solution in Eq.~\ref{eqn:affine_approximation}, 
\begin{equation}
	\frac{\text{d} \bm{x}_i}{\text{d} t}=\tilde{\bm{u}}_i+\delta \tilde{\bm{u}}_i.
	\label{eqn:mPMOR_particleAdvection}
\end{equation}
Note that $\delta \tilde{\bm{u}}_i$ is the particle shifting computed via Eq.~\ref{eq:shifting}, however, it is derived using the PMOR approximate state vector. To evolve the low-dimensional system in Eq.~\ref{eqn:mPMOR} and Eq.~\ref{eqn:mPMOR_particleAdvection} in time, both equations can be integrated separately by the RK4 method presented in Eq.~\ref{eqn:sph_rungekutta}.  

\subsection{Reference space}

\begin{figure}[t!]
\includegraphics[trim = -2.5cm 8cm 0cm 7cm, scale=0.5]{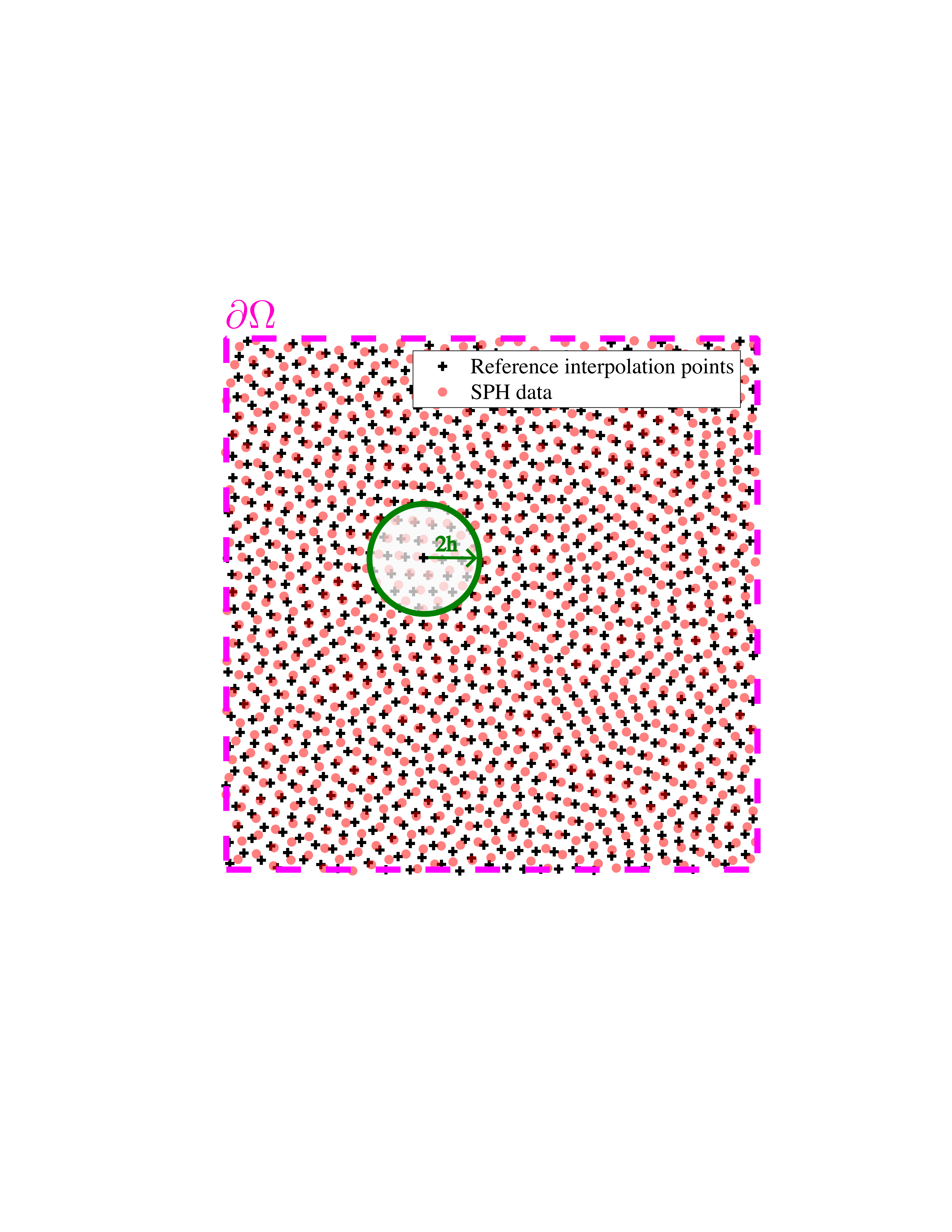}
	\caption{Illustration of the SPH reference space on a periodic domain. Field quantities from the Lagragian space (i.e., the SPH data) are mapped onto the reference space.} 
	\centering
	\label{fig:LEMmapping}
\end{figure}

To construct the trial basis we employ the POD. Traditionally, data snapshots of the solution vector are collected and stored in a matrix, $\bm{\mathcal{S}}\in\mathbb{R}^{N_{\bar{d}}\times N_{\mathcal{S}}}$, where  $N_{\mathcal{S}}\in\mathbb{N}$ denotes the number of snapshots stored. In the current work, temporal snapshots at some uniform spacing are considered, where $N_{\mathcal{S}}\in[1, N_t]$. For instance, the snapshot matrix for a uniform spacing of one time step reads, 
\begin{equation}
	\bm{\mathcal{S}}:=\left[\bm{\omega}^1, \bm{\omega}^2, \ldots, \bm{\omega}^{N_{t}-1}, \bm{\omega}^{N_{t}} \right].
	\label{eqn:snapshot_normal}
\end{equation}
However, because the numerical topology of the solution snapshots evolves dynamically and mixes, the proposed work introduces a composition operator, $\mathcal{G}:\mathbb{R}^{nN} \times \mathbb{R}^{N_{\bar{d}}}\rightarrow\mathbb{R}^{N_{\bar{d}}}$, that maps the snapshots onto a fixed, discrete, meshless domain, defined as the reference space. This space is defined by interpolation points belonging to the set, $\mathcal{N}_{\mathcal{G}}$, with position vector, $\bm{x}_{\mathcal{G}}\in\mathbb{R}^{nN}$. The configuration of the reference space can in principle be arbitrary, but the proposed work chooses either the initial configuration of SPH particles from the snapshot data or any ``relaxed" configuration, i.e., $\bm{x}_{\mathcal{G}}=\bm{x}^{0}$ or $\bm{x}_{\mathcal{G}}=\bm{x}^{\tilde{n}}$. The snapshot matrix in Eq.~\ref{eqn:snapshot_normal} can then be expressed in the reference domain as
\begin{equation}
	\bm{\mathcal{S}}_{\mathcal{G}}:=\left[\mathcal{G}(\bm{x}_{\mathcal{G}},\bm{\omega})^1, \mathcal{G}(\bm{x}_{\mathcal{G}},\bm{\omega})^2, \ldots, \mathcal{G}(\bm{x}_{\mathcal{G}},\bm{\omega})^{N_{t}-1}, \mathcal{G}(\bm{x}_{\mathcal{G}},\bm{\omega})^{N_{t}} \right].
	\label{eqn:snapshot_reference}
\end{equation}
The composition operator is defined by an interpolation of the FOM state vector, $\bm{\omega}$, onto the reference space via the SPH discrete approximation in Eq.~\ref{eqn:smoothing_function_interpolation} \cite{magargal2022lagrangian}. For instance, let, $\mathcal{G}(\bm{x}_{i,\mathcal{G}},\bm{\omega}_i):=\{\rho_{i,\mathcal{G}}, \bm{u}_{i,\mathcal{G}}^T \}^T$, where
\begin{equation}
	\begin{split}
		\rho_{i,\mathcal{G}}:=&\frac{\sum_j\rho_jW(\lVert \bm{x}_{i,\mathcal{G}}-\bm{x}_{j}\rVert, h)\Delta x^n}{\sum_jW(\lVert \bm{x}_{i,\mathcal{G}}-\bm{x}_{j}\rVert, h)\Delta x^n},\\
		\bm{u}_{i,\mathcal{G}}:=&\frac{\sum_j \bm{u}_jW(\lVert \bm{x}_{i,\mathcal{G}}-\bm{x}_{j}\rVert, h)\Delta x^n}{\sum_jW(\lVert \bm{x}_{i,\mathcal{G}}-\bm{x}_{j}\rVert, h)\Delta x^n},
	\end{split}
	\label{eqn:reference_operator}
\end{equation}
where the Shepard filter is employed to normalize particles with incomplete support, which normally exist near boundaries. Here, $\square{}_{i,\mathcal{G}}$ refers to the quantity (e.g., $\rho$, $\bm{u}$, or $\bm{x}$) at particle $i$ in the reference space, and $i\in\mathcal{N}_{\mathcal{G}}$ and $j\in\mathcal{N}$. Notice that Eq.~\ref{eqn:reference_operator} employs a fixed $n$-dimensional weight, $\Delta x^n$, equal to the subdomain diameter of the SPH discretization chosen for the reference domain. An illustration of the reference space mapping is shown in Fig.~\ref{fig:LEMmapping}. A trial basis can now be defined by factoring the reference space snapshot matrix via the singular value decomposition, 

\begin{equation}
	\bm{\mathcal{S}}_{\mathcal{G}}=\bm{U}\bm{\Sigma}\bm{V}^T,
\end{equation}
where $\bm{U}\in\mathcal{V}_{N_{\bar{d}}}(\mathbb{R}^{N_{\bar{d}}})$, $\bm{\Sigma}\equiv\text{diag}(\sigma_i)\in\mathbb{R}^{N_{\bar{d}}\times N_{\mathcal{S}}}$ has monotonocally decreasing diagonal entries, and $\bm{V}\in\mathcal{V}_{N_{\mathcal{S}}}(\mathbb{R}^{N_{\mathcal{S}}})$. Finally, the trial basis is defined by taking the $M$ left singular vectors of $\bm{U}$, such that $M\ll\text{min}(N_{\bar{d}},N_{\mathcal{S}})$, where Eq.~\ref{eqn:trial_basis} is equivalent to $\bm{\Phi}\equiv\lbrack\bm{U}^1,\dots,\bm{U}^M\rbrack$. Algorithm \ref{alg:reference space trial basis} provides the pseudo-code for the reference space mapping.

\begin{algorithm}
	\caption{Constructing a trial basis from reference space}\label{alg:reference space trial basis}
	\SetAlgorithmName{Algorithm}{}{}
	\SetKwInput{Input}{Input}
	\SetKwInput{Output}{Output}
	\SetKwFunction{FMain}{PolyharmonicSpline}
	\SetKwProg{Fn}{function}{:}{}
	\SetKwArray{State}{name}
	\Input{SPH simulation data, $\bm{\rho}\:\in\mathbb{R}^N$, $\bm{x},\bm{u}\:\in\mathbb{R}^{nN}$, and reference space domain, $\bm{x}_{\mathcal{G}}\in\mathbb{R}^{nN}$.}
	\Output{Reference trial basis, $\bm{\Phi}$}
	\begin{doublespace}
		\vspace*{-.1cm}
		\tcp{Map SPH data onto reference space \vspace*{-.2cm}}
		\For
		{$n=0$ \KwTo $N_t$}{
			\For{$i=0$ \KwTo $N$} 
			{
				\ForEach{$j$ \textbf{in} $\mathcal{N}_{\Omega_s}^i$} 
				{
				\begin{equation*}
					\begin{split}
						\rho_{i,\mathcal{G}}:=&\frac{\sum_j\rho_jW(\lVert \bm{x}_{i,\mathcal{G}}-\bm{x}_{j}\rVert, h)\Delta x^n}{\sum_jW(\lVert \bm{x}_{i,\mathcal{G}}-\bm{x}_{j}\rVert, h)\Delta x^n}\\
					\bm{u}_{i,\mathcal{G}}:=&\frac{\sum_j \bm{u}_jW(\lVert \bm{x}_{i,\mathcal{G}}-\bm{x}_{j}\rVert, h)\Delta x^n}{\sum_jW(\lVert \bm{x}_{i,\mathcal{G}}-\bm{x}_{j}\rVert, h)\Delta x^n}\\
					\end{split}
				\end{equation*}		
									\vspace*{-0.8cm} 
				}
			}
	     
	    }
	\tcp{Collect reference snapshot matrix}
	$\bm{\mathcal{S}}_{\mathcal{G}}:=\left[\mathcal{G}(\bm{x}_{\mathcal{G}},\bm{\omega})^1, \ldots, \mathcal{G}(\bm{x}_{\mathcal{G}},\bm{\omega})^{N_{t}-1}, \mathcal{G}(\bm{x}_{\mathcal{G}},\bm{\omega})^{N_{t}} \right],$

    $\text{where, } \mathcal{G}(\bm{x}_{i,\mathcal{G}},\bm{\omega}_i):=\lbrace\rho_{i,\mathcal{G}}, \bm{u}_{i,\mathcal{G}}^T \rbrace ^T.$
    \vspace*{0.125cm}
	
	\tcp{Compute singular value decomposition of reference snapshot matrix}
	$\bm{U}\bm{\Sigma}\bm{V}^T=\textup{SVD}\left(\bm{\mathcal{S}}_{\mathcal{G}}\right)$.\vspace*{0.125cm}
	
	\tcp{Define low-dimensional reference trial basis}
	$\bm{\Phi}\equiv\lbrack\bm{U}^1,\dots,\bm{U}^M\rbrack.$
	\end{doublespace}

\end{algorithm}

\subsection{Approximation of the trial basis}
The present work leverages the notion that the empirical trial basis becomes a continuous function of space (i.e., there is an interpolation point at every point in space) as $N\rightarrow\infty$. Clearly, it is not possible to construct a trial basis with infinitely many particles or interpolation points. So, the present work instead approximates the infinite dimensional trial basis as a function of space. Specifically, scattered data approximation \cite{wendland2004scattered} via polyharmonic spline interpolation (PSI) is adopted, but a myriad meshless interpolants can be employed such as the moving least-squares method or radial basis functions, and even deep learning methods could be used to learn spatially continuous functions. The proposed trial basis approximation enables the unstructured, dynamic, and mixing numerical topology to inherit modal quantities from reference space. On a final note, the PSI approach is employed over an SPH interpolation since its weights can be pre-computed, further reducing any overhead in PMOR computations.  

To construct a polyharmonic spline, data from the reference space trial basis is used to derive the interpolant weights. For each column, $m$, of the trial basis at the reference space point location, ${\bm{\phi}}^m(\bm{x}_{i, \mathcal{G}})$, the following interpolation is defined:

\begin{equation}
	{\bm{\phi}}^m(\bm{x}_{i, \mathcal{G}}):=\sum_{j}^{N}\xi_j^m\gamma(\lVert \bm{x}_{i, \mathcal{G}}- \bm{x}_{j, \mathcal{G}}\rVert)+\bm{\eta}^m\cdot\bm{q}_i,
	\label{eqn:PSI}
\end{equation}
where, $m\in\mathbb{N}(M)$. Here, $i\in\mathcal{N}_{\mathcal{G}}$ and $j\in\mathcal{N}_{\mathcal{G}}$. The scalar values $\xi_j^m \in \mathbb{R}$ are the unknown interpolation weights of the radial basis function,  $\bm{\eta}^m:=\lbrace \eta_1^m,\ldots, \eta_{n+1}^m \rbrace^T$ is a $(n+1)$-tuple of the unknown real-valued weights of the polynomial basis, and $\bm{q}_i:=\lbrace 1, \bm{x}_{i,\mathcal{G}}^T \rbrace^T$ is the polynomial basis constructed from the interpolation points in reference space. The function, $\gamma: \lbrack0,\infty)\rightarrow\mathbb{R}$, is taken to be a polyharmonic radial basis function, 
\begin{equation}
	\gamma(\lVert \bm{x}_{i, \mathcal{G}}- \bm{x}_{j, \mathcal{G}}\rVert)=
	\begin{cases}
		\lVert \bm{x}_{i, \mathcal{G}}- \bm{x}_{j, \mathcal{G}}\rVert^k & \text{when} \:\: k=1,3,5, \ldots\\
		\lVert \bm{x}_{i, \mathcal{G}}- \bm{x}_{j, \mathcal{G}}\rVert^k \ln(	\lVert \bm{x}_i- \bm{x}_{j, \mathcal{G}}\rVert)& \text{when} \:\: k=2,4,6, \ldots
	\end{cases}.
\end{equation}
Obtaining the unknown interpolant weights, $\bm{\xi}^m:=\lbrace\xi_1^m,\ldots,\xi_{N}^m\rbrace^T$ and $\bm{\eta}^m$, in Eq.~\ref{eqn:PSI} requires solving a linear system of equations in the reference space for known modal quantities at the locations of the reference particles. The linear system reads as 

\begin{equation}
	\left[
	\begin{matrix}
		\bm{\Gamma} & \bm{{Q}} \\ 
		\bm{{Q}}^T &\bm{0}
	\end{matrix}
	\right]
	\left[
	\begin{matrix}
		\bm{\xi}^m \\ 
		\bm{\eta}^m 
	\end{matrix}
	\right]=
	\left[
	\begin{matrix}
		\bm{\phi}^m \\ 
		\bm{0}			
	\end{matrix}
	\right],
	\label{eqn:PSI_linear}
\end{equation}
where $\bm{\Gamma}_{ij}:=\gamma(\lVert \bm{x}_{i, \mathcal{G}}- \bm{x}_{j, \mathcal{G}}\rVert)$, $\bm{Q}:=\lbrace \bm{q}_1, \ldots, \bm{q}_N  \rbrace^T$, and $\bm{\phi}^m\equiv \bm{\phi}^m(\bm{x}_{\mathcal{G}})$. Once the weights are defined for each mode, the full trial basis can be approximated in \emph{meshless space}, that is, $\tilde{\bm{\Phi}}\approx{\bm{\Phi}}$, and $\tilde{\bm{\Phi}}(\bm{x}):=\lbrack \tilde{\bm{\phi}}(\bm{x})^1, \ldots, \tilde{\bm{\phi}}(\bm{x})^M\rbrack$. To approximate modal quantities, one simply calls Eq.~\ref{eqn:PSI} for each mode at arbitrary locations in the \emph{meshless space}, $\bm{x}_{i}$, where $i\in\mathcal{N}$, such that 

\begin{equation}
	\tilde{\bm{\phi}}^m(\bm{x}_{i}):=\sum_{j}^{N}\xi_j^m\gamma(\lVert \bm{x}_{i}- \bm{x}_{j, \mathcal{G}}\rVert)+\bm{\eta}^m\cdot\bm{q}_i.
	\label{eqn:PSI_approximate}
\end{equation}

\begin{figure}[tbh]
  \includegraphics[trim = -2.5cm 8cm 0 7cm, scale=0.5]{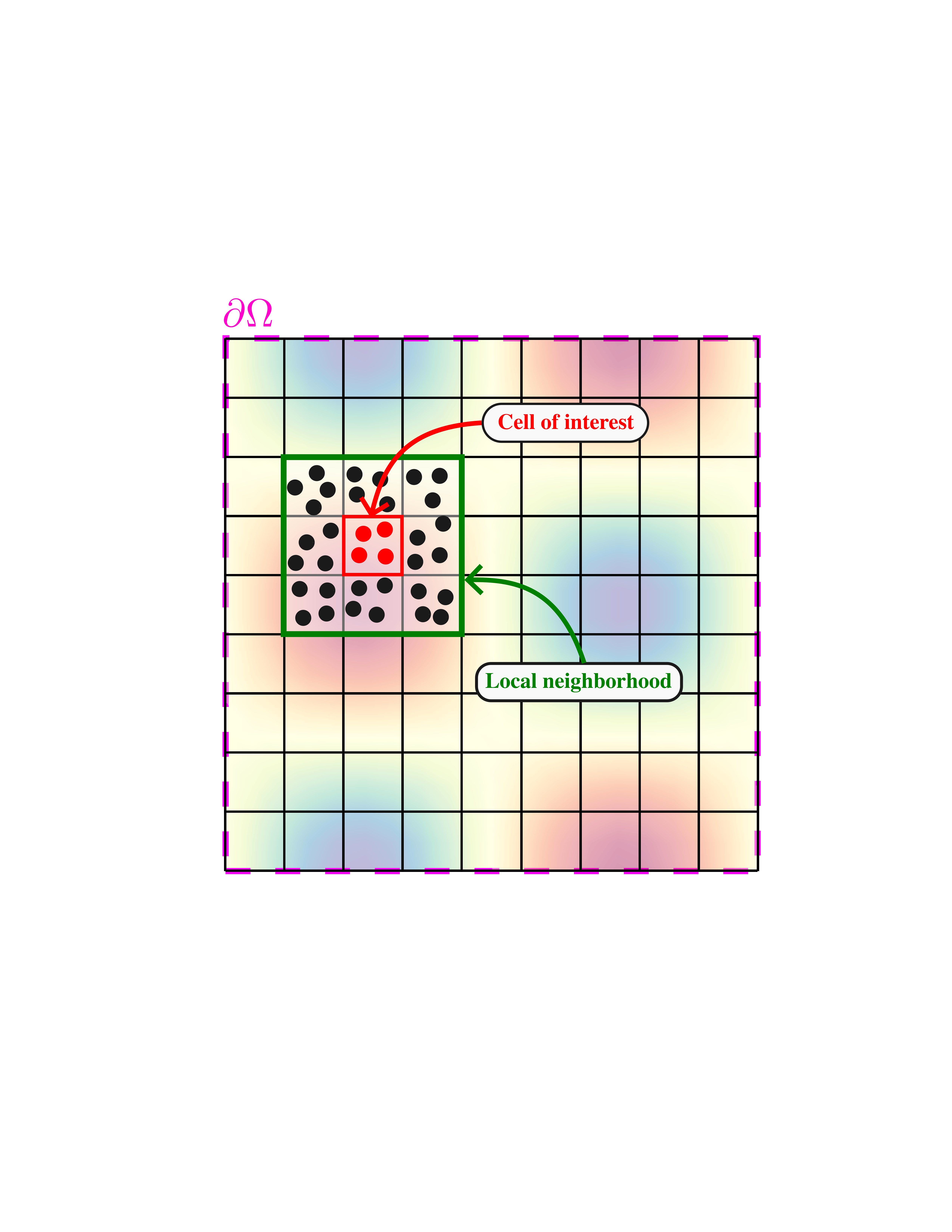}
	\caption{Illustration of a cell containing particles of interest (red) and particles inside neighboring cells (black particles). Local polyharmonic splines are constructed for each local neighborhood belonging to each cell of interest.} 
	\centering
	\label{fig:phsCells}
\end{figure}

	\vspace{0.25cm}
    \noindent \textbf{Remark.} \normalfont{Computing the weights in Eq.~\ref{eqn:PSI_linear} for the full $N$-dimensional reference domain is computationally expensive. The presented work avoids computing the dense linear system of equations over the global set of particles by leveraging the cell linked-list data structure used in the neighbor search algorithm for all simulations to construct local PSI systems. Specifically, local data points within the neighboring cells of a particle $i$ contribute to the PSI construction. Several local PSI systems are then constructed rather than one large and expensive PSI. Furthermore, adopting a local approach reduces the summation cost in Eq.~\ref{eqn:PSI} for online computations, where the summation is no longer done over the whole $N$-particle reference space but is instead performed online over a compact support of data points. For convenience, the cell linked-list algorithm employed is presented in Algorithm \ref{alg:cell linked-list}. An illustration and pseudo-code of the local PSI approach are also given in Fig.~\ref{fig:phsCells} and Algorithm \ref{alg:polyharmonic spline}, respectively.}

\begin{algorithm}[h!]
	\caption{Cell linked-list \cite{viccione2007fast}}\label{alg:cell linked-list}
	\SetAlgorithmName{Algorithm}{}{}
	\SetKwInput{Input}{Input}
	\SetKwInput{Output}{Output}
	\Input{Particle positions, $\bm{x}$; cell linked-list data structure, $\mathcal{\bm{C}}\in\mathbb{N}^K$ and $\mathcal{\bm{C}}^k\subset \mathcal{\bm{C}}$, where $K$ is the number of cells in the data structure and $k$ is the index of the $k$\textsuperscript{th} cell in the data structure. Note: the width of each cell is twice the smoothing length, i.e., $\alpha=2h$.}
	\Output{Populated cell data structure, $\mathcal{\bm{C}}$.}
	\begin{doublespace}
	\vspace*{-.1cm}
	\tcp{Populating cells with particle indices	\vspace*{-.2cm}}
			\For{$i=0$ \KwTo $N$} {
			{$\mathcal{\bm{C}}(i)=K\cdot\texttt{int}\left(\frac{x_i}{\alpha}\right)+\texttt{int}\left(\frac{y_i}{\alpha}\right)+1$}\\
			$\mathcal{\bm{C}}^k\leftarrow \mathcal{\bm{C}}(i)$ \Comment{Update the $k$\textsuperscript{th} cell index with the $i$\textsuperscript{th} particle} index}	
	\end{doublespace}   
{Note: to execute an efficient neighbor search, only search for neighboring particles in neighboring cells of the cell containing particle $i$.}
\end{algorithm}

\begin{algorithm}
	\caption{Trial basis polyharmonic spline construction}\label{alg:polyharmonic spline}
	\SetAlgorithmName{Algorithm}{}{}
	\SetKwInput{Input}{Input}
	\SetKwInput{Output}{Output}
	\SetKwFunction{FMain}{PolyharmonicSpline}
	\SetKwProg{Fn}{function}{:}{}
	\Input{Reference space particle positions, $\bm{x}_{\mathcal{G}}$; cell linked-list data structure, $\mathcal{\bm{C}}$, reference space trial basis, $\bm{\Phi}$.}
	\Output{Interpolant weights, $\bm{\xi}$ and $\bm{\eta}$}
	\begin{doublespace}
		\vspace*{-.1cm}
		\tcp{Construct polyharmonic spline interpolants \vspace*{-.2cm}}
	\ForEach{Cell \normalfont{\textbf{in}} {$\mathcal{C}$}}
	{
		\For{$m=0$ \KwTo $M$} 
		{
			Solve Eq.~\ref{eqn:PSI_linear} using reference neighboring points in neighboring cells
		}
	}
						\vspace*{-0.25cm}
	\end{doublespace}
\end{algorithm}

\subsection{A meshless Galerkin POD and adjoint Petrov--Galerkin PMOR}
The presented work adopts the traditional Galerkin POD (GPOD) and the adjoint Petrov--Galerkin (APG) PMOR \cite{parish2020adjoint}. In this work, the former employs the trial basis as the test basis itself, i.e., $\bm{\Psi}=\bm{\Phi}$. By employing the trial basis approximation, $ \tilde{\bm{\Phi}}\approx\bm{\Phi}$, the proposed meshless PMOR in Galerkin POD form reads as
\begin{equation}
	\frac{\partial \hat{\bm{\omega}}}{\partial t}=	 \tilde{\bm{\Phi}}^+\bar{\bm{f}}(\tilde{\bm{\Phi}}\hat{\bm{\omega}},t;\bm{\mu}).
	\label{eqn:mGPMOR}
\end{equation}
Here, the superscript, $+$, denotes the Moore-Penrose pseudo-inverse. 

For the adjoint Petrov--Galerkin PMOR, the test basis presented in \cite{parish2020adjoint} is adopted, where $\bm{\Psi}=\left[\left(\bm{I}-\tau\bm{\Pi}^{\prime T} \bm{J}^T\right)\bm{\Phi}\right]$, where $\bm{J}:=\frac{\partial \bar{\bm{f}}}{\partial \tilde{\bm{\omega}}}$ is the Jacobian of the SPH functional in the restructured dynamical system, of Eq.~\ref{eqn:reference_dynamical_system}, $\bm{\Pi}^{\prime}:=\left(\bm{I}-\bm{\Phi}\bm{\Phi}^{+}\right)$, and $\tau\in\mathbb{R}^+$ is a memory-length coefficient heuristically chosen. By employing the trial basis approximation, $ \tilde{\bm{\Phi}}\approx\bm{\Phi}$, the proposed meshless PMOR in adjoint Petrov--Galerkin form reads as
\begin{equation}
	\frac{\partial \hat{\bm{\omega}}}{\partial t}=	\tilde{\bm{\Phi}}^+\left(\bar{\bm{f}}(\tilde{\bm{\Phi}}\hat{\bm{\omega}},t;\bm{\mu})+\tau \bm{J}\bm{\Pi}^{\prime } \bar{\bm{f}}(\tilde{\bm{\Phi}}\hat{\bm{\omega}},t;\bm{\mu}) \right)
	\label{eqn:mAPGPMOR}
\end{equation}
The present work approximates the \emph{action} of the Jacobian and orthogonal projection onto the SPH funcational via $ \bm{J}\bm{\Pi}^{\prime } \bar{\bm{f}}(\tilde{\bm{\omega}})\approx \frac{1}{\varepsilon}\left[\bar{\bm{f}}(\tilde{\bm{\omega}}+\varepsilon\bm{\Pi}^{\prime } \bar{\bm{f}}(\tilde{\bm{\omega}}))-\bar{\bm{f}}(\tilde{\bm{\omega}})\right]$, where $\varepsilon\sim O(10^{-5})$ is a perturbation, and the temporal and parametric arguments have been omitted for notational convenience \cite{parish2020adjoint,an2011finite}. This approximation enables tractable computations of the APG method that would otherwise be expensive for an SPH framework, even with an analytical Jacobian, due to the non-local support of the SPH kernel. 

For both Galerkin and Petrov--Galerkin approaches, the particle advection is enforced by the velocity vector, $\tilde{\bm{u}}_i$, obtained from the back-projected approximate solution in Eq.~\ref{eqn:affine_approximation}, 
\begin{equation}
	\frac{\text{d} \bm{x}_i}{\text{d} t}=\tilde{\bm{u}}_i+\delta \tilde{\bm{u}}_i.
	\label{eqn:mGPMOR_particleAdvection}
\end{equation}
Again, to evolve the low-dimensional system in Eqs.~\ref{eqn:mGPMOR}, \ref{eqn:mAPGPMOR}, and Eq.~\ref{eqn:mGPMOR_particleAdvection} in time, the RK4 method presented in Eq.~\ref{eqn:sph_rungekutta} can be employed. Finally, Eq.~\ref{eqn:mGPMOR}/\ref{eqn:mAPGPMOR} and Eq.~\ref{eqn:mGPMOR_particleAdvection} together present a \emph{meshless} PMOR framework, which is the \textbf{\emph{main contribution}} of the presented work. Algorithm \ref{alg:meshlessPMOR} provides pseudo-code for the presented framework.

\vspace{0.25cm}
	\noindent \textbf{Remark.} \normalfont{The operator, $\bar{\bm{f}}$, has preserved all SPH operations in their meshless form. The operator $\tilde{\bm{\Phi}}$, derived from a reference space, also serves as a scattered-data interpolant, which is a meshless construct. It is important to note that the advection of the particles plays a significant role in enabling inherent numerical adaptivity, which is why the system maps the low-dimensional system in Eq.~\ref{eqn:mGPMOR} back to the meshless state via Eq.~\ref{eqn:affine_approximation}. However, the operational complexity to map back to the meshless state scales with $N_{\bar{d}}$ and will require hyper-reduction to avoid expensive back-projection. Unfortunately, hyper-reduction techniques for meshless methods are also severely underdeveloped and are beyond the scope of the presented work. Current efforts to develop these techniques are underway and will be the focus for future work.}

\begin{algorithm}
	\caption{Meshless PMOR algorithm}\label{alg:meshlessPMOR}
	\SetAlgorithmName{Algorithm}{}{}
	\SetKwInput{Input}{Input}
	\SetKwInput{Output}{Output}
	\SetKwFunction{FMain}{ProjectDynamics}
	\SetKwFunction{FMainBasis}{TrialBasisApproximation}
	\SetKwProg{Fn}{function}{(te):}{}
	\Input{Trial basis approximation, $\tilde{\bm{\Phi}}$, initial conditions $\hat{\bm{\omega}}^0$.}
\Output{Low-dimensional state history $\hat{\bm{\omega}}^1, \hat{\bm{\omega}}^2, \ldots, \hat{\bm{\omega}}^{N_t}$ }
\begin{doublespace}
	\vspace*{-.2cm}
	\For
	{$n=0$ \KwTo $N_t$}{
			\tcp{RK 4 Stage 1}
			{1. Get approximate trial basis, $\tilde{\bm{\Phi}}(\bm{x}^{\tilde{n},0})$, via Eq.~\ref{eqn:PSI_approximate}.}\\
			{2. Compute the full SPH state vector: $\tilde{\bm{w}}^{\tilde{n},0}=\tilde{\bm{\Phi}}\hat{\bm{\omega}}^{\tilde{n},0}$.}\\
			{3. Compute the SPH functional: $\bar{\bm{f}}(\bm{x}^{\tilde{n},0}, \tilde{\bm{w}}^{\tilde{n},0})$.}\\
			{4. $\hat{\bm{f}}(\hat{\bm{\omega}}^{\tilde{n},0})\:$= \FMain{$\tilde{\bm{\Phi}}$,$\bar{\bm{f}}(\bm{x}^{\tilde{n},0}, \tilde{\bm{w}}^{\tilde{n},0})$} See Alg.~\ref{alg:ProjectDynamics}.}\\
			{5. Stage 1 integration: $\hat{\bm{\omega}}^{\tilde{n},1}=\hat{\bm{\omega}}^{\tilde{n},0}+\frac{\Delta t}{2}\hat{\bm{f}}(\hat{\bm{\omega}}^{\tilde{n},0})$ and ${\bm{x}}^{\tilde{n},1}={\bm{x}}^{\tilde{n},0}+\frac{\Delta t}{2}\left(\tilde{\bm{u}}+\delta\tilde{\bm{u}}\right)^{\tilde{n},1}$}.
			
			\tcp{RK 4 Stage 2}
			{1. Get approximate trial basis, $\tilde{\bm{\Phi}}(\bm{x}^{\tilde{n},1})$, via Eq.~\ref{eqn:PSI_approximate}.}\\
			{2. Compute the full SPH state vector: $\tilde{\bm{w}}^{\tilde{n},1}=\tilde{\bm{\Phi}}\hat{\bm{\omega}}^{\tilde{n},1}$.}\\
			{3. Compute the SPH functional: $\bar{\bm{f}}(\bm{x}^{\tilde{n},1}, \tilde{\bm{w}}^{\tilde{n},1})$.}\\
			{4. $\hat{\bm{f}}(\hat{\bm{\omega}}^{\tilde{n},1})\:$= \FMain{$\tilde{\bm{\Phi}}$,$\bar{\bm{f}}(\bm{x}^{\tilde{n},1}, \tilde{\bm{w}}^{\tilde{n},1})$} See Alg.~\ref{alg:ProjectDynamics}.}\\
			{5. Stage 2 integration: $\hat{\bm{\omega}}^{\tilde{n},2}=\hat{\bm{\omega}}^{\tilde{n},0}+\frac{\Delta t}{2}\hat{\bm{f}}(\hat{\bm{\omega}}^{\tilde{n},1})$ and ${\bm{x}}^{\tilde{n},2}={\bm{x}}^{\tilde{n},0}+\frac{\Delta t}{2}\left(\tilde{\bm{u}}+\delta\tilde{\bm{u}}\right)^{\tilde{n},1}$}.
			
				\tcp{RK 4 Stage 3}
			{1. Get approximate trial basis, $\tilde{\bm{\Phi}}(\bm{x}^{\tilde{n},2})$, via Eq.~\ref{eqn:PSI_approximate}.}\\
			{2. Compute the full SPH state vector: $\tilde{\bm{w}}^{\tilde{n},2}=\tilde{\bm{\Phi}}\hat{\bm{\omega}}^{\tilde{n},2}$.}\\
			{3. Compute the SPH functional: $\bar{\bm{f}}(\bm{x}^{\tilde{n},2}, \tilde{\bm{w}}^{\tilde{n},2})$.}\\
			{4. $\hat{\bm{f}}(\hat{\bm{\omega}}^{\tilde{n},2})\:$= \FMain{$\tilde{\bm{\Phi}}$,$\bar{\bm{f}}(\bm{x}^{\tilde{n},2}, \tilde{\bm{w}}^{\tilde{n},2})$} See Alg.~\ref{alg:ProjectDynamics}.}\\
			{5. Stage 3 integration: $\hat{\bm{\omega}}^{\tilde{n},3}=\hat{\bm{\omega}}^{\tilde{n},0}+{\Delta t}\hat{\bm{f}}(\hat{\bm{\omega}}^{\tilde{n},2})$ and ${\bm{x}}^{\tilde{n},3}={\bm{x}}^{\tilde{n},0}+{\Delta t}\left(\tilde{\bm{u}}+\delta\tilde{\bm{u}}\right)^{\tilde{n},2}$}.
			
				\tcp{RK 4 Stage 4}
			{1. Get approximate trial basis, $\tilde{\bm{\Phi}}(\bm{x}^{\tilde{n},3})$, via Eq.~\ref{eqn:PSI_approximate}.}\\
			{2. Compute the full SPH state vector: $\tilde{\bm{w}}^{\tilde{n},3}=\tilde{\bm{\Phi}}\hat{\bm{\omega}}^{\tilde{n},3}$.}\\
			{3. Compute the SPH functional: $\bar{\bm{f}}(\bm{x}^{\tilde{n},3}, \tilde{\bm{w}}^{\tilde{n},3})$.}\\
			{4. $\hat{\bm{f}}(\hat{\bm{\omega}}^{\tilde{n},3})\:$= \FMain{$\tilde{\bm{\Phi}}$,$\bar{\bm{f}}(\bm{x}^{\tilde{n},3}, \tilde{\bm{w}}^{\tilde{n},3})$} See Alg.~\ref{alg:ProjectDynamics}.}\\
			{5. Stage 4 integration:\\ $\hat{\bm{\omega}}^{\tilde{n}+1}=\hat{\bm{\omega}}^{\tilde{n},0}+\frac{\Delta t}{6}\left[\hat{\bm{f}}(\hat{\bm{\omega}}^{\tilde{n},0}) + 2\hat{\bm{f}}(\hat{\bm{\omega}}^{\tilde{n},1}) + 2\hat{\bm{f}}(\hat{\bm{\omega}}^{\tilde{n},2}) + \hat{\bm{f}}(\hat{\bm{\omega}}^{\tilde{n},3})\right]$ and \\
				 ${\bm{x}}^{\tilde{n}+1}={\bm{x}}^{\tilde{n},0}+\frac{\Delta t}{6}\left[\left(\tilde{\bm{u}}+\delta\tilde{\bm{u}}\right)^{\tilde{n},0} + 2\left(\tilde{\bm{u}}+\delta\tilde{\bm{u}}\right)^{\tilde{n},1} + 2\left(\tilde{\bm{u}}+\delta\tilde{\bm{u}}\right)^{\tilde{n},2} + \left(\tilde{\bm{u}}+\delta\tilde{\bm{u}}\right)^{\tilde{n},3}\right]$ }.
	
	}
	\end{doublespace}
	\vspace*{-0.5cm}
\end{algorithm}

\begin{algorithm}
	\caption{Algorithm projecting SPH equations onto a low-dimensional basis}\label{alg:ProjectDynamics}
	\SetAlgorithmName{Algorithm}{}{}
	\SetKwInput{Input}{Input}
	\SetKwInput{Output}{Output}
	\SetKwFunction{FMain}{ProjectDynamics}
	\SetKwFunction{FMainBasis}{TrialBasisApproximation}
	\SetKwProg{Fn}{function}{($\tilde{\bm{\Phi}}$, $\bar{\bm{f}}$):}{}
	\Input{Trial basis approximation, $\tilde{\bm{\Phi}}$ and SPH functional, $\bar{\bm{f}}$.}
	\Output{Projected low-dimensional SPH functional, $\hat{\bm{f}}$.}
	\begin{doublespace}
		\vspace*{-.1cm}
		
			\Fn\FMain{
					\eIf{\normalfont{\texttt{Galerkin projection}}}
					{
						{Projected low-dimensional SPH functional: $\hat{\bm{f}}=\tilde{\bm{\Phi}}^+\bar{\bm{f}}$}
					}
					(\normalfont{\texttt{Adjoint Petrov-Galerkin projection}}\cite{parish2020adjoint})
					{				
						{1. Compute projection of the SPH functional, $\tilde{\bm{\Pi}}\bar{\bm{f}}:=\tilde{\bm{\Phi}}\tilde{\bm{\Phi}}^+\bar{\bm{f}}$.}\\
						{2. Compute orthogonal projection of the SPH functional, ${\bm{\Pi}}^{\prime}\bar{\bm{f}}:=\bar{\bm{f}}-\tilde{\bm{\Pi}}\bar{\bm{f}}$.}\\
						{3. Compute the approximation, $ \bm{J}\bm{\Pi}^{\prime } \bar{\bm{f}}(\tilde{\bm{\omega}})\approx \frac{1}{\varepsilon}\left[\bar{\bm{f}}(\tilde{\bm{\omega}}+\varepsilon\bm{\Pi}^{\prime } \bar{\bm{f}}(\tilde{\bm{\omega}}))-\bar{\bm{f}}(\tilde{\bm{\omega}})\right]$.}\\
						{4. Projected low-dimensional SPH functional, $\hat{\bm{f}}:={\bm{\Phi}}^{+}\left(\bar{\bm{f}}+\tau\bm{J}{\bm{\Pi}}^{\prime}\bar{\bm{f}}\right)$.}\\
         				\vspace*{-0.5cm}
					}
					}
	\end{doublespace}
	\vspace*{-0.35cm}
\end{algorithm}

\section{Numerical experiments} \label{Section:NumericalExperiments}
The proposed meshless PMOR is tested on three benchmark studies: 1) the Taylor--Green vortex, for which there exists an analyitical solution; 2) the lid-driven cavity with no-slip boundary conditions; and 3) flow past an open cavity. It is important to reiterate that to the best of the authors' knowledge, the current work marks a first step enabling intrusive meshless PMOR. Therefore, the meshless PMOR is exercised on the aforementioned canonical examples where solutions and behaviors are well understood. Future work will consider problems where meshless methods provide invaluable capabilities, such as free-surface flows and multiphase flows. For all numerical experiments, the percent relative discrepancy is employed to measure error between FOMs and PMORs. Specifically, for a collection of measurements, $\bm{c}\in\mathbb{R}^N$, at a time step $\tilde{n}$, the percent relative discrepancy is measured as,

\begin{equation}
	{\text{percent relative discrepancy}}=100\times\frac{\frac{1}{N}\sum\limits_{i=1}^{N}\lvert\bm{c}_{i,\text{FOM}}^{\tilde{n}}-\bm{c}_{i,\text{PMOR}}^{\tilde{n}} \rvert}{\lvert{\text{max}}(\bm{c}_{\text{FOM}}^{\tilde{n}})-{\text{min}}(\bm{c}_{\text{FOM}}^{\tilde{n}})\rvert}.
    \label{eq:PRD}
\end{equation}
To quantify the level of dimensionality reduction, a compression factor is employed and is defined as 
\begin{equation}
	CF=\frac{M}{N_d}.
    \label{eq:CF}
\end{equation}
Finally, the cumulative energy contribution is employed to quantify the cumulative contribution of a set of $M$ modes to the system's energy and is defined as
\begin{equation}
    \text{percent cumulative energy}=100\times\frac{\sum\limits_i^M\sigma_i}{\sum\limits_i^{N_{\bar{d}}} \sigma_i}.
\end{equation}
%
\subsection{Taylor--Green Vortex}
The Taylor--Green vortex (TGV) is modeled analytically by the following expressions as defined in \cite{antuono2020tri}:
\begin{equation}
	\begin{split}
		&p=-\frac{1}{4}\rho\left[\cos\left(4\pi x\right)+\cos\left(4\pi y\right)\right]\rm{e}^{-4\nu\kappa^2t},\\
		&u_x=\sin\left(2\pi x\right)\cos\left(2\pi y\right)\rm{e}^{-2\nu\kappa^2t},\\
		&u_y=-\cos\left(2\pi x\right)\sin\left(2\pi y\right)\rm{e}^{-2\nu\kappa^2t},\\
	\end{split}
	\label{eqn:TG_analytical}
\end{equation}
where $\kappa=2\pi$, $\nu$ is the kinematic viscosity varied in the present experiments to define the Reynolds number, $p$, $u_x$, and $u_y$ are the pressure and the $x$ and $y$ components of the velocity field, respectively. To model the Taylor--Green vortex, Eq.~\ref{eqn:TG_analytical} is employed to provide initial conditions to the SPH equations at $t=0$ s. A periodic domain is defined by the unit square, $\Omega=[0, 1] \times [0, 1]$ m$^2$. All simulations, including PMOR results, run for a physical time of $T_f=1$ s and employ a smoothing length of $h=4\Delta x$, a particle resolution of $N=300\times 300$, and a time step of $\Delta t=  5 \times 10^{-4}$ s. Future work will investigate a rigorous error bounds analysis tied to particle resolution and smoothing length for both the ROM and reference spaces. The one-dimensional Reynolds number parametric space of interest is defined by $Re$=[100, 250], where training data samples are selected at $Re=100,\:150,\:200,\:250$. The reference density of $\rho_0=1000$ kg/m$^3$ is employed for all TGV simulations and the Reynolds number is defined by varying viscosity. The Mach number, $M_a={U_{\textup{ref}}}/{c_0}$, selected is $M_a=0.1$, with a reference speed of $U_{\textup{ref}}=1$ m/s. Finally, the packing algorithm presented in \cite{colagrossi2012particle} is employed to generate the initial particle configuration. 

\subsubsection{Reconstructive results}

\begin{figure}[t!]
	\centering
	\begin{subfigure}[t]{0.5\textwidth}
		\centering
		\includegraphics[trim = {0cm 0cm 0 0},  scale=0.325]{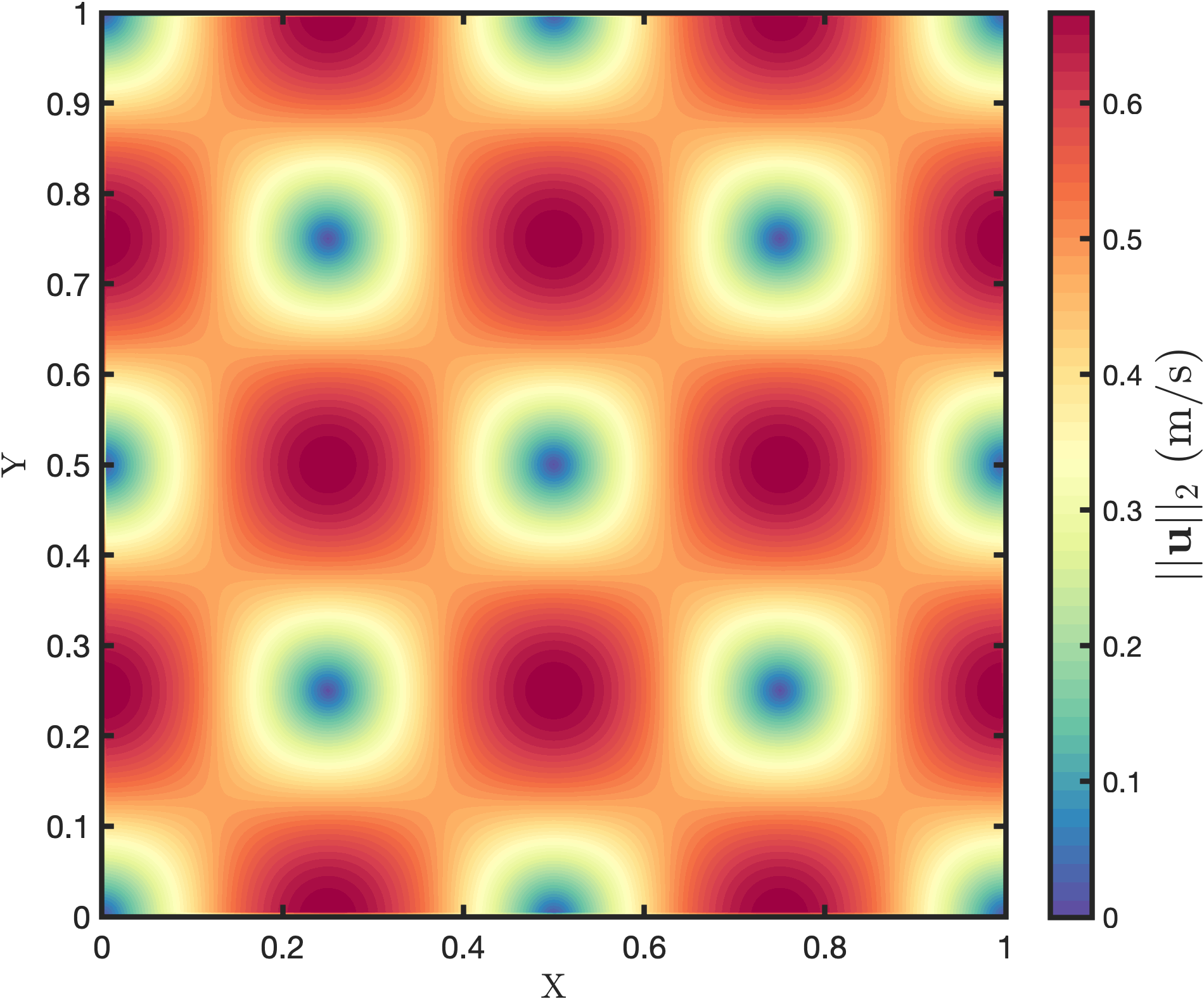}
		\caption{Velocity field}
			     \label{fig:TGV_FOM_results_velocity}
	\end{subfigure}%
	\hfill
	\begin{subfigure}[t]{0.5\textwidth}
		\centering
		\includegraphics[ trim={0cm 0cm 0 0}, clip, scale=0.325]{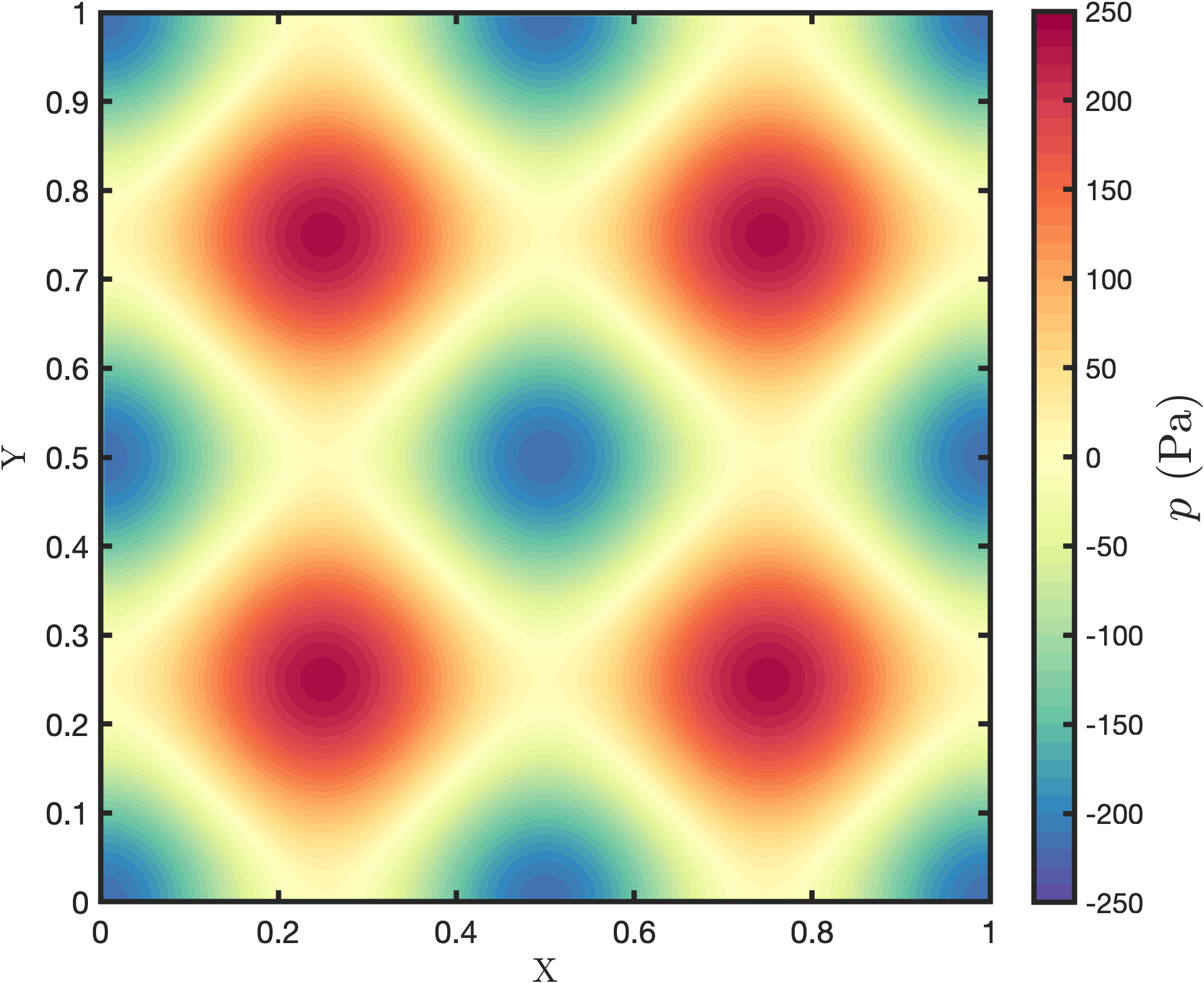}
		\caption{Pressure field}
	     \label{fig:TGV_FOM_results_pressure}
	\end{subfigure}
	\caption{TGV at $Re=100$; FOM snapshot at $t=0.5$ s.}
	\label{fig:TGV_FOM_results}
\end{figure}

\begin{figure}[t!]
	\centering
\centering
	\begin{subfigure}[t]{0.5\textwidth}
		\centering
		\includegraphics[trim = {0cm 0cm 0 0},  scale=0.325]{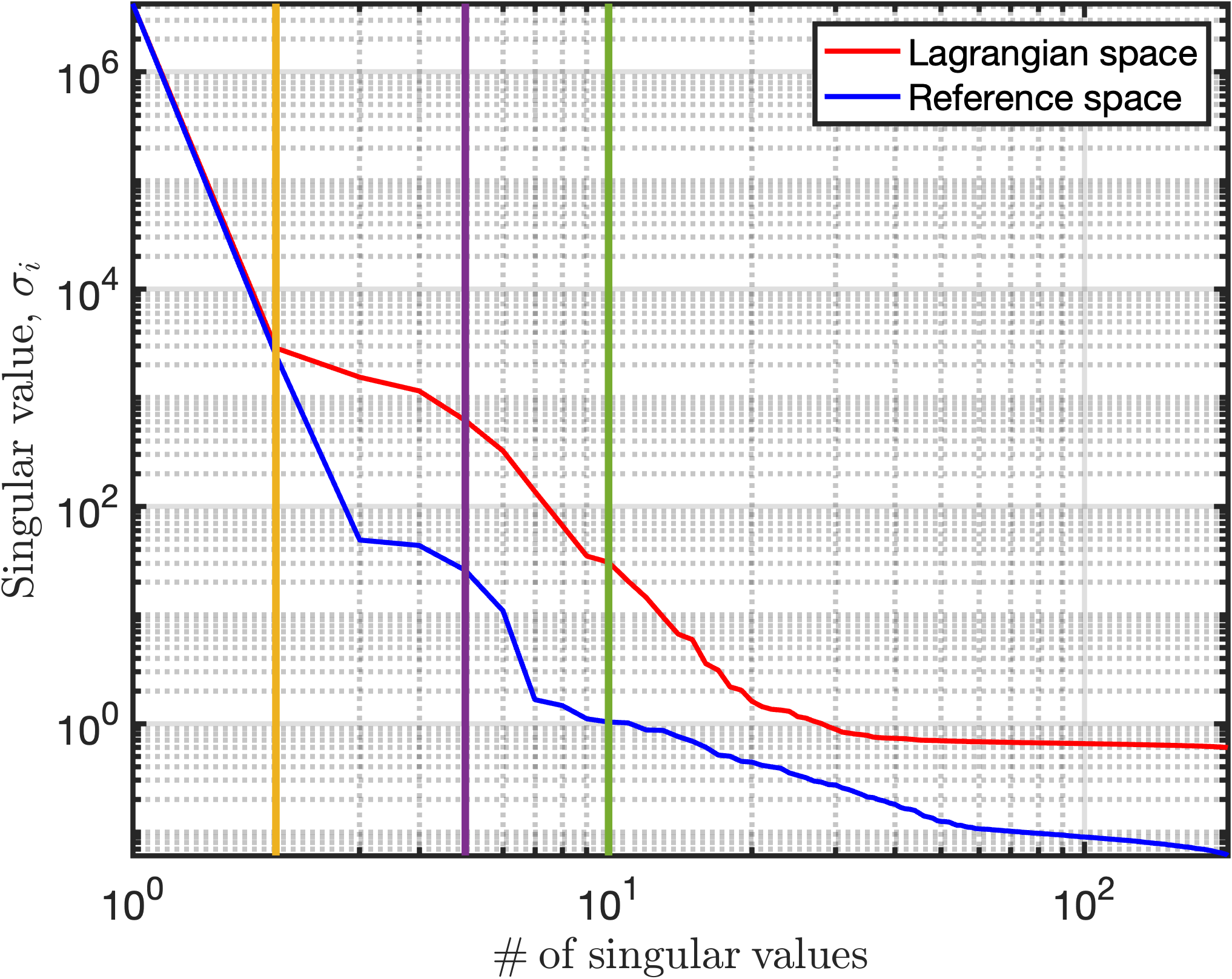}
		\caption{Singular value decay}		
	\end{subfigure}%
	\hfill
	\begin{subfigure}[t]{0.5\textwidth}
		\centering
		\includegraphics[ trim={0cm 0cm 0 0}, clip, scale=0.325]{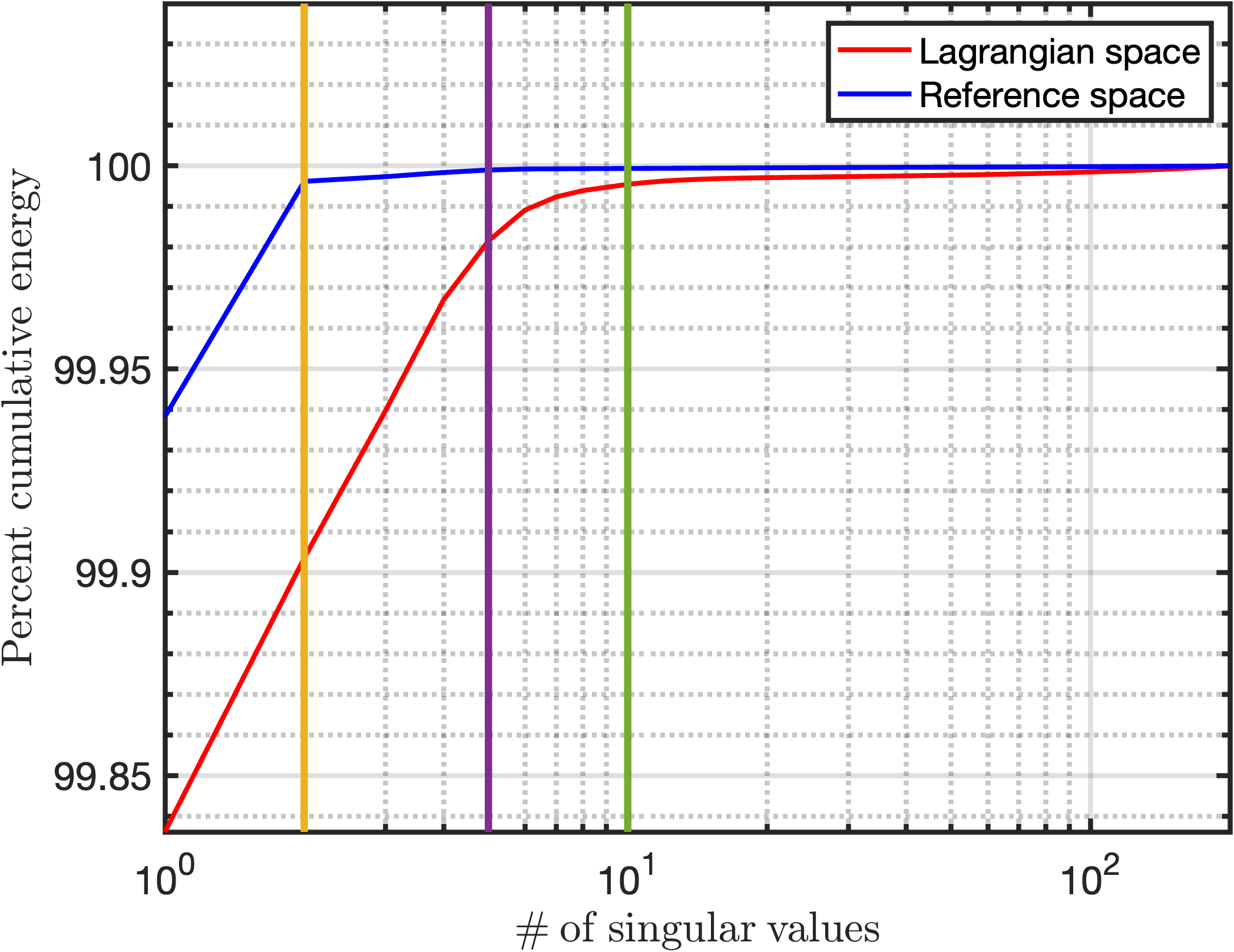}
		\caption{Cumulative energy contribution}
	\end{subfigure}
	\caption{Singular values derived from Lagrangian and reference space snapshot matrices. Vertical colored lines are meant to highlight the differences in singular value decay and cumulative energy between Lagrangian and reference space at $M=2$ (yellow), $M=5$ (purple), and $M=10$ (green).} 
	\centering
\label{fig:TGV_LagrangianVReferenceSingularValues}
\end{figure}

The meshless model-reduction method proposed is first tested on reconstructing training data for $Re=100$. Figure \ref{fig:TGV_FOM_results} presents snapshots of the FOM velocity field and pressure field at $t=0.5$ s. 
Snapshots of the FOM data are collected at an interval of every 10 timesteps, for a total of 200 snapshots. The SPH snapshots are directly stored into a matrix representing a \emph{Lagrangian} snapshot matrix, and are also mapped onto reference space and stored in matrix form. The POD modes are extracted from both the Lagrangian and reference spaces and their corresponding singular value decays and cumulative energy contributions are presented in Fig.~\ref{fig:TGV_LagrangianVReferenceSingularValues}. It is clear to see that singular values decay at a faster rate when the SPH data is mapped onto the reference space. Similarly, fewer modes are needed in the reference space to contribute to the system's dynamics to fully represent the underlying energy of the data. For instance, at $M=2$ the Lagrangian space retains only 99.9\% of the system's energy from the FOM, while the reference space retains 99.996\%. At five modes, $M=5$, singular values from the reference space retain 99.999\% of the energy, while in Lagrangian space, only 99.98\% is retained. At $M=10$, the reference space retains 99.9994\% and the Lagrangian space retains 99.995\%. 

\begin{figure}[t!]
	\centering
	\includegraphics[trim = 0cm 0cm 0 0, scale=0.4]{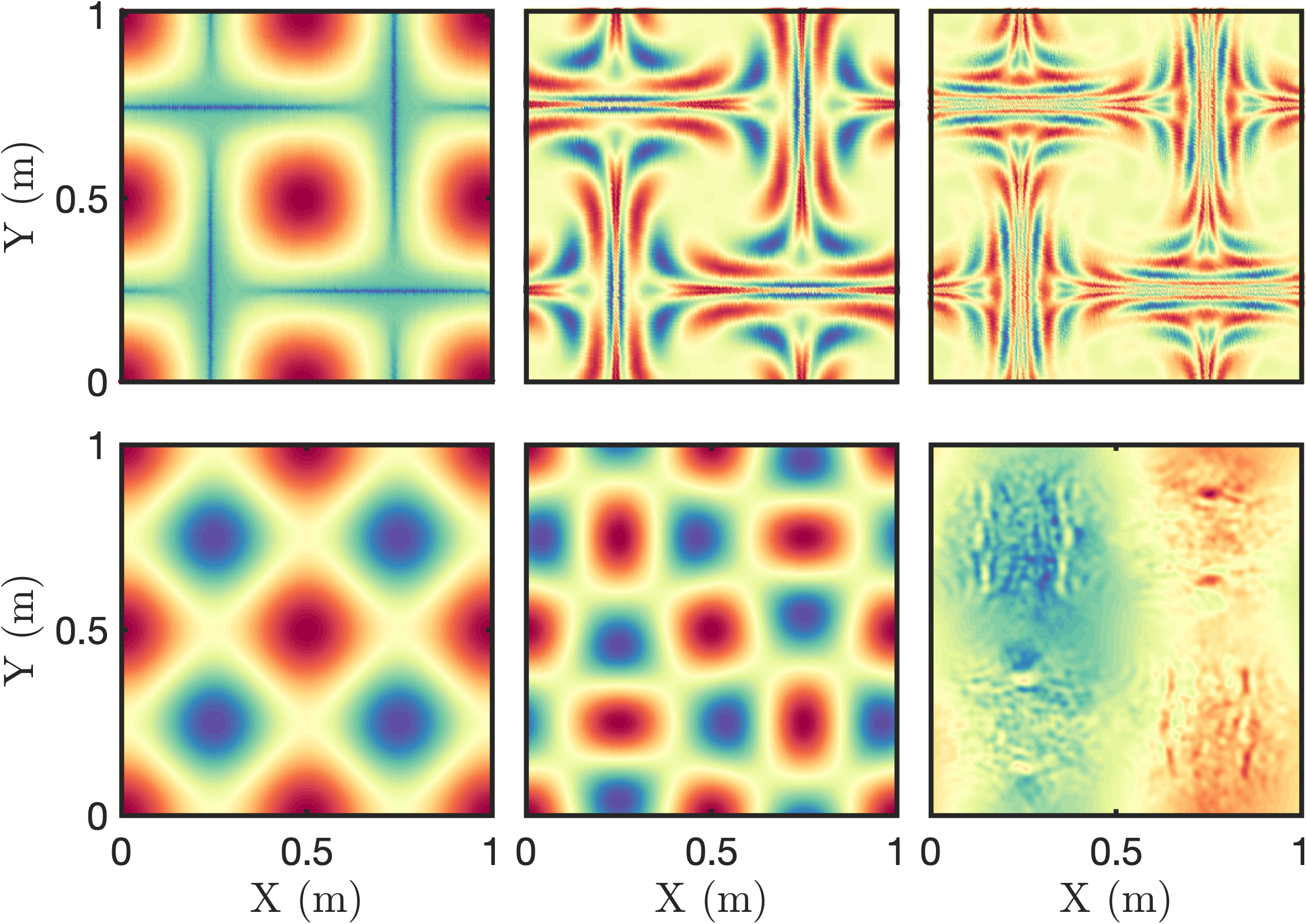}
	\caption{Density field modes of the TGV at $Re=100$, from left to right $M=1, 5, {\text{and}\:} 10$. Top row: Lagrangian space; Bottom row: Reference space.} 
	\centering
	\label{fig:TGV_ModalRho}
\end{figure}

\begin{figure}[t!]
	\centering
	\includegraphics[trim = 0cm 0cm 0 0, scale=0.4]{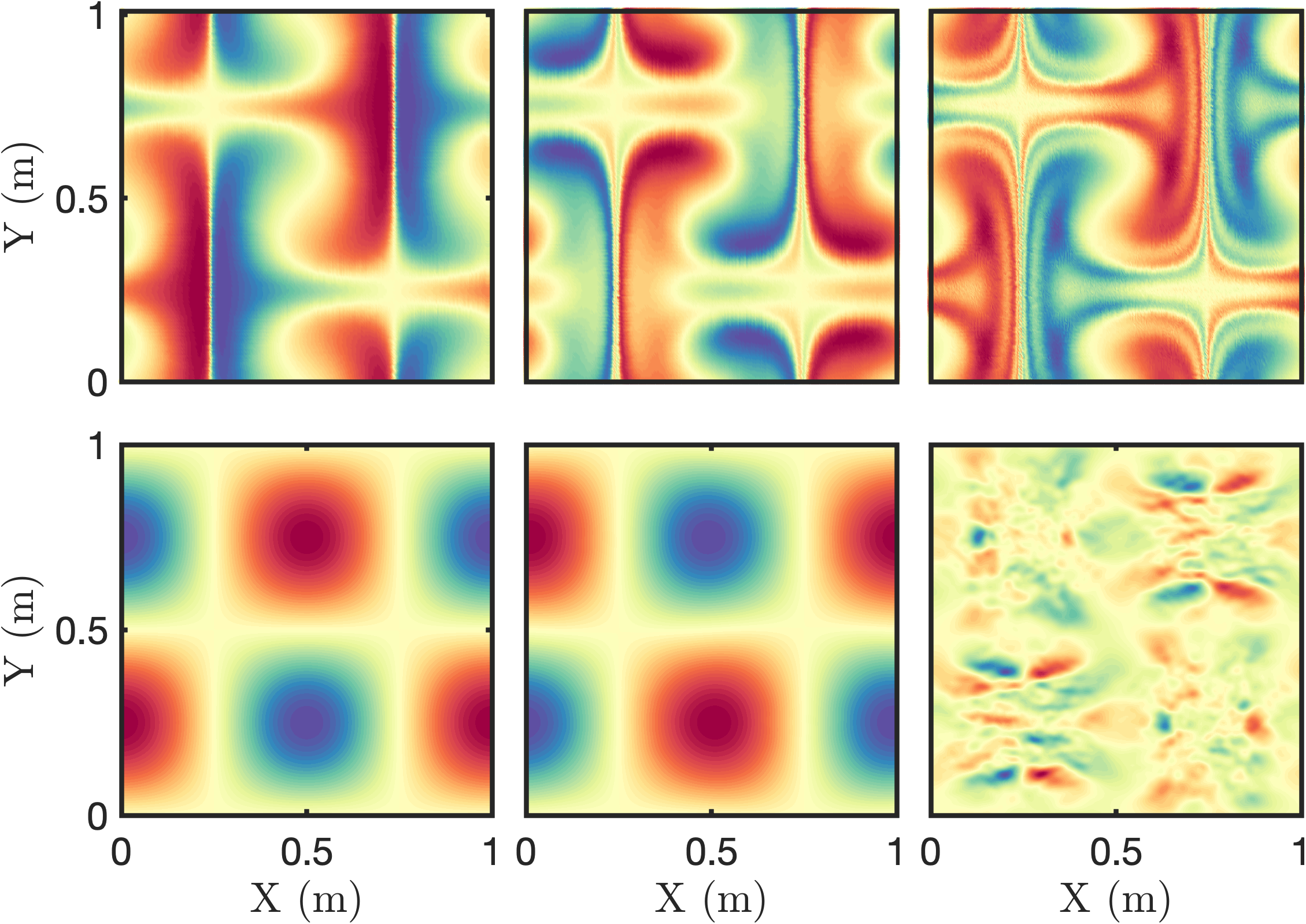}
	\caption{Velocity field $x-$component modes of the TGV at $Re=100$, from left to right $M=1, 5, {\text{and}\:} 10$. Top row: Lagrangian space; Bottom row: Reference space.}
	\centering
		\label{fig:TGV_ModalX}
\end{figure}

\begin{figure}[t!]
	\centering
	\includegraphics[trim = 0cm 0cm 0 0, scale=0.4]{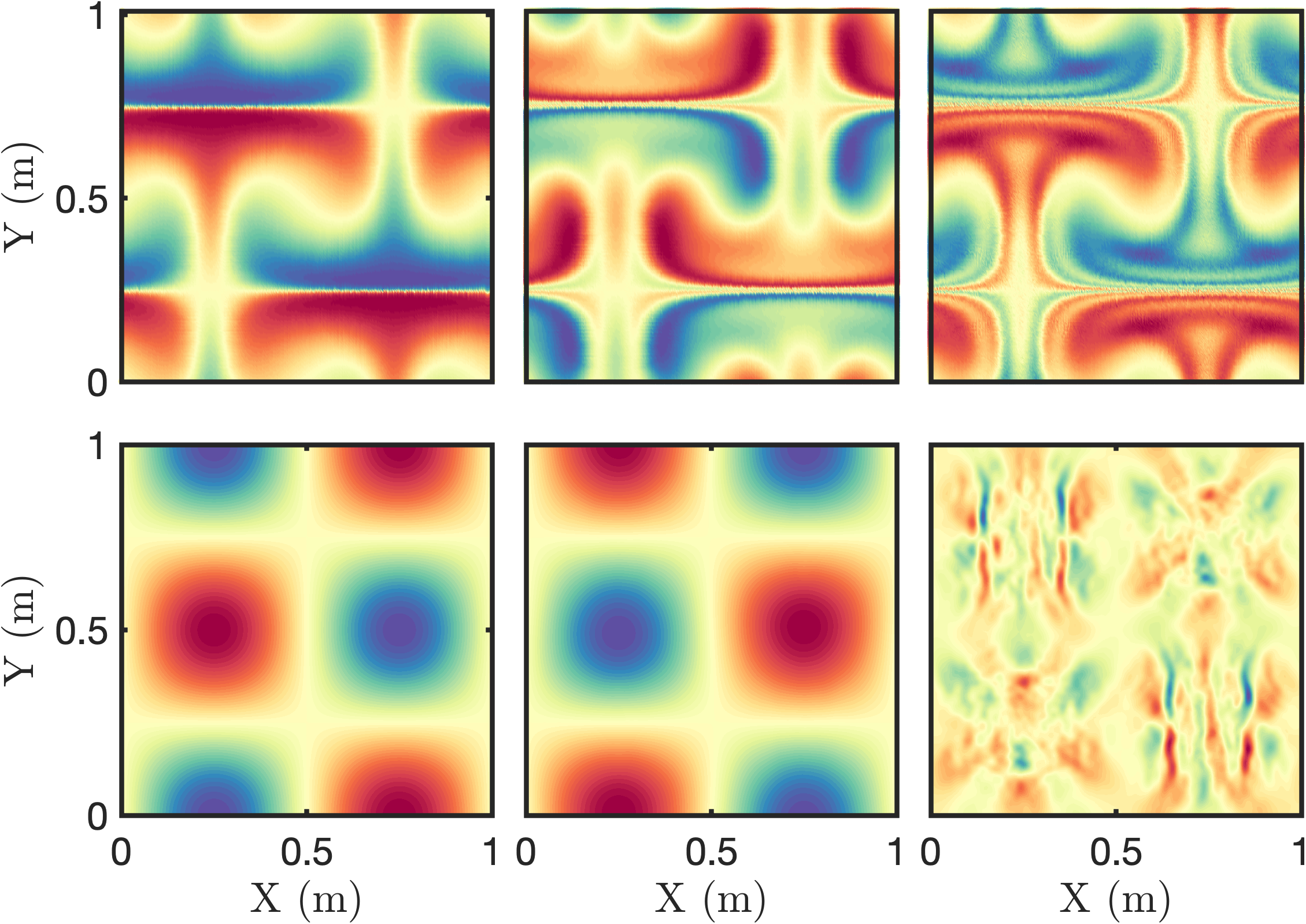}
	\caption{Velocity field $y-$component modes of the TGV at $Re=100$, from left to right $M=1, 5, {\text{and}\:} 10$. Top row: Lagrangian space; Bottom row: Reference space.}
	\centering
    \label{fig:TGV_ModalY}
\end{figure}

Singular values from the Lagrangian space are derived in a traditional sense, where the numerical topology is assumed to be fixed in space. Thereby, for strongly mixing flows, the singular value decomposition is tasked with drawing correlations between particles at different locations in space that may not be strongly correlated in field quantities. On the other hand, mapping the SPH data onto a reference frame guarantees that the singular value decomposition is tasked with drawing correlations between local regions of physical space. Figures \ref{fig:TGV_ModalRho}-\ref{fig:TGV_ModalY} show modes derived from Lagrangian and reference space for modes $M=1, 5, {\text{and}\:} 10$. For Lagrangian topology, modes exhibit mixing dynamics, while in reference space, modes exhibit modal behavior aligned with the expected TGV behavior. 

\begin{figure}[t!]
			\centering
\includegraphics[trim = 0cm 0cm 0 0, scale=0.45]{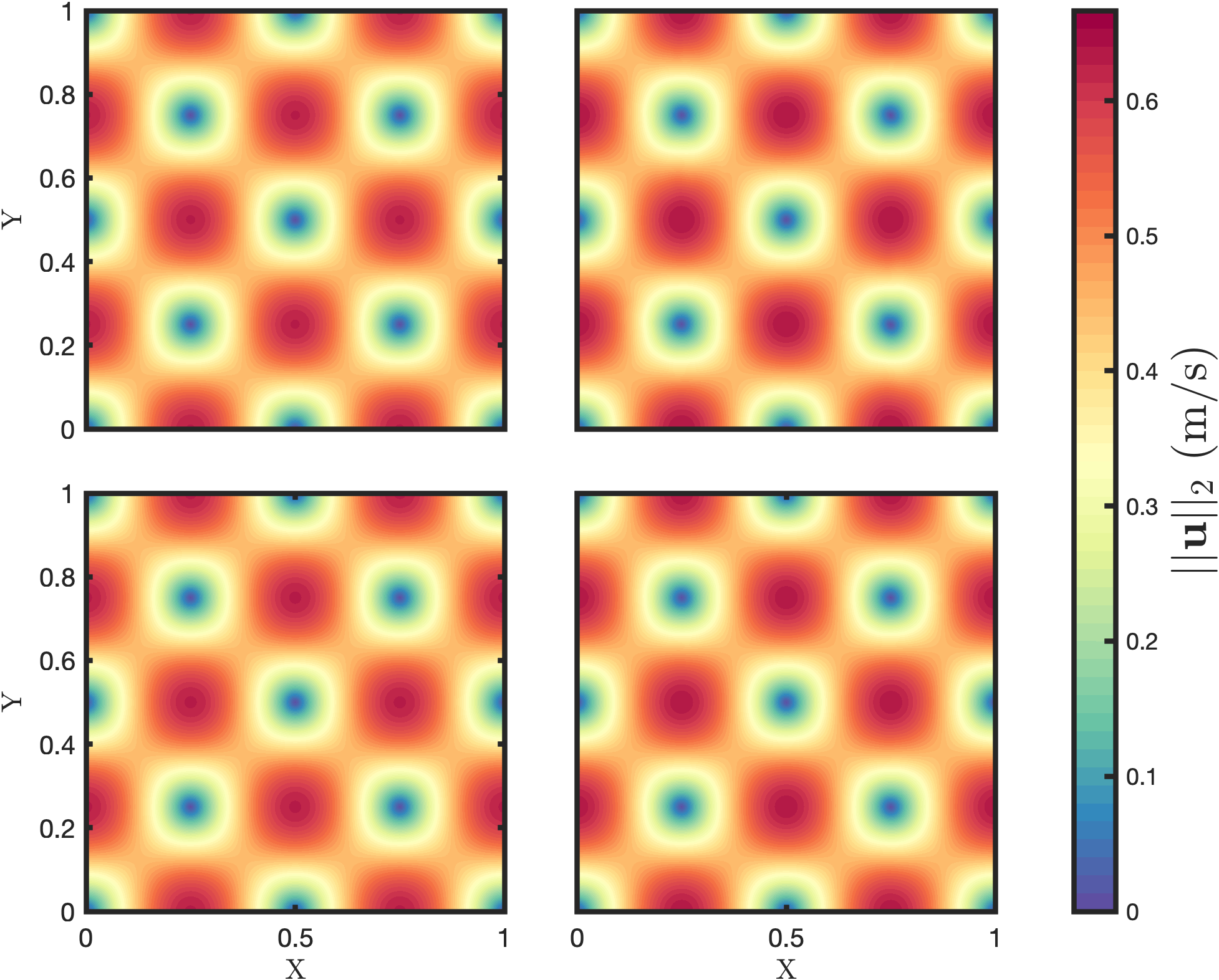}
	\caption{Velocity fields read from left to right; top row: GPOD Case 1 and Case 2; bottom row: APG Case 1 and Case 2.} 
	\centering
	\label{fig:TGV_reconstructive_velocity}
\end{figure}

\begin{figure}[t!]
			\centering
\includegraphics[trim = 0cm 0cm 0 0, scale=0.45]{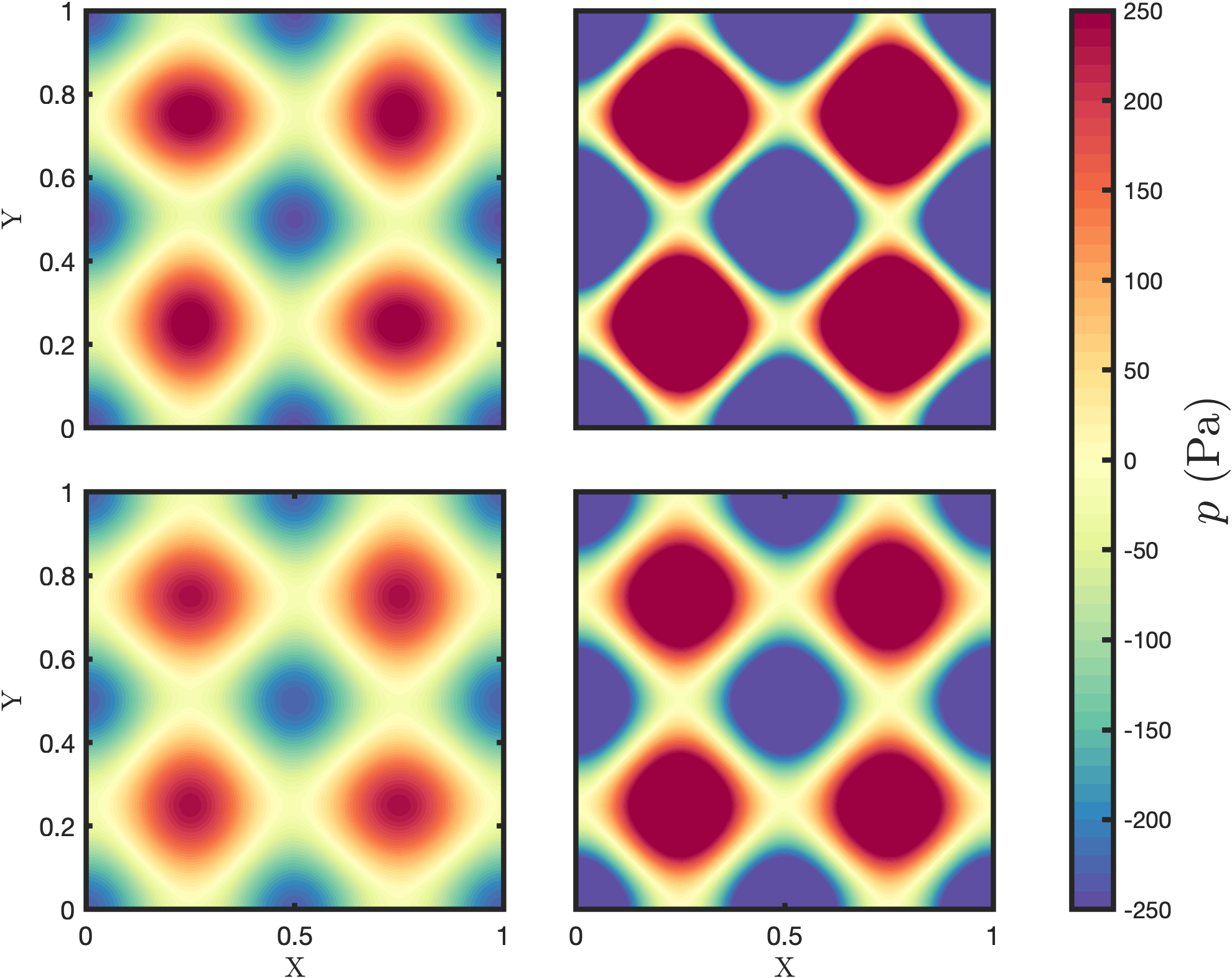}
\caption{Pressure fields read from left to right; top row: GPOD Case 1 and Case 2; bottom row: APG Case 1 and Case 2.} 
	\centering
	\label{fig:TGV_reconstructive_pressure}
\end{figure}

\begin{figure*}[t!]
	\centering
	\begin{subfigure}[t]{0.5\textwidth}
		\centering
		\includegraphics[trim = {0cm 0cm 0 0},  scale=0.325]{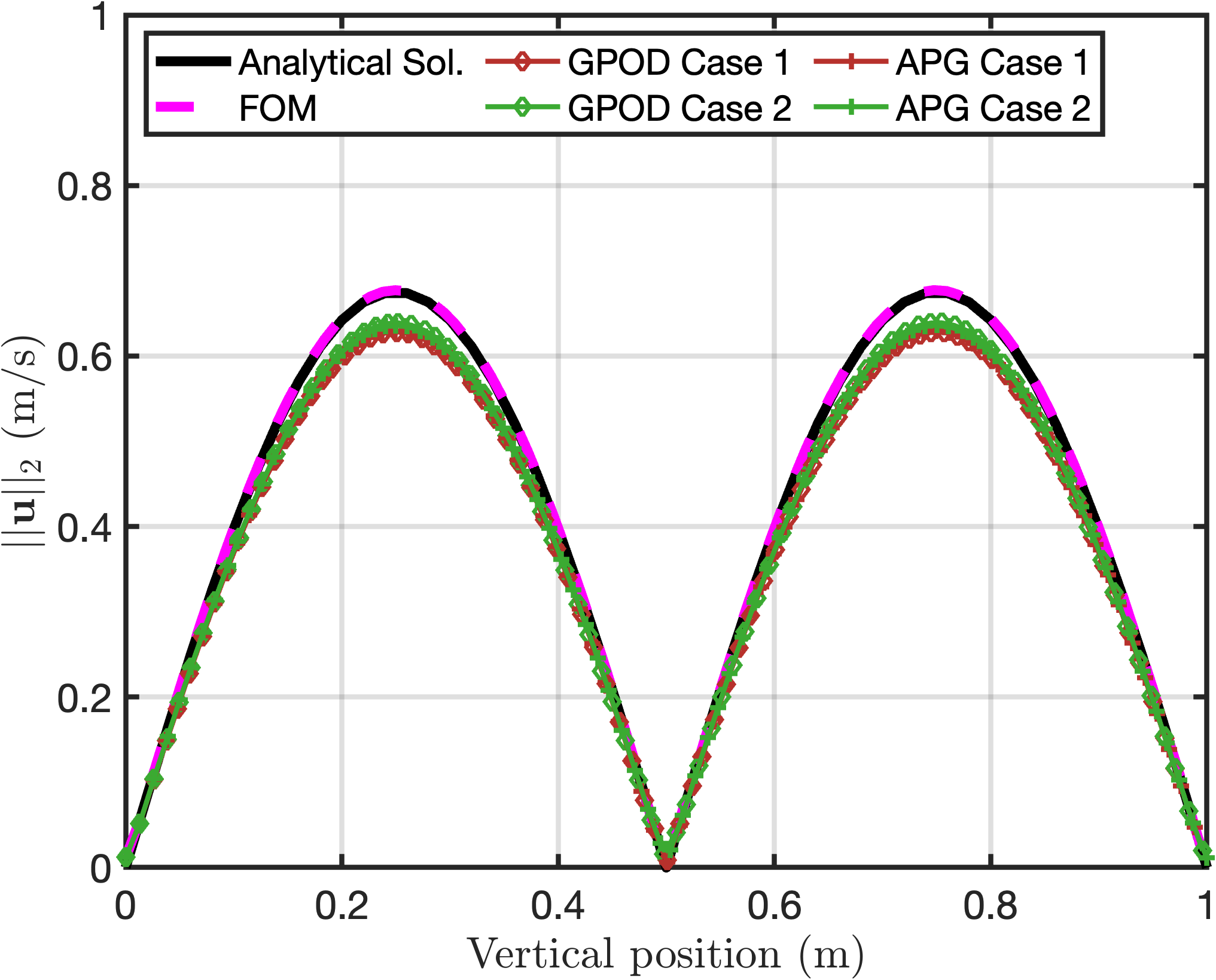}
		\caption{Velocity field at vertical centerline}
		\label{fig:TGV_reconstructive_velocity_centerline}
	\end{subfigure}%
	\hfill
	\begin{subfigure}[t]{0.5\textwidth}
		\centering
		\includegraphics[ trim={0cm 0cm 0 0}, clip, scale=0.325]{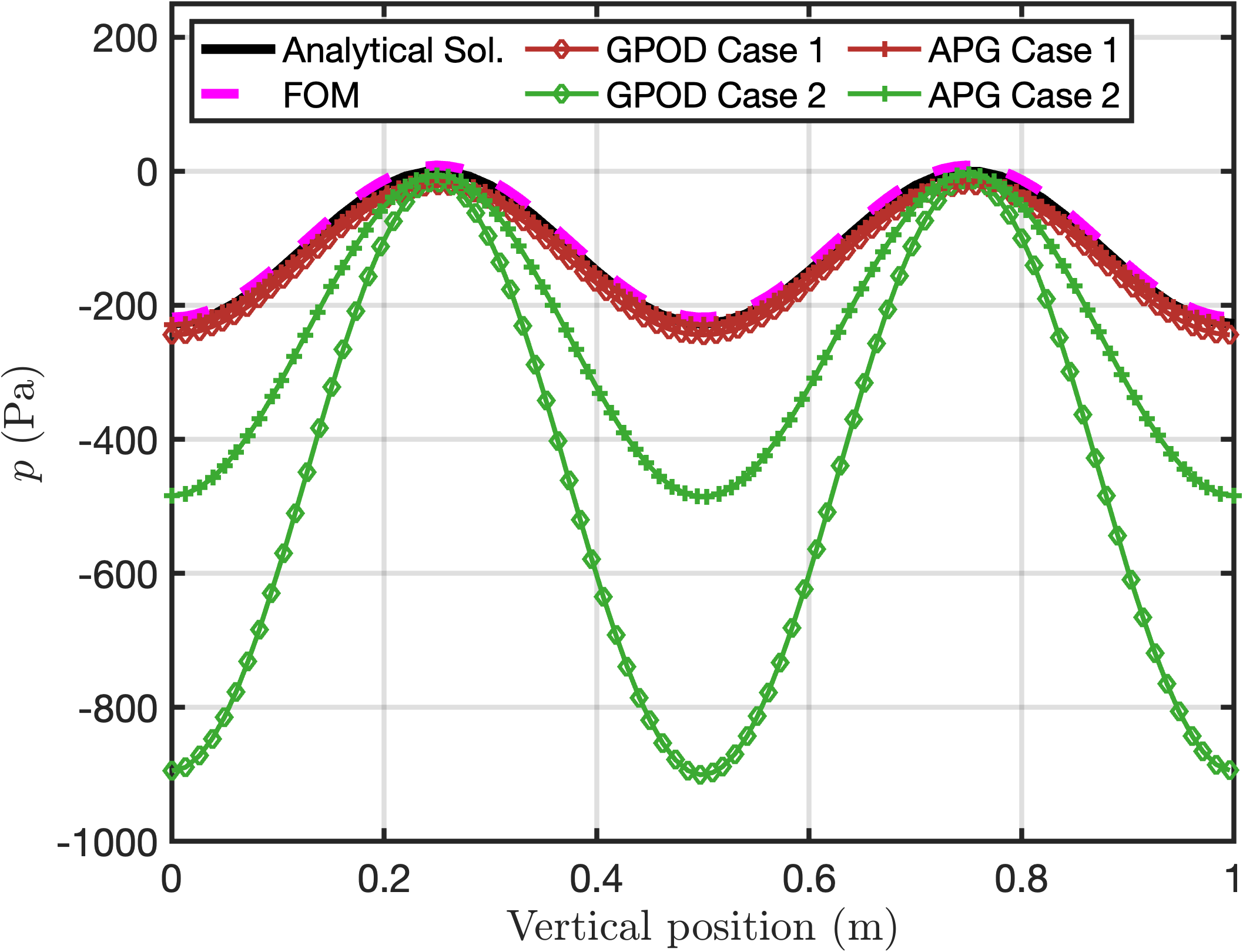}
		\caption{Pressure field at vertical centerline}
		\label{fig:TGV_reconstructive_pressure_centerline}
	\end{subfigure}
	\caption{Centerline snapshots at $t=0.5$ s.}
\end{figure*}

Four numerical experiments were run to test the proposed model-reduction framework in a reconstructive setting: GPOD Case 1) with a basis dimension of $M=5$; GPOD Case 2) with a basis dimension of $M=10$; APG Case 1) with a basis dimension of $M=5$ and memory length of $\tau=10^{-4}$ s; and APG Case 2) with a basis dimension of $M=10$ and memory length of $\tau=10^{-4}$ s. This memory length was chosen empirically to arrive at the most accurate and stable results. Future work will focus on defining a heuristic approach for selecting the memory length for the presented meshless method as was done in \cite{parish2020adjoint}. Finally, the selected basis dimensions result in compression factors for GPOD/APG Case 1 and Case 2 of $CF=90,000$ and $CF=45,000$, respectively. Figures \ref{fig:TGV_reconstructive_velocity} and \ref{fig:TGV_reconstructive_pressure} show the velocity and pressure fields for all four cases, respectively. There is a marginal qualitative difference between  Fig.~\ref{fig:TGV_reconstructive_velocity} and the FOM in Fig.~\ref{fig:TGV_FOM_results_velocity} for both GPOD and APG methods. To qualitatively compare the PMOR directly against the FOM and analytical solutions, the centerline velocity across the vertical position of the domain is recorded at snapshot $t=0.5$ s, shown in Fig.~\ref{fig:TGV_reconstructive_velocity_centerline}. These results show that the velocity profiles closely match both the FOM and the analytical solution, but underpredict the velocity peaks of the vortices. It is interesting to highlight that both Case 1 and Case 2 for GPOD and APG align well with the FOM and retain $99.999\%$ and $99.9994\%$ percent of the statistical energy, respectively, but still marginally underpredict vortex peaks. It is likely that mapping the SPH data onto the reference space attenuated the modal information necessary to reconstruct vortex peaks exactly.  

On the other hand, the pressure field in Fig.~\ref{fig:TGV_reconstructive_pressure} for Case 2 in both methods shows amplified trends that do not align with the FOM or analytical solution. This behavior is further highlighted in Fig.~\ref{fig:TGV_reconstructive_pressure_centerline}, where the centerline pressure distribution for Case 2 exhibits similar sinusoidal characteristics seen in the true solution, but at higher amplitudes. The amplification of the pressure field likely stems from the projection error of the density field and its role in the stiff equation of state employed in the weakly-compressible SPH framework. Specifically, in the present SPH setting the weakly-compressible assumption chooses a speed of sound that limits the density field to a deviation of at most 1\% from the reference density. This assumption would require the projection error of the density field to be within the same range as the weakly-compressible assumption to ensure an accurate reconstruction of the pressure field. However, it is important to highlight that Case 1 with $M=5$, predicts the pressure field to a better degree of accuracy than Case 2 with $M=10$. It is likely that Case 1 truncates higher-frequency content that exacerbates the ability of the POD affine basis to approximate the dynamics of the weakly-compressible assumption.

\begin{figure*}[t!]
	\centering
	\begin{subfigure}[t]{0.5\textwidth}
		\centering
		\includegraphics[trim = {0cm 0cm 0 0},  scale=0.325]{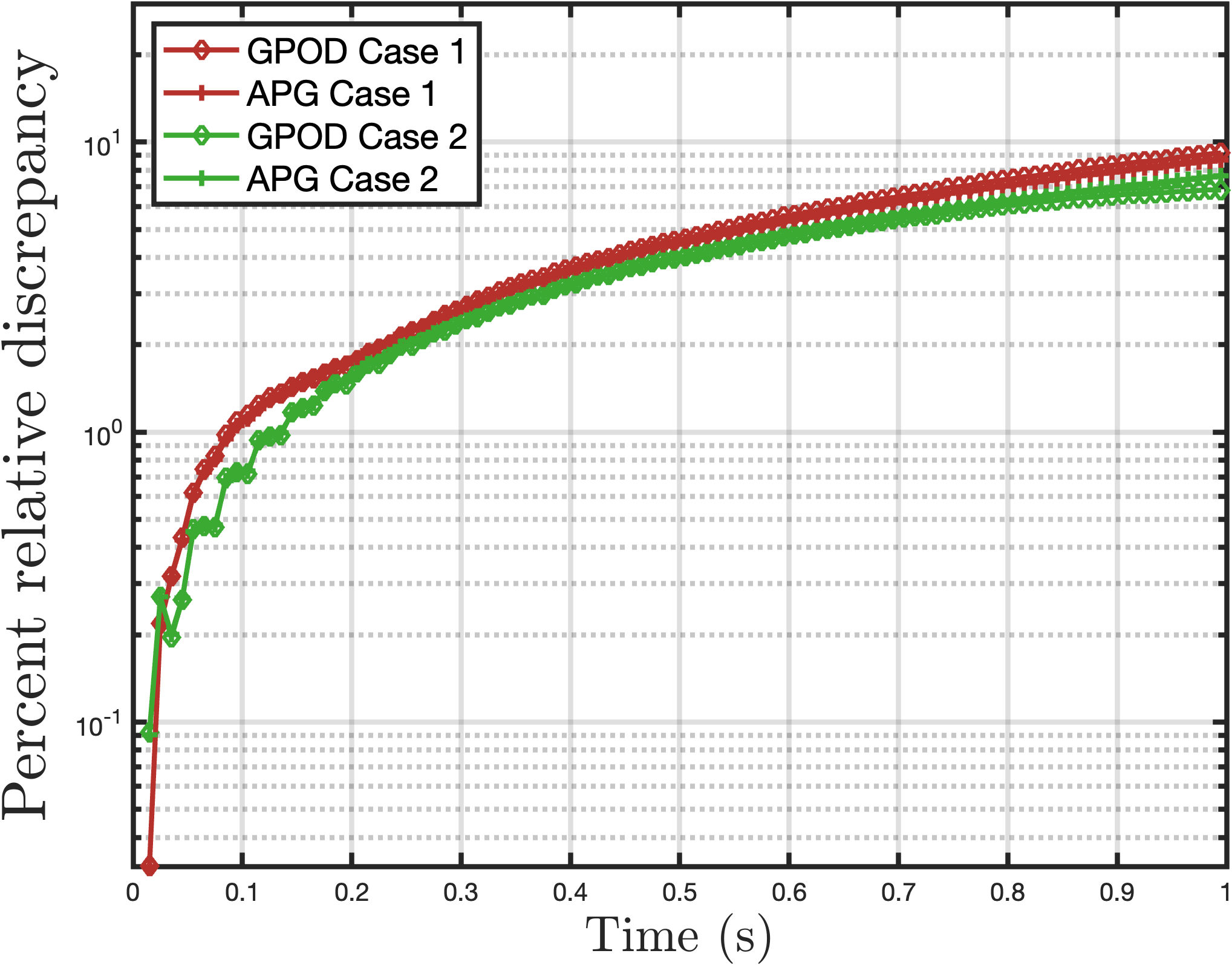}
		\caption{Velocity field relative discrepancy}
    	\label{fig:TGV_reconstructive_velocity_centerline_error}
	\end{subfigure}%
	\hfill
	\begin{subfigure}[t]{0.5\textwidth}
		\centering
		\includegraphics[ trim={0cm 0cm 0 0}, clip, scale=0.325]{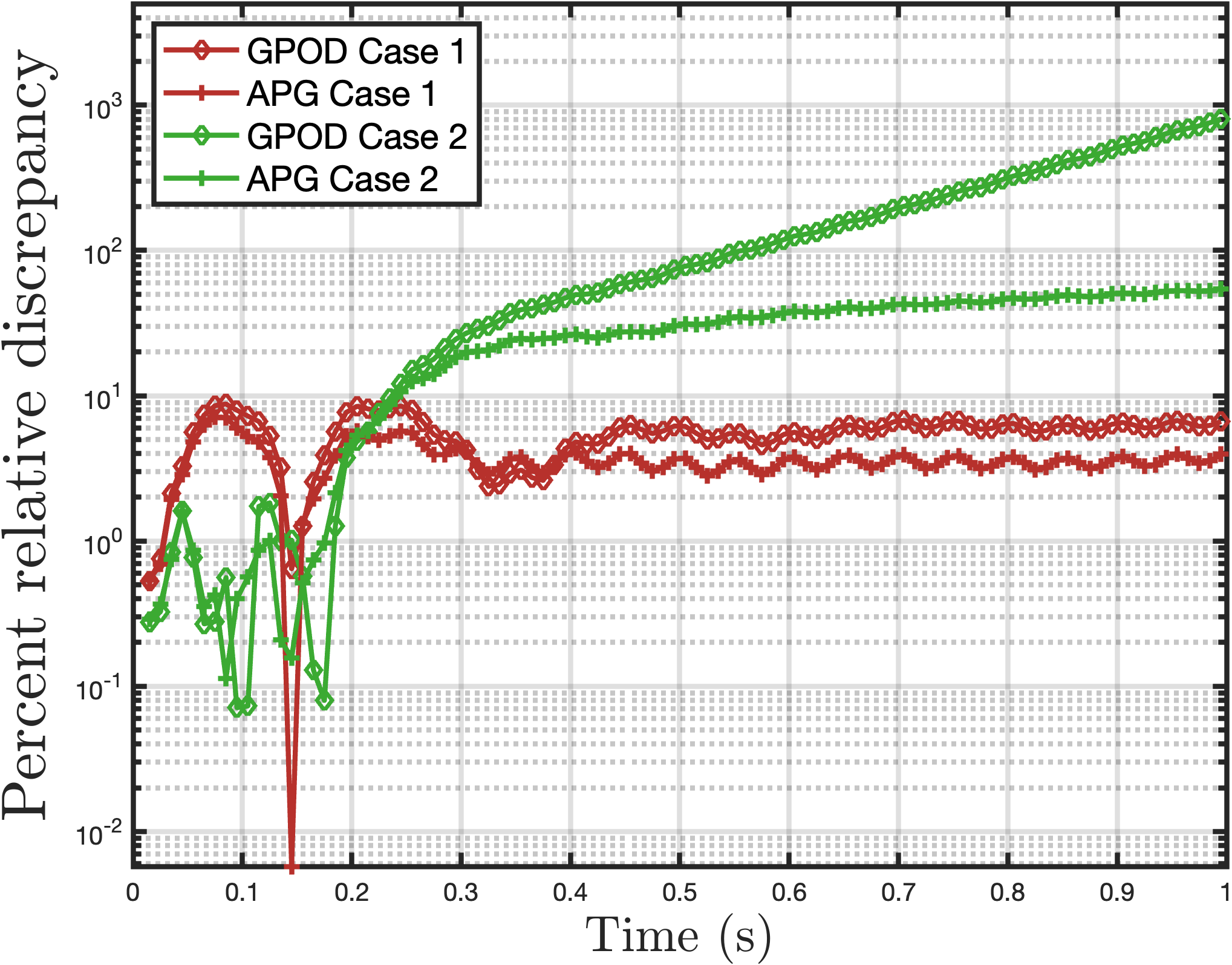}
		\caption{Pressure field relative discrepancy}
			\label{fig:TGV_reconstructive_pressure_centerline_error}
	\end{subfigure}
	\caption{Time histories of centerline relative discrepancy errors (Eq.~\ref{eq:PRD}).}
	\label{fig:TGV_reconstructive_centerline_error}
\end{figure*}

Finally, time histories of the relative discrepancy error between PMOR and FOM are presented in Fig.~\ref{fig:TGV_reconstructive_centerline_error}.  Results show consistent agreement in the velocity norm, $u^n:=\lVert \mathbf{u}^n\rVert_2$. Figure \ref{fig:TGV_reconstructive_velocity_centerline_error} shows peak errors occurring toward the end of the simulation near 10\% in Case 1. Note that Case 2, in both GPOD and APG cases, provides improved accuracy compared to Case 1 in the velocity norm, with errors between 7 and 8\%. However, Case 2 generates unstable pressure field results with GPOD and errors on the order of 50\% with APG, shown in Fig. \ref{fig:TGV_reconstructive_pressure_centerline_error}. On the other hand, Case 1 reaches peak error values near 6\% for GPOD and 4\% for APG. It is important to highlight that while reconstructive errors reach approximately 10\% in the velocity norm and 6\% in the pressure field, the framework is capable of reaching stable results and good qualitative agreement with a relatively small dimensional basis, namely, $M=5$. Future work will focus on attenuating or bypassing high-frequency modes derived from weakly compressible data to improve reconstruction. 

\subsubsection{Parametric results}

\begin{figure}[t!]
	\centering
	\includegraphics[trim = 0cm 0cm 0 0, scale=0.5]{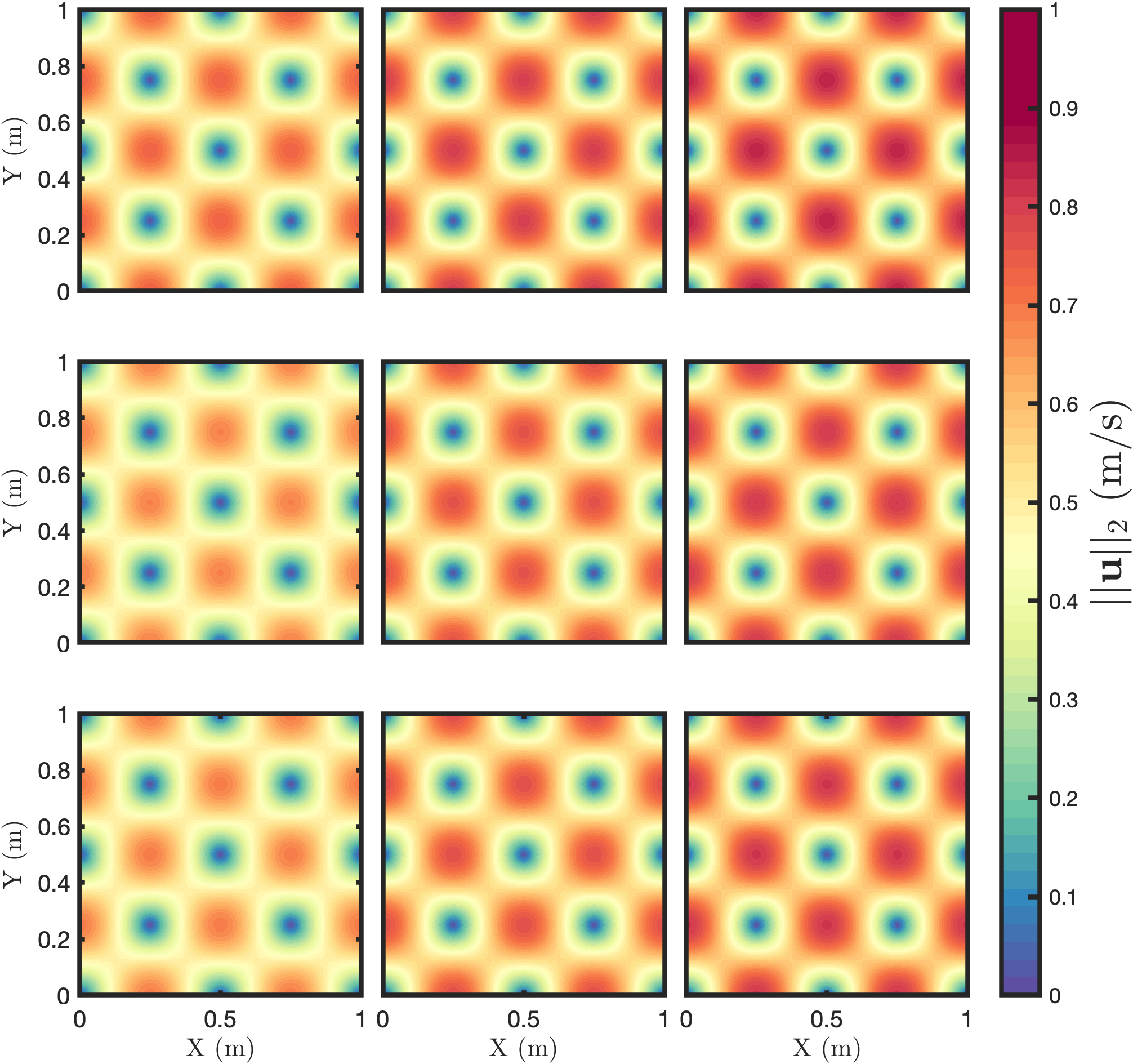}
	\caption{Taylor-Green vortex velocity fields. Columns from left to right: $Re=125, 175, 225$. First row: FOM; Second row: GPOD; Third row: APG.} 
	\centering
	\label{fig:TGV_parametric_snapshot_velocity}
\end{figure}

\begin{figure}[t!]
	\centering
	\includegraphics[trim = 0cm 0cm 0 0, scale=0.5]{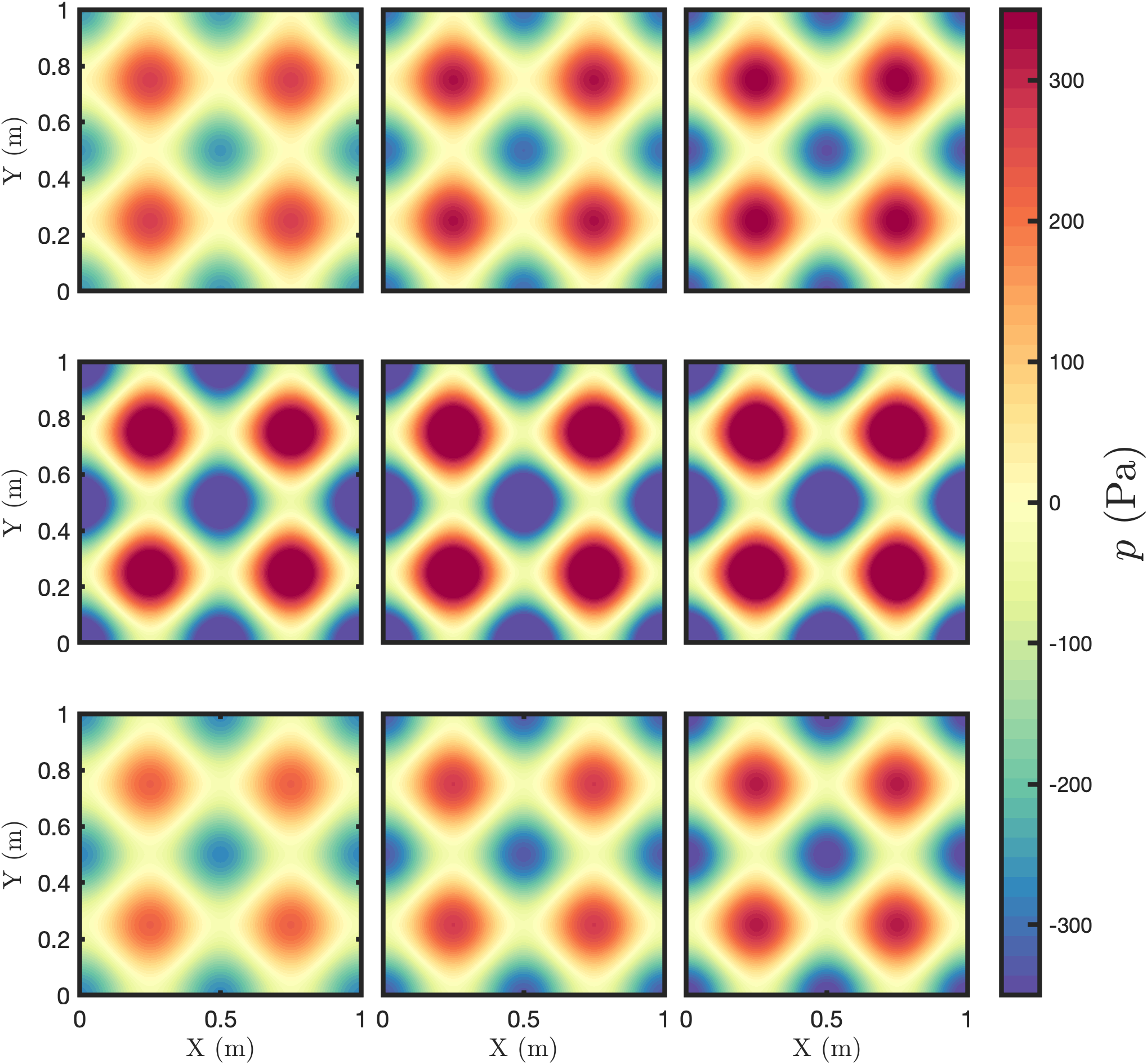}
	\caption{Taylor-Green vortex pressure fields. Columns from left to right: $Re=125, 175, 225$. First row: FOM; Second row: GPOD; Third row: APG.}
	\centering
		\label{fig:TGV_parametric_snapshot_pressure}
\end{figure}

A predictive parametric study of the Taylor-Green vortex was conducted for Reynolds numbers, $Re=125, 175, 225$. The meshless GPOD and APG approaches were employed, both with a basis dimension of $M=5$, which was heuristically chosen based on the most stable results in the previous reconstructive experiments. Figures \ref{fig:TGV_parametric_snapshot_velocity} and \ref{fig:TGV_parametric_snapshot_pressure} show snapshots of the velocity and pressure fields at time, $t=0.5$ s. Like prior reconstructive results, both predictive PMOR frameworks aligns closely with FOM velocity fields and provide stable results across all Reynolds numbers. On the other hand, pressure fields derived from GPOD show significant qualitative deviation from FOM results across all Reynolds numbers. APG results, however, showcase stable results that are close in qualitative behavior to the FOM across all Reynolds numbers.

\begin{figure}[t!]
	\centering
	\begin{subfigure}{.4\textwidth}
		\centering
		\includegraphics[width=.9\linewidth]{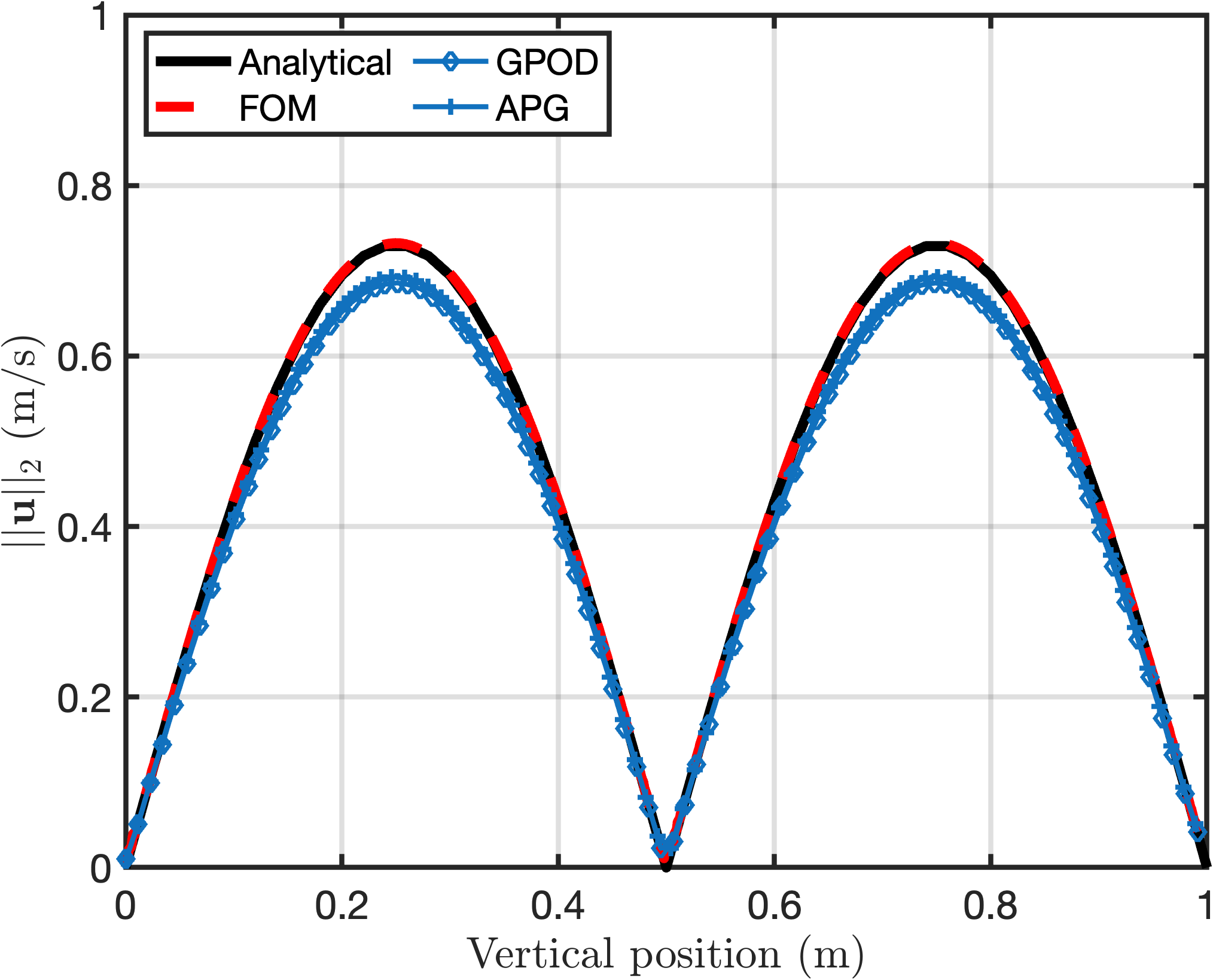}  
		\caption{$Re=125$}
		\label{fig:parametric_prediction_velocity_TGV_Re125}
	\end{subfigure}
	\hfill
	\begin{subfigure}{.4\textwidth}
		\centering
		\includegraphics[width=.9\linewidth]{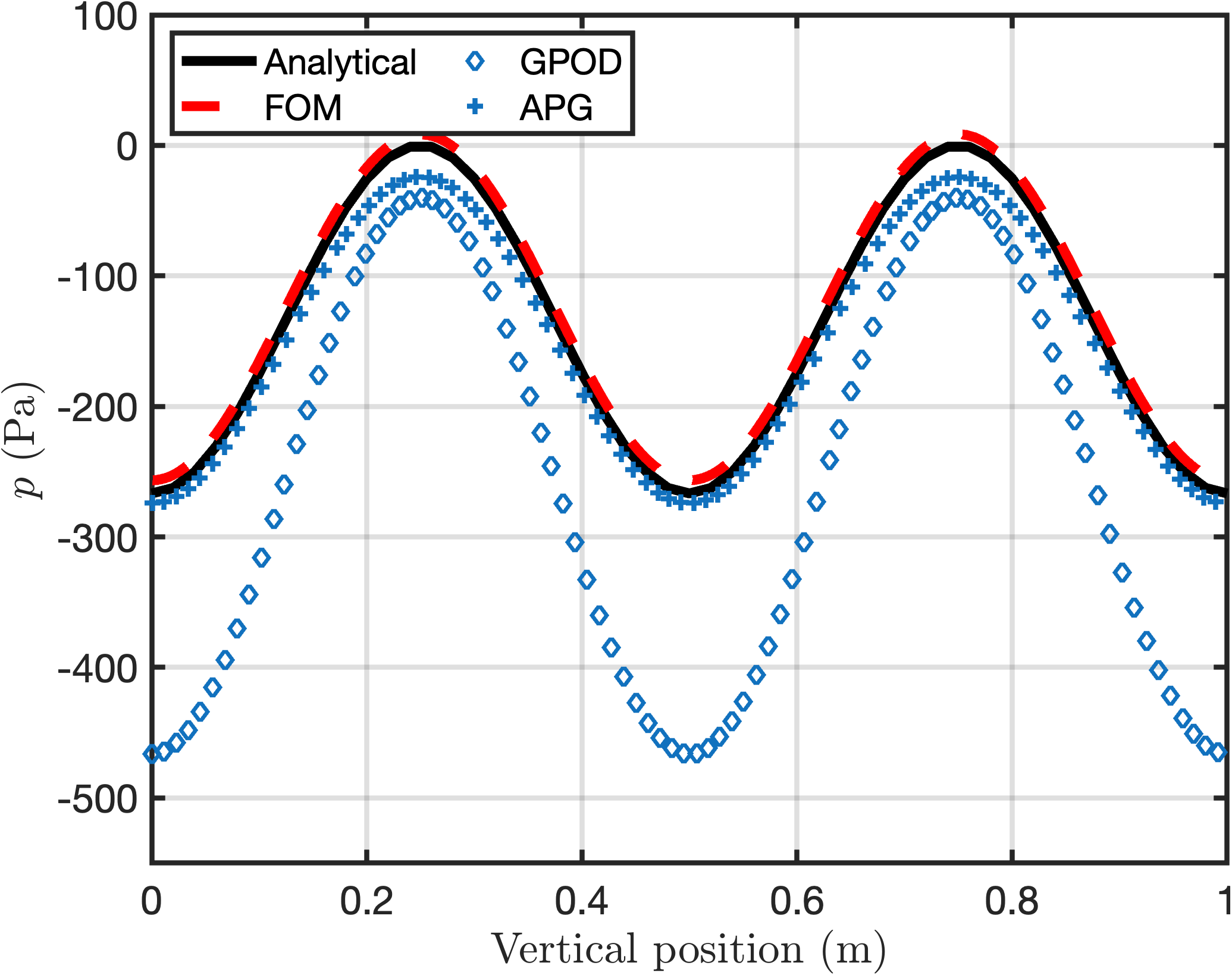}  
		\caption{$Re=125$}
		\label{fig:parametric_prediction_pressure_TGV_Re125}
	\end{subfigure}
	\begin{subfigure}{.4\textwidth}
		\centering
		\includegraphics[width=.9\linewidth]{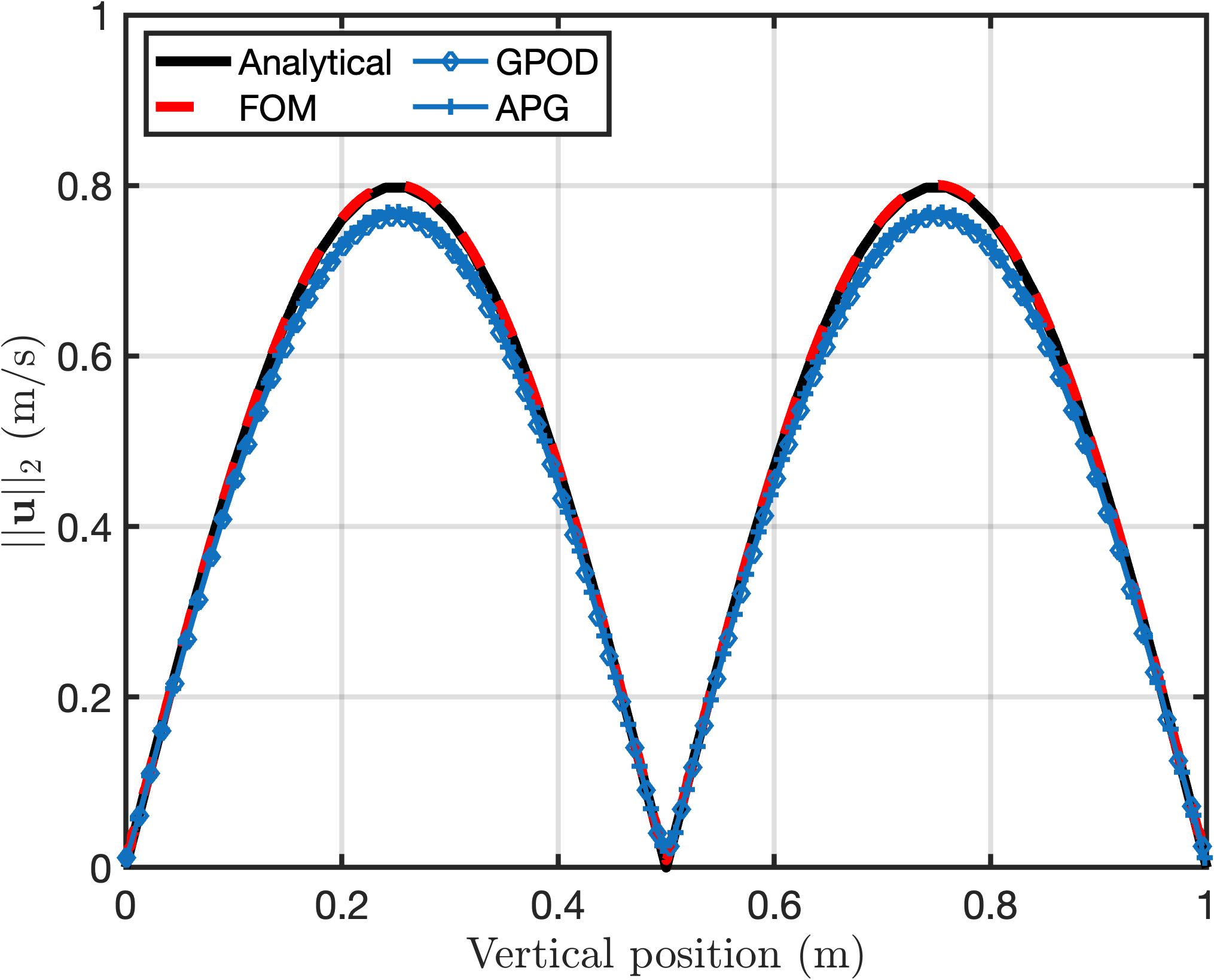}  
		\caption{$Re=175$}
		\label{fig:parametric_prediction_velocity_TGV_Re175}
	\end{subfigure}
	\hfill
	\begin{subfigure}{.4\textwidth}
		\centering
		\includegraphics[width=.9\linewidth]{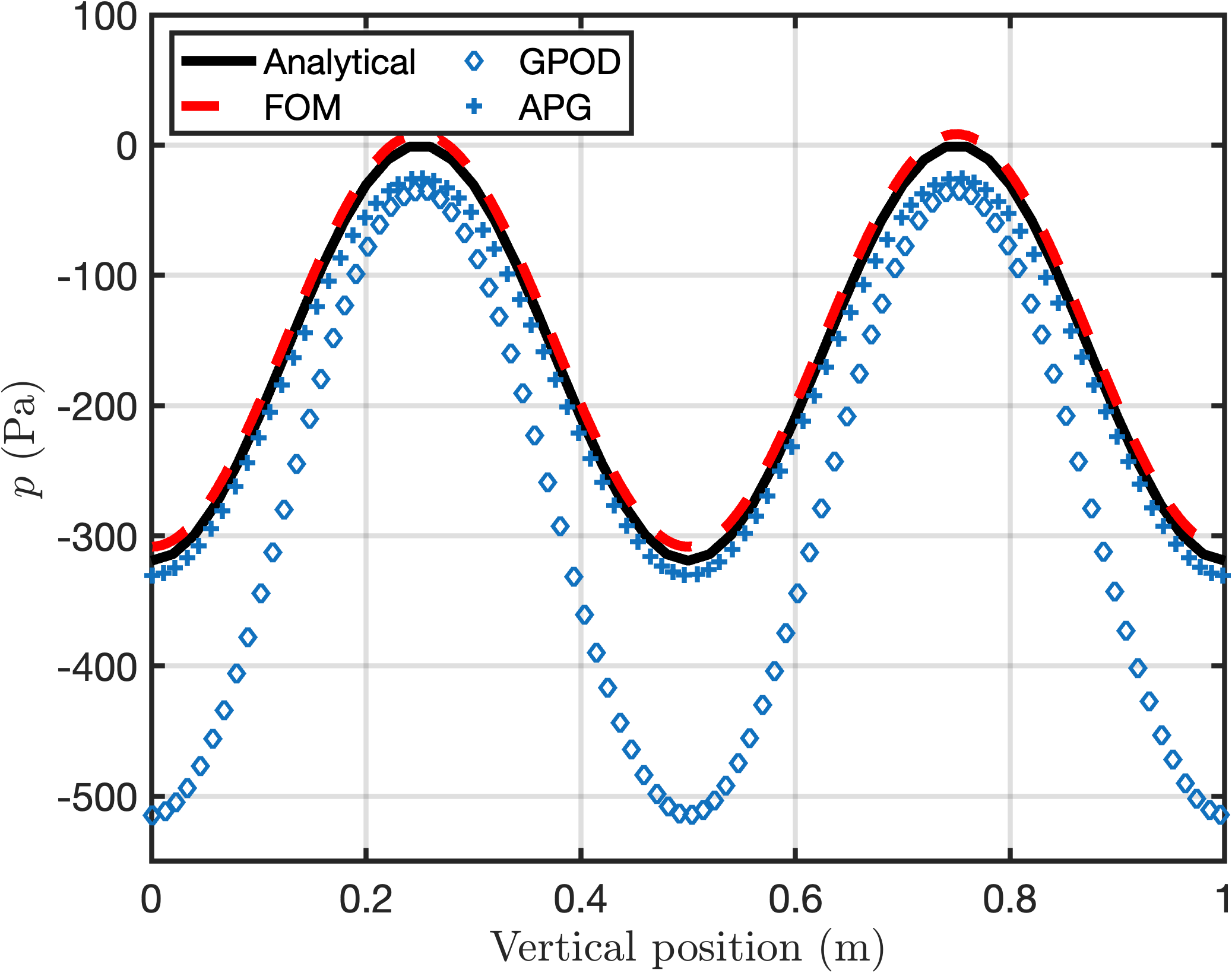}  
		\caption{$Re=175$}
		\label{fig:parametric_prediction_pressure_TGV_Re175}
	\end{subfigure}
		\begin{subfigure}{.4\textwidth}
		\centering
		\includegraphics[width=.9\linewidth]{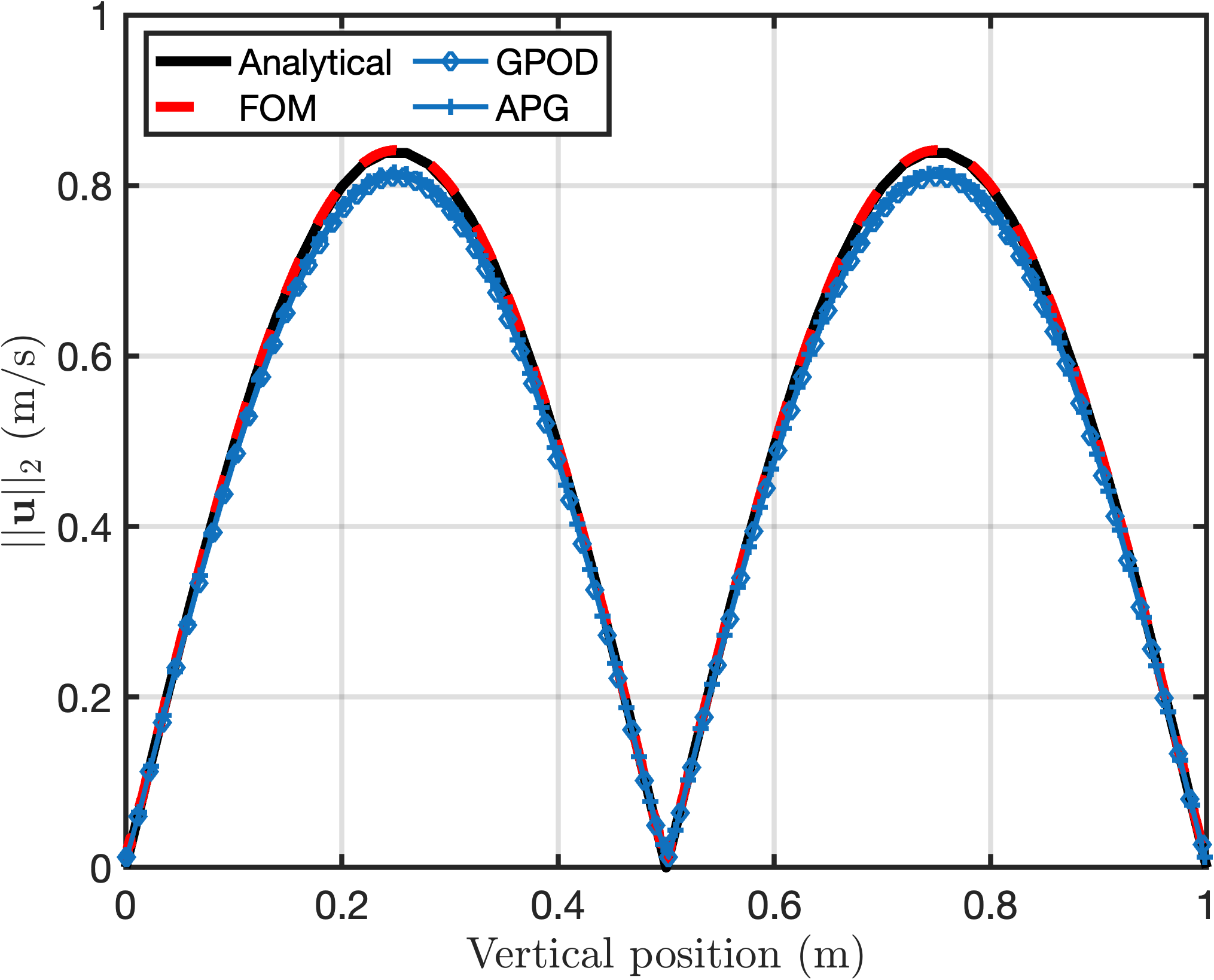}  
		\caption{$Re=225$}
		\label{fig:parametric_prediction_velocity_TGV_Re225}
	\end{subfigure}
	\hfill
	\begin{subfigure}{.4\textwidth}
		\centering
		\includegraphics[width=.9\linewidth]{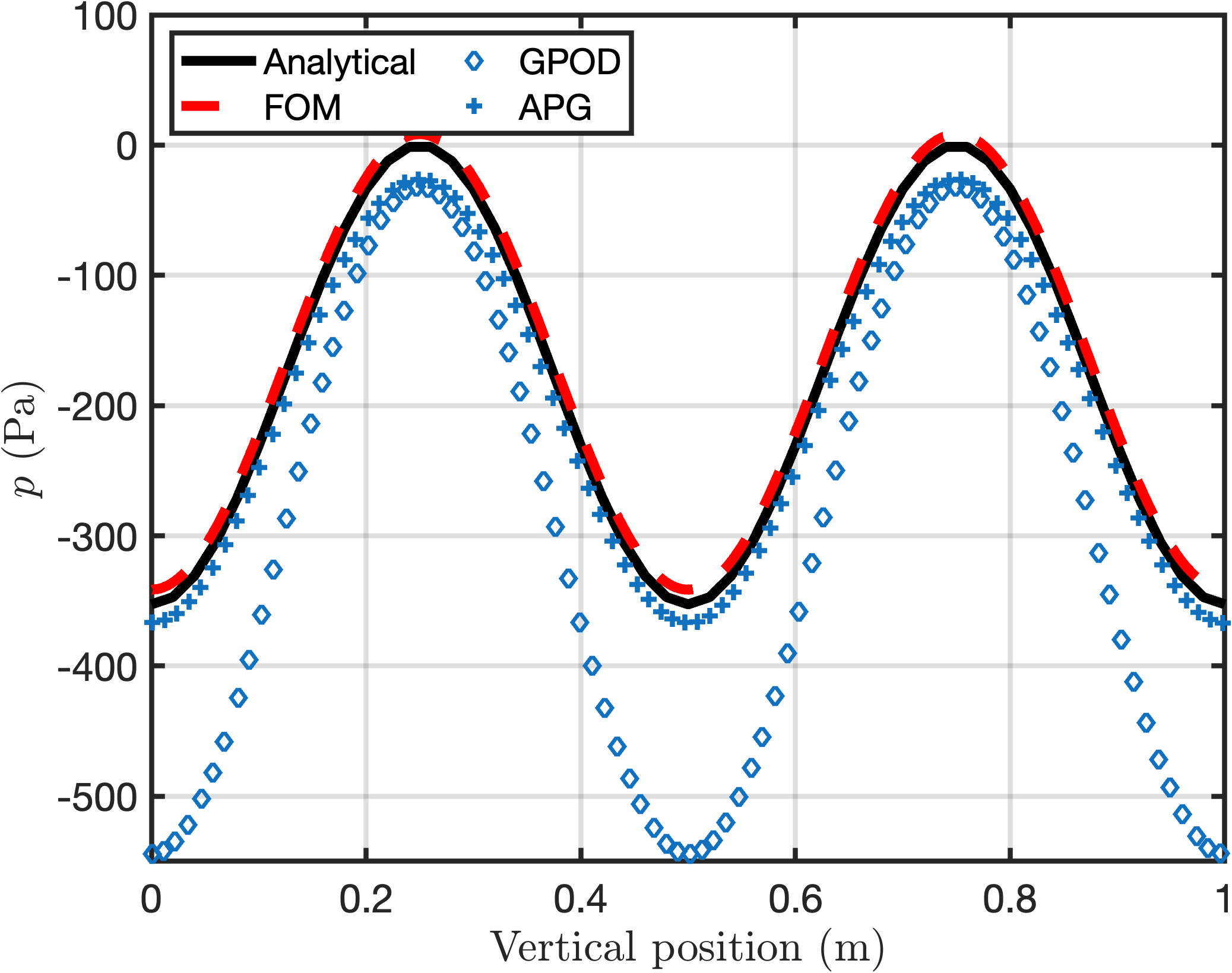}  
		\caption{$Re=225$}
		\label{fig:parametric_prediction_pressure_TGV_Re225}
	\end{subfigure}
	\caption{Centerline parametric predictions of the Taylor-Green vortex at $t=0.5$ s. Left column: velocity profile. Right column: pressure profile.}
	\label{fig:parametric_prediction_TGV}
\end{figure}

Vertical centerline velocity and pressure field profiles are presented in Fig.~\ref{fig:parametric_prediction_TGV} and are directly compared against the FOM and analytical solutions. These results reflect the full field solution from Figs.~\ref{fig:TGV_parametric_snapshot_velocity} and \ref{fig:TGV_parametric_snapshot_pressure} and highlight that the velocity field derived from GPOD and APG methods agree qualitatively with FOM and analytical solutions. Here, the large deviations from the GPOD pressure fields are further highlighted, while APG results good qualitative agreement.

\begin{figure*}[t!]
	\centering
	\begin{subfigure}[t]{0.5\textwidth}
		\centering
		\includegraphics[trim = {0cm 0cm 0 0},  scale=0.325]{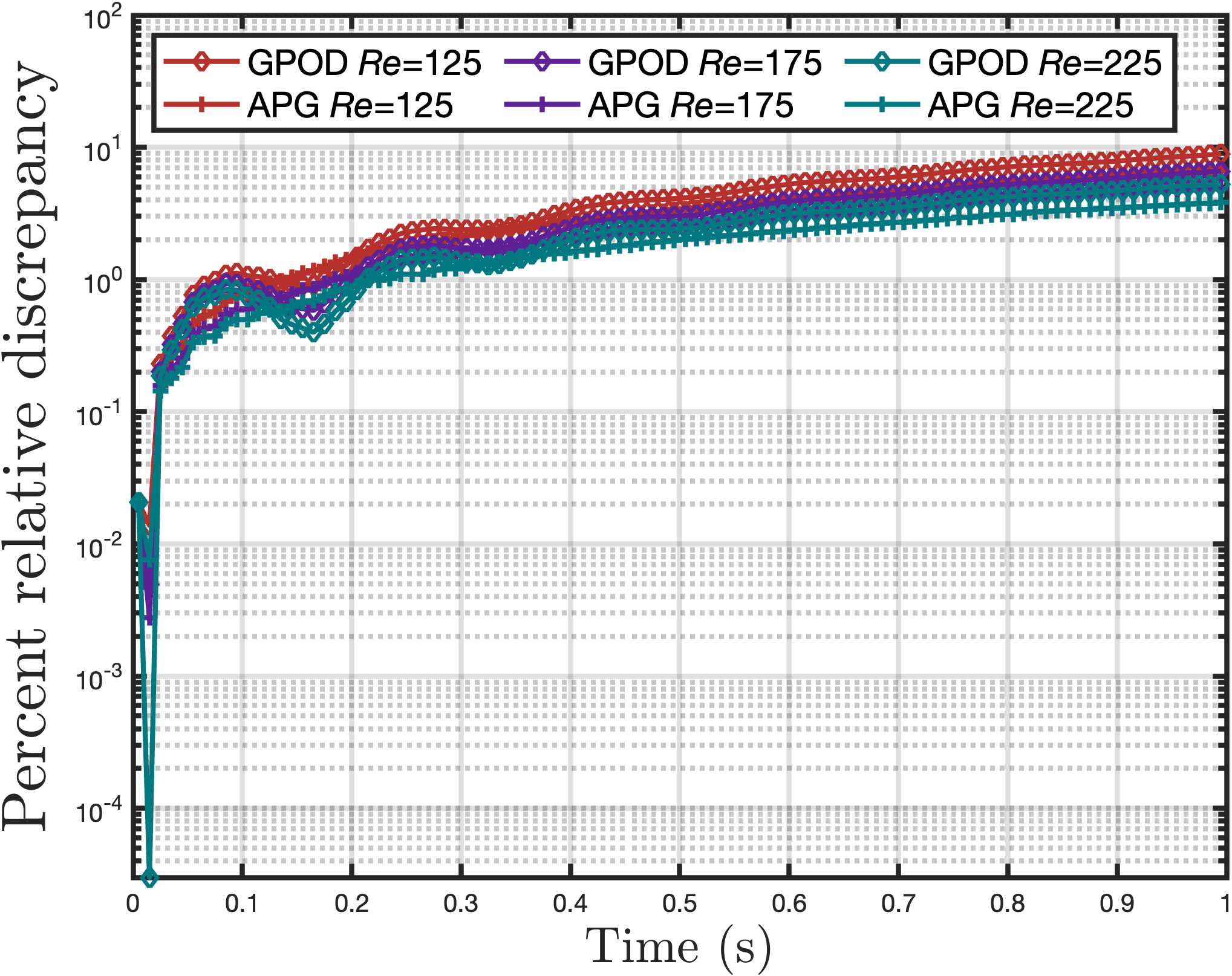}
		\caption{Velocity field relative discrepancy}
		\label{fig:parametric_errror_TGV_velocity}
	\end{subfigure}%
	\hfill
	\begin{subfigure}[t]{0.5\textwidth}
		\centering
		\includegraphics[ trim={0cm 0cm 0 0}, clip, scale=0.325]{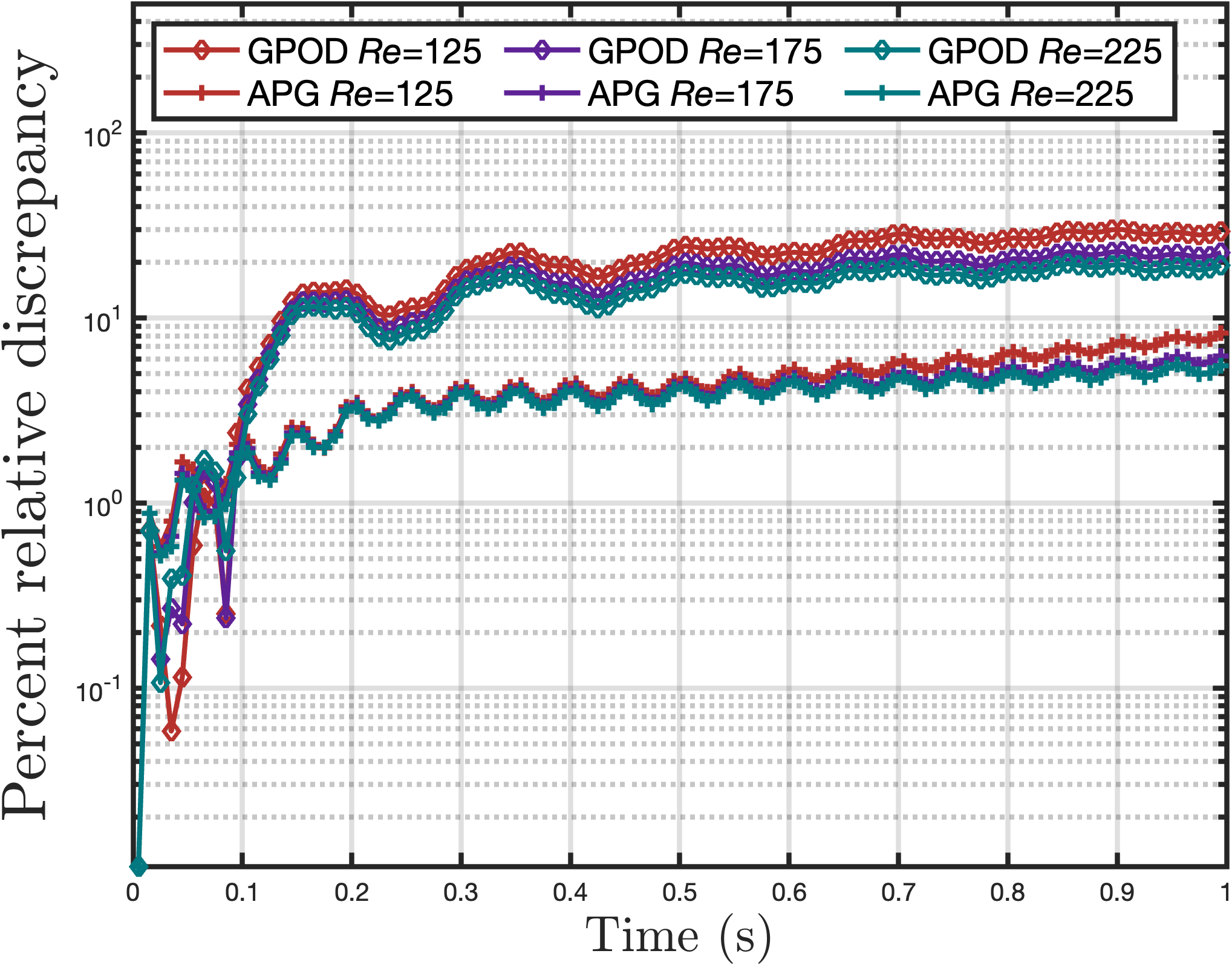}
		\caption{Pressure field relative discrepancy}
		\label{fig:parametric_errror_TGV_pressure}
	\end{subfigure}
	\caption{Time histories of centerline relative discrepancy errors.}
		\label{fig:parametric_errror_TGV}
\end{figure*}

Finally, percent relative discrepancy errors are shown in Fig.~\ref{fig:parametric_errror_TGV}. Velocity field errors in Fig.~\ref{fig:parametric_errror_TGV_velocity} showcase results with a peak around 10\% at $Re=125$. Discrepancy errors for $Re=175$ and $Re=225$ show peaks at around 4\% and 8\%. In all velocity cases, APG marginally outperforms GPOD. It is important to highlight that there is an exponential relationship between decay rates of the vortex dynamics and Reynolds numbers. Therefore, for the physical time window of interest in the current experiments (1 physical second), as the Reynolds numbers increase, the behavior of the decay rate approaches a linear decay-rate. Thereby, the POD subspace performs better for slow-decaying dynamics since it is better suited to embed linear dynamics in its span. Pressure field errors in Fig.~\ref{fig:parametric_errror_TGV_pressure} show peak discrepancies near 30\% for GPOD at $Re=125$ and around 20\% for both $Re=175$ and $Re=225$. On the other hand, the APG approach outperforms GPOD for all cases with discrepancies errors below 10\%. Pressure field errors again indicate that POD subspaces are sensitive to the weakly compressible dynamics assumed in the current SPH framework. It is important to note that although a dimensionality of $M=5$ attenuated higher-frequency content in the density field for the reconstructive case, the GPOD framework does not sufficiently reconcile the variations in the density field within the five-dimensional subspace. However, the APG framework provides ``adjoint stabilization" \cite{parish2020adjoint} that results in significant improvements in pressure field predictions with $M=5$ in the parametric setting. Overall, the proposed meshless PMOR provides good performance in a predictive and parametric setting, and there is ongoing work to more rigorously study the impact of weak compressibility on the spectral content of subspace embeddings.

\subsection{Lid-driven cavity}
The lid-driven cavity benchmark problem is employed to test the proposed meshless PMOR approach. The domain is defined by the unit square, $\Omega=[0, 1] \times [0, 1]$ m$^2$. All simulations, including PMOR results, run for a physical time of $T_f=10$ s and employ a smoothing length of $h=2\Delta x$, an interior particle resolution of $200\times200$, three layers of ghost particles, and a time-step of $\Delta t=2\times 10^{-4}$ s. No-slip boundary conditions are enforced with ghost particles on the lateral and bottom walls. The top wall consists of fixed ghost particles with the following velocity profile to avoid singularities at the corners, $U=(1-(2x-1)^{14})^2$ m/s \cite{lind2016high}. Here, the Reynolds number is varied by changing fluid viscosity and a reference density of $1$ kg/m$^3$ is employed. A Mach number of $M_a=0.1$ is chosen with a reference speed of $U_{\textup{ref}}=1$ m/s.

\subsubsection{Reconstructive results}
\begin{figure*}[t!]
	\centering
	\begin{subfigure}[t]{0.5\textwidth}
		\centering
		\includegraphics[trim = {0cm 0cm 0 0},  scale=0.325]{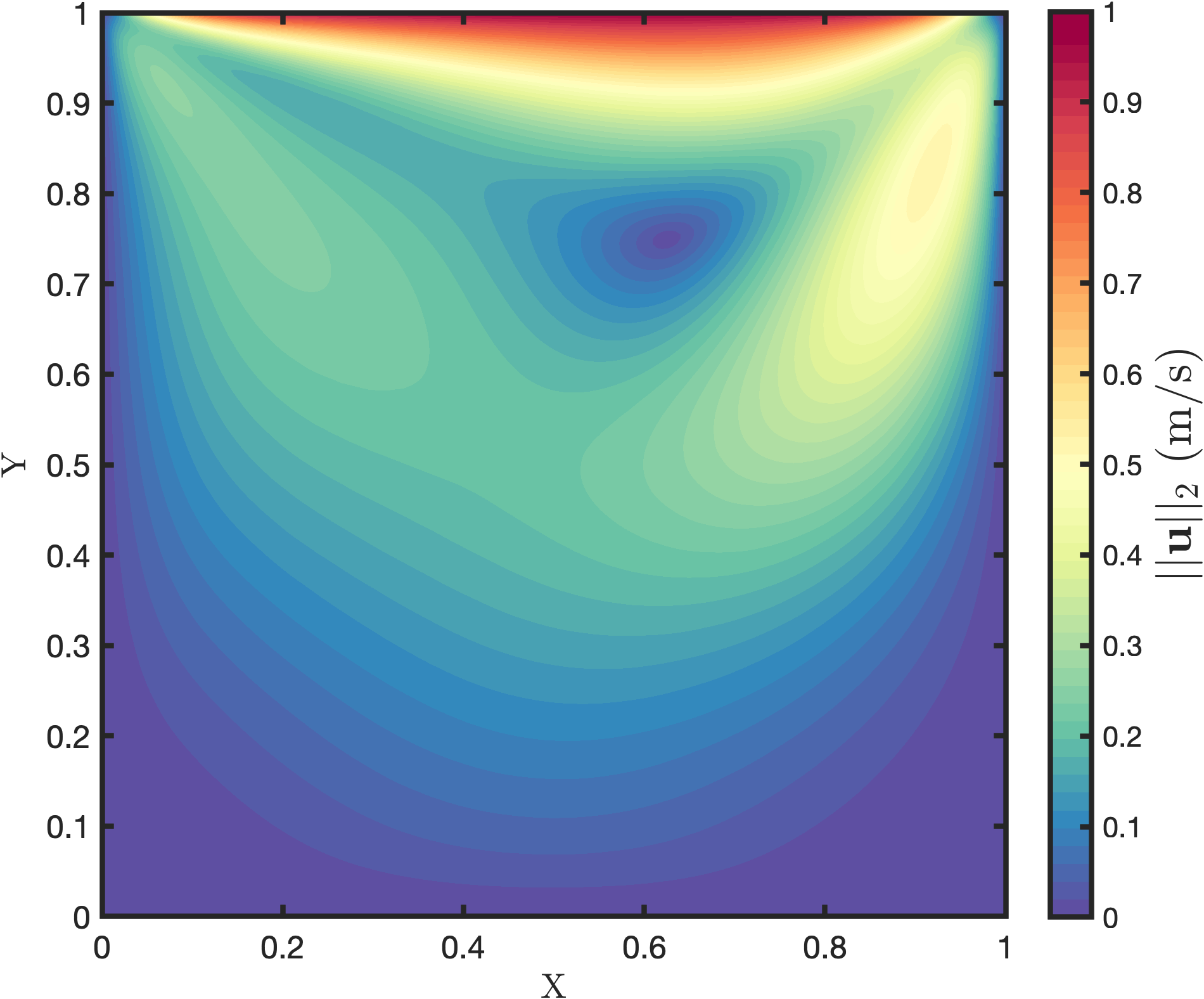}
		\caption{FOM velocity field}
	\end{subfigure}%
	\hfill
	\begin{subfigure}[t]{0.5\textwidth}
		\centering
		\includegraphics[ trim={0cm 0cm 0 0}, clip, scale=0.325]{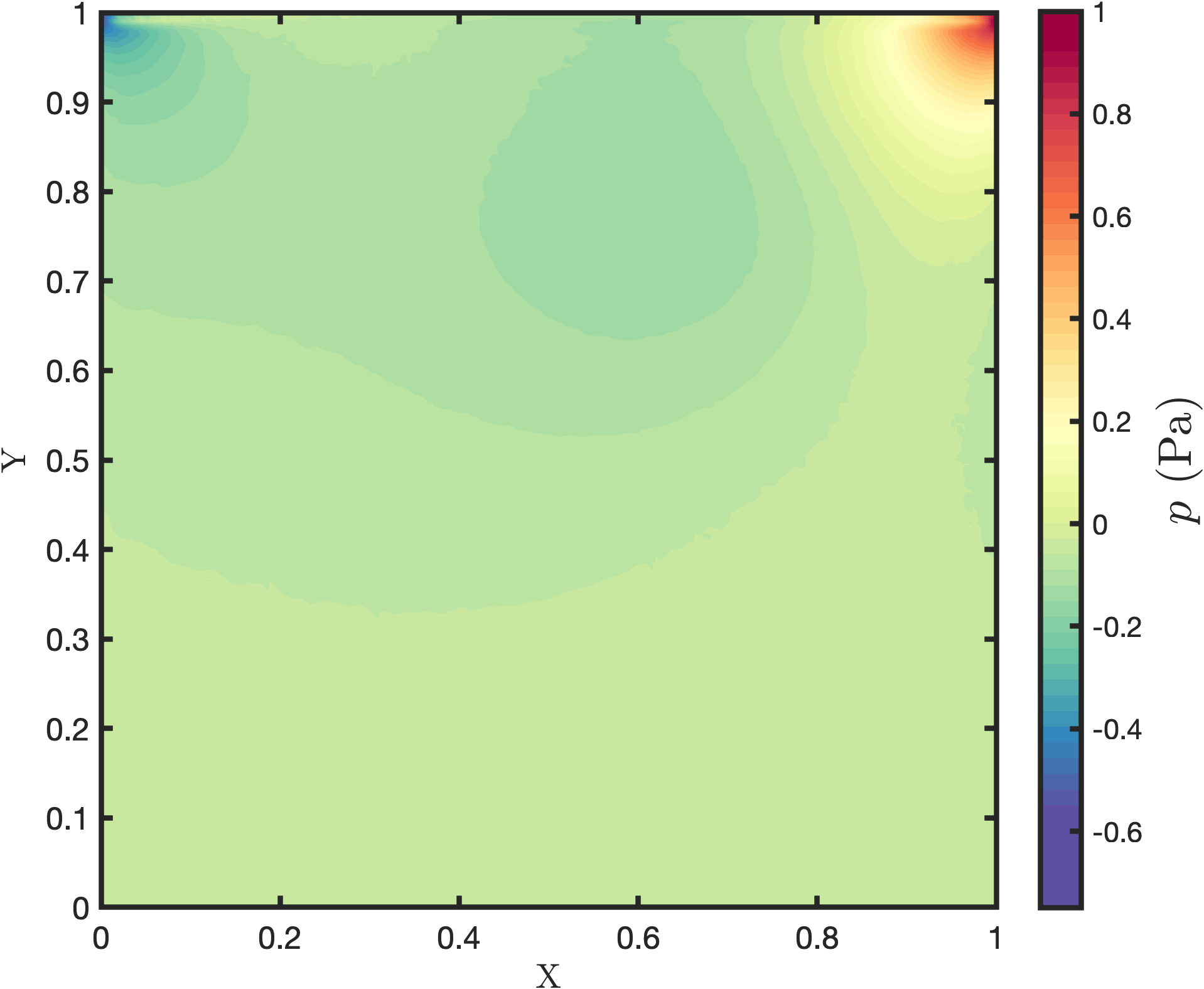}
		\caption{FOM pressure field}
	\end{subfigure}
	\caption{Lid-driven cavity at $Re=100$. FOM snapshot at $t=5$ s.}
	\label{fig:ldc_Re100}
\end{figure*}

\begin{figure}[t!]
	\centering
\centering
	\begin{subfigure}[t]{0.5\textwidth}
		\centering
		\includegraphics[trim = {0cm 0cm 0 0},  scale=0.325]{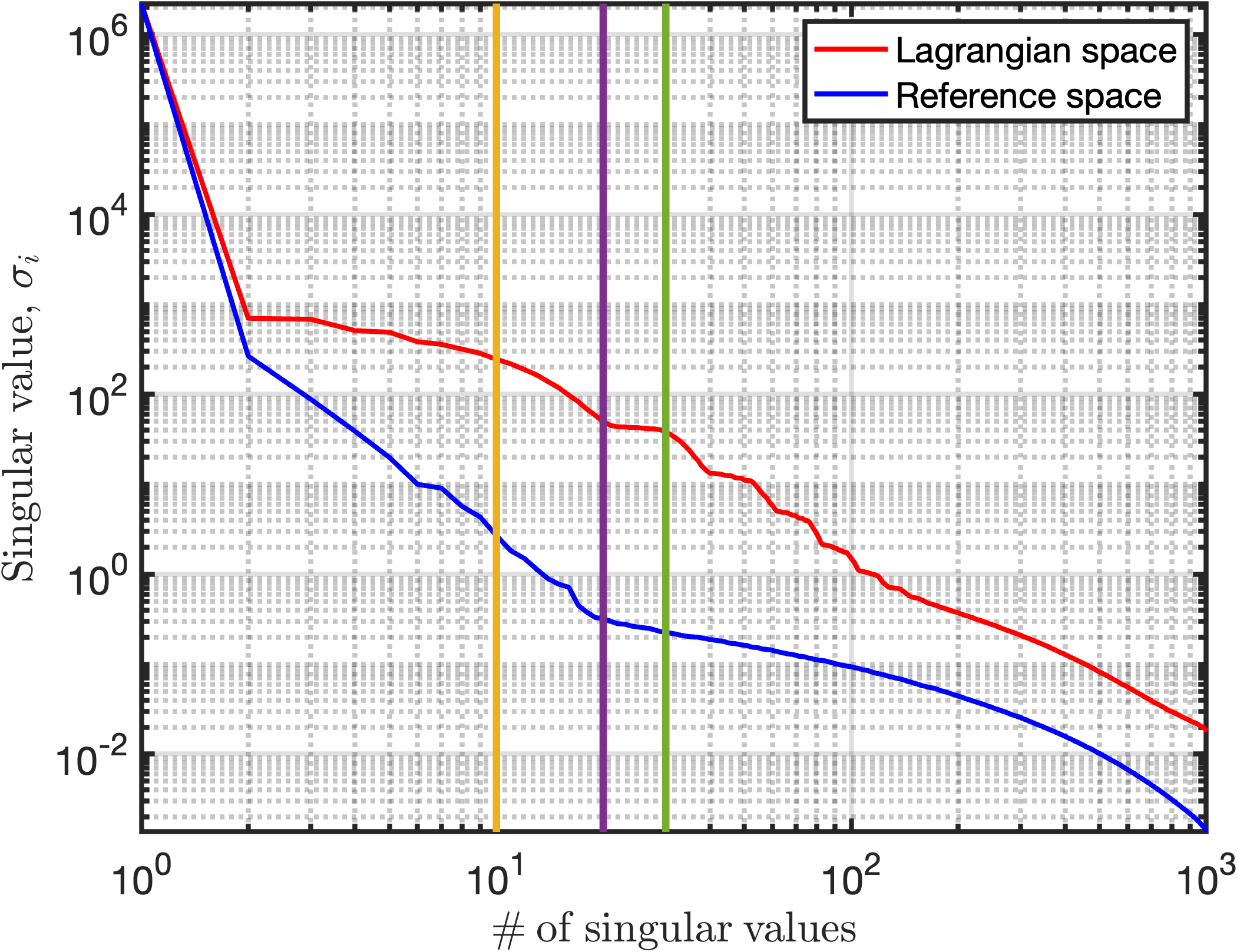}
		\caption{Singular value decay}		
	\end{subfigure}%
	\hfill
	\begin{subfigure}[t]{0.5\textwidth}
		\centering
		\includegraphics[ trim={0cm 0cm 0 0}, clip, scale=0.325]{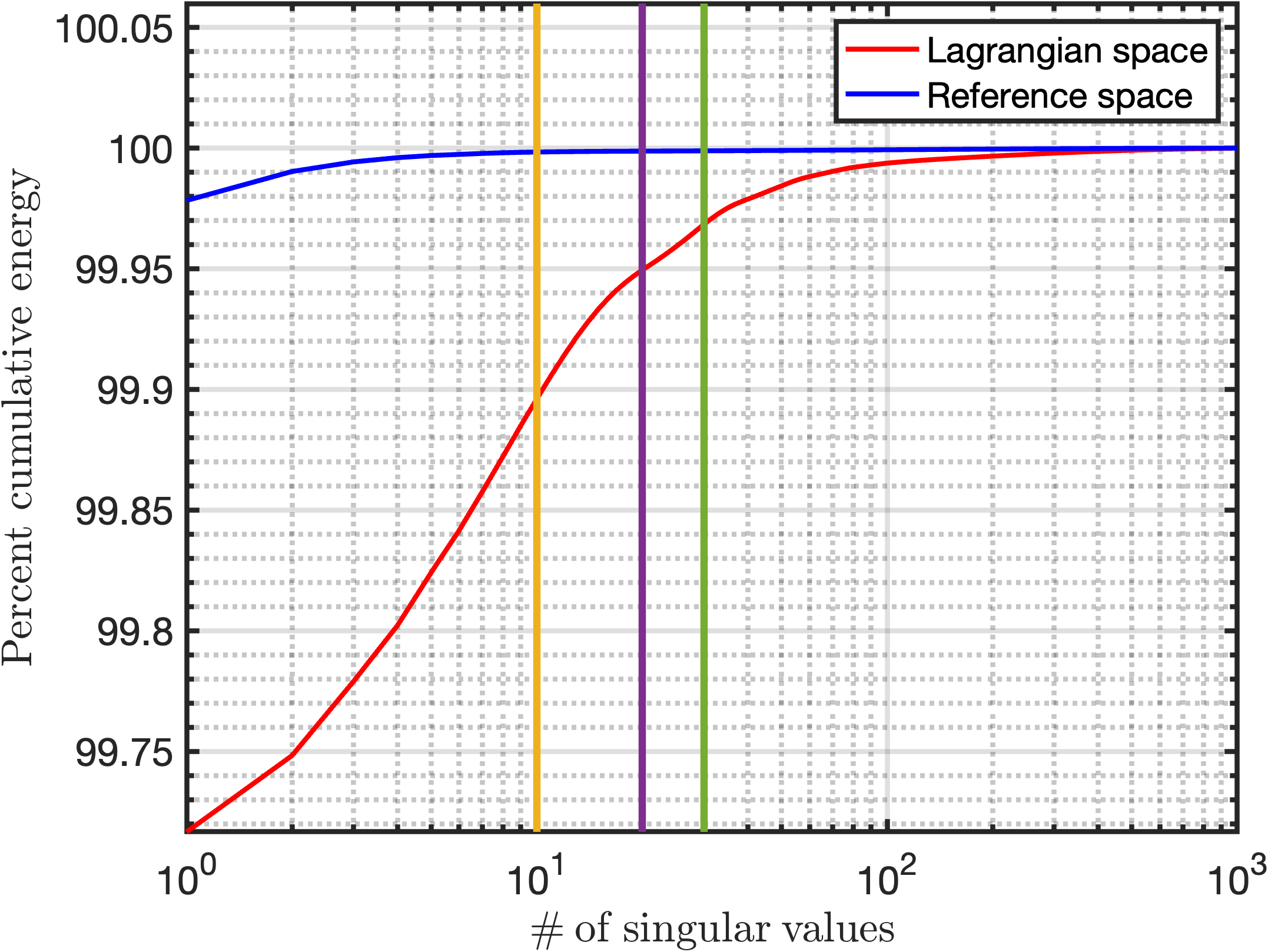}
		\caption{Cumulative energy contribution}
	\end{subfigure}
	\caption{Singular values derived from Lagrangian and reference space snapshot matrices. Vertical colored lines are meant to highlight the differences in singular value decay and cumulative energy between Lagrangian and reference space at $M=10$ (yellow), $M=20$ (purple), and $M=30$ (green).} 	\label{fig:LDC_LagrangianVReferenceSingularValues}
	\centering
\end{figure}

Full-order model velocity and pressure field results for $Re=100$ are shown in Fig.~\ref{fig:ldc_Re100}. Snapshots of the FOM are collected at an interval of 50 for a total of 1000 snapshots. Dimensionality reduction is performed on $Re=100$ for the reconstructive results, and the singular value decay for Lagrangian and reference spaces is shown in Fig.~\ref{fig:LDC_LagrangianVReferenceSingularValues}. There is a large discrepancy between singular value decay rates and corresponding cumulative energy contributions derived from Lagrangian and reference spaces. For instance, by mode 10 $99.9984\%$ of the energy content is retained in the reference space, while in the Lagrangian space only $99.89\%$ is retained. By mode 20, $99.9988\%$ is held in the reference space, while the Lagrangian space only contains $99.94\%$. At mode 30, $99.9989\%$ of the energy is represented in reference space and only $99.9684\%$ is in the Lagrangian space.

\begin{figure}[t!]
	\centering
	\includegraphics[trim = 0cm 0cm 0 0, scale=0.4]{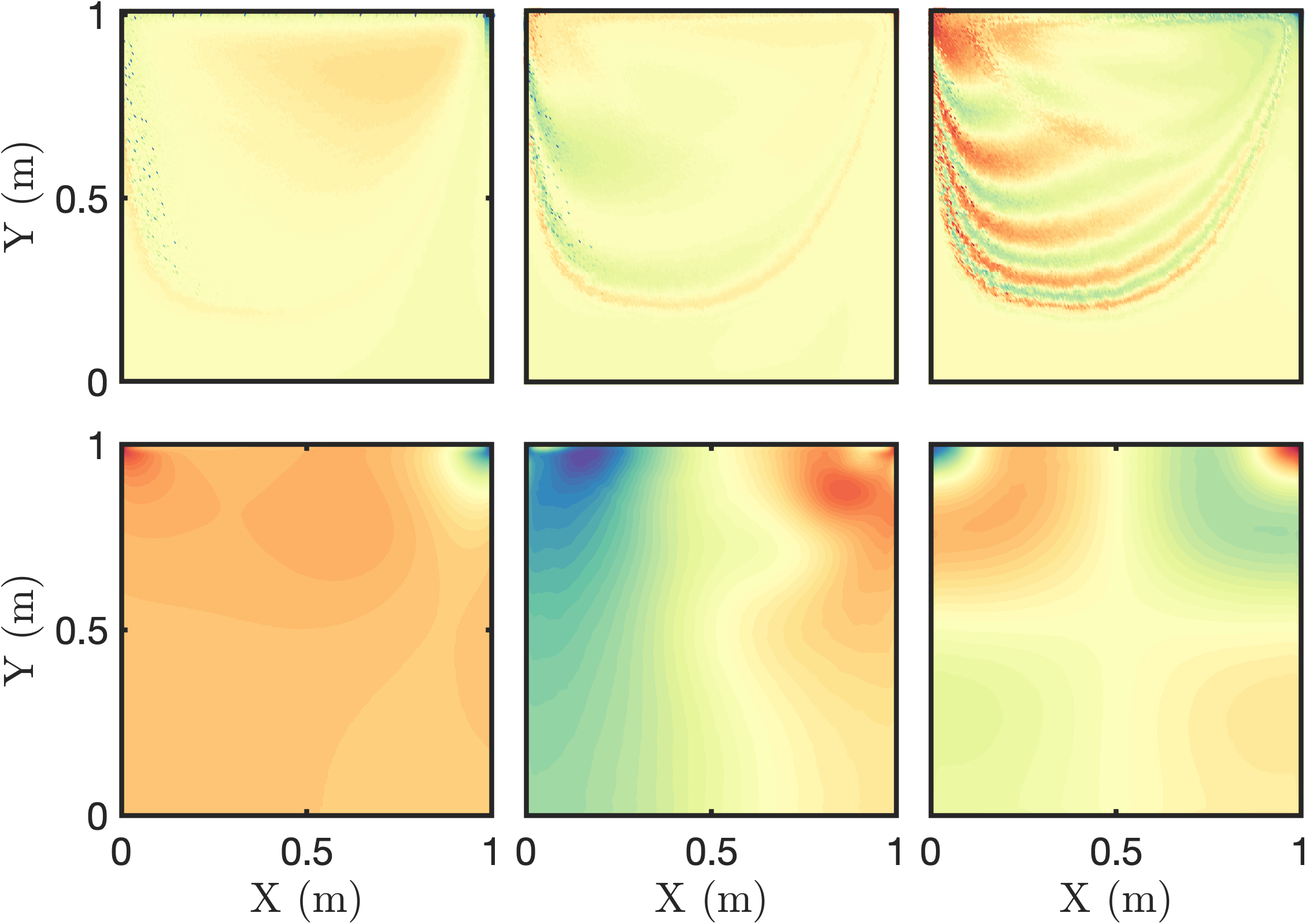}
	\caption{Density field modes, from left to right $M=1, 5, {\text{and}\:} 10$. Top row: Lagrangian space; Bottom row: Reference space.} 
	\label{fig:LDC_rhomode}
	\centering
\end{figure}

\begin{figure}[t!]
	\centering
	\includegraphics[trim = 0cm 0cm 0 0, scale=0.4]{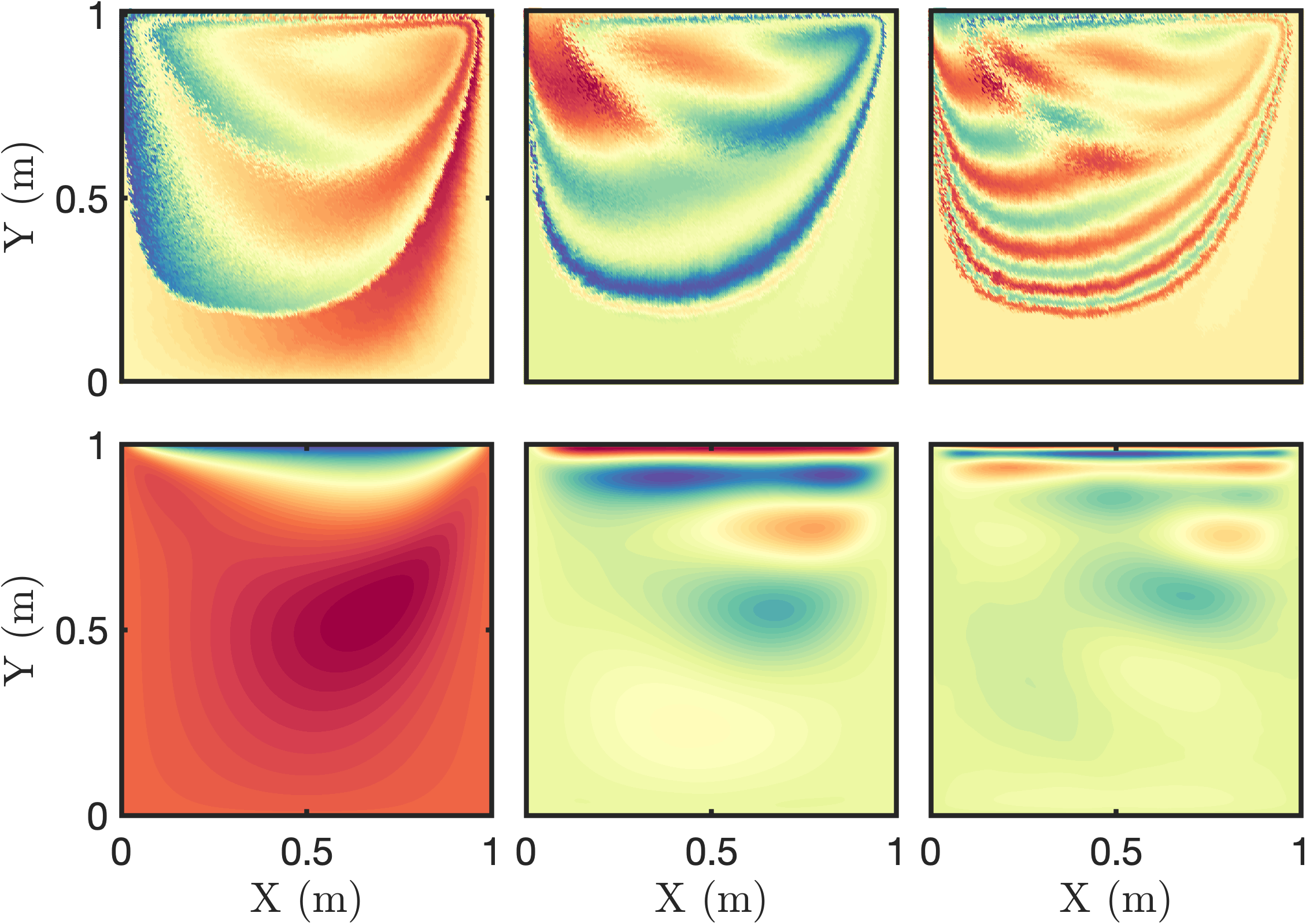}
	\caption{Velocity field $x-$component modes, from left to right $M=1, 5, {\text{and}\:} 10$. Top row: Lagrangian space; Bottom row: Reference space.} 
	\label{fig:LDC_xmode}
	\centering
\end{figure}

\begin{figure}[t!]
	\centering
	\includegraphics[trim = 0cm 0cm 0 0, scale=0.4]{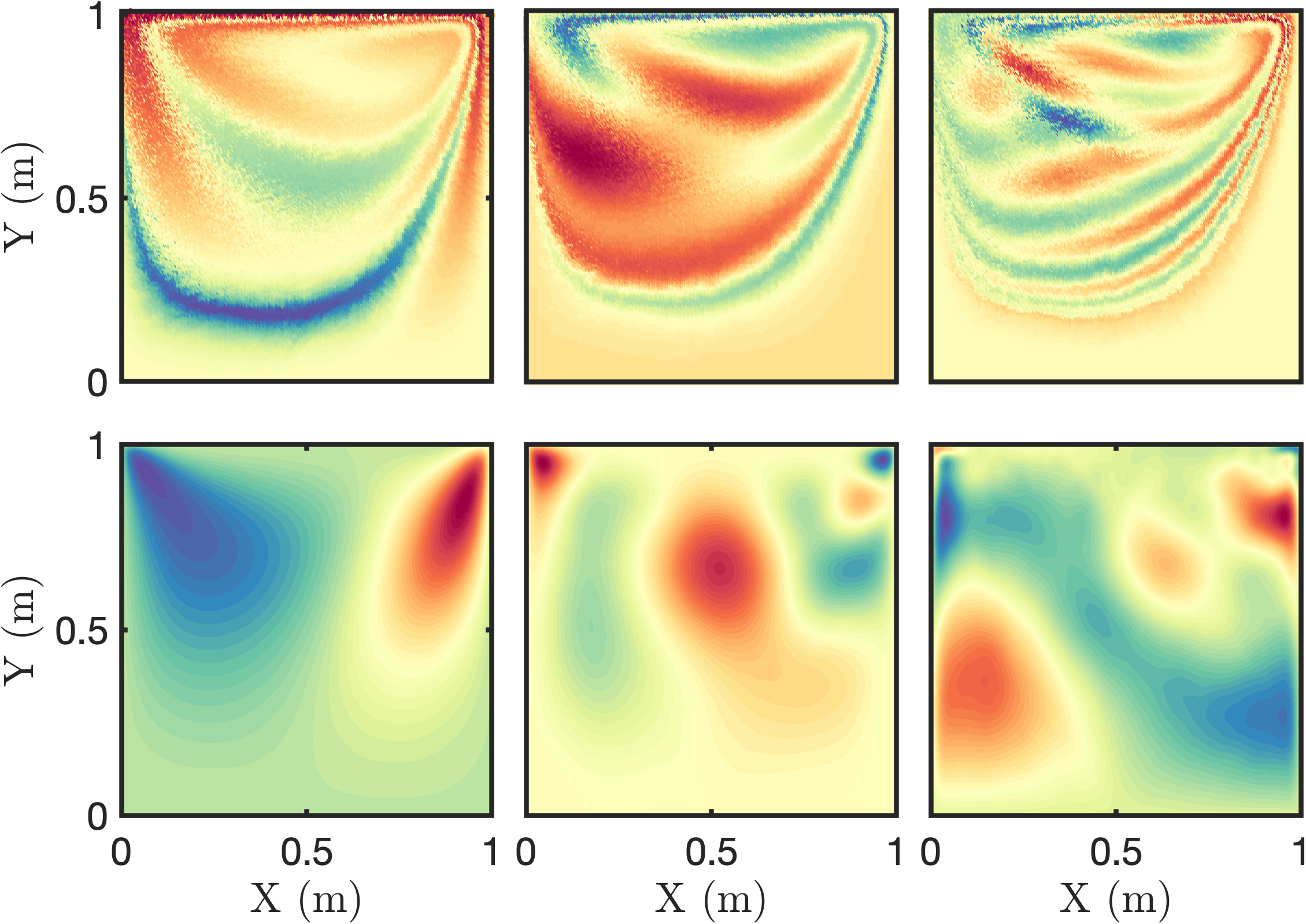}
	\caption{Velocity field $y-$component modes, from left to right $M=1, 5, {\text{and}\:} 10$. Top row: Lagrangian space; Bottom row: Reference space.} 
	\label{fig:LDC_ymode}
	\centering
\end{figure}

Density and velocity field modes $M=1, 5, {\text{and}\:} 10$ are shown in Figs.~\ref{fig:LDC_rhomode}-\ref{fig:LDC_ymode}. Note that the first mode in the density field derived in Lagrangian space, shown in Fig.~\ref{fig:LDC_rhomode} exhibits large variations in modal content scales which could lead to ill-conditioned subspaces. In contrast, the reference space counterpart in the density field exhibits smooth behavior that aligns closely with the mean density field of the FOM. Higher density modes in Lagrangian space strongly exhibit the mixing of the numerical topology in their modal quantities as opposed to the reference space modes, which exhibit modal behavior aligned with expected flow field characteristics from the lid-driven cavity problem. Similar mixing characteristics are observed in the $x$ and $y$ velocity modal quantities.

\begin{figure}[t!]
	\centering
	\includegraphics[trim = 0cm 0cm 0 0, scale=0.5]{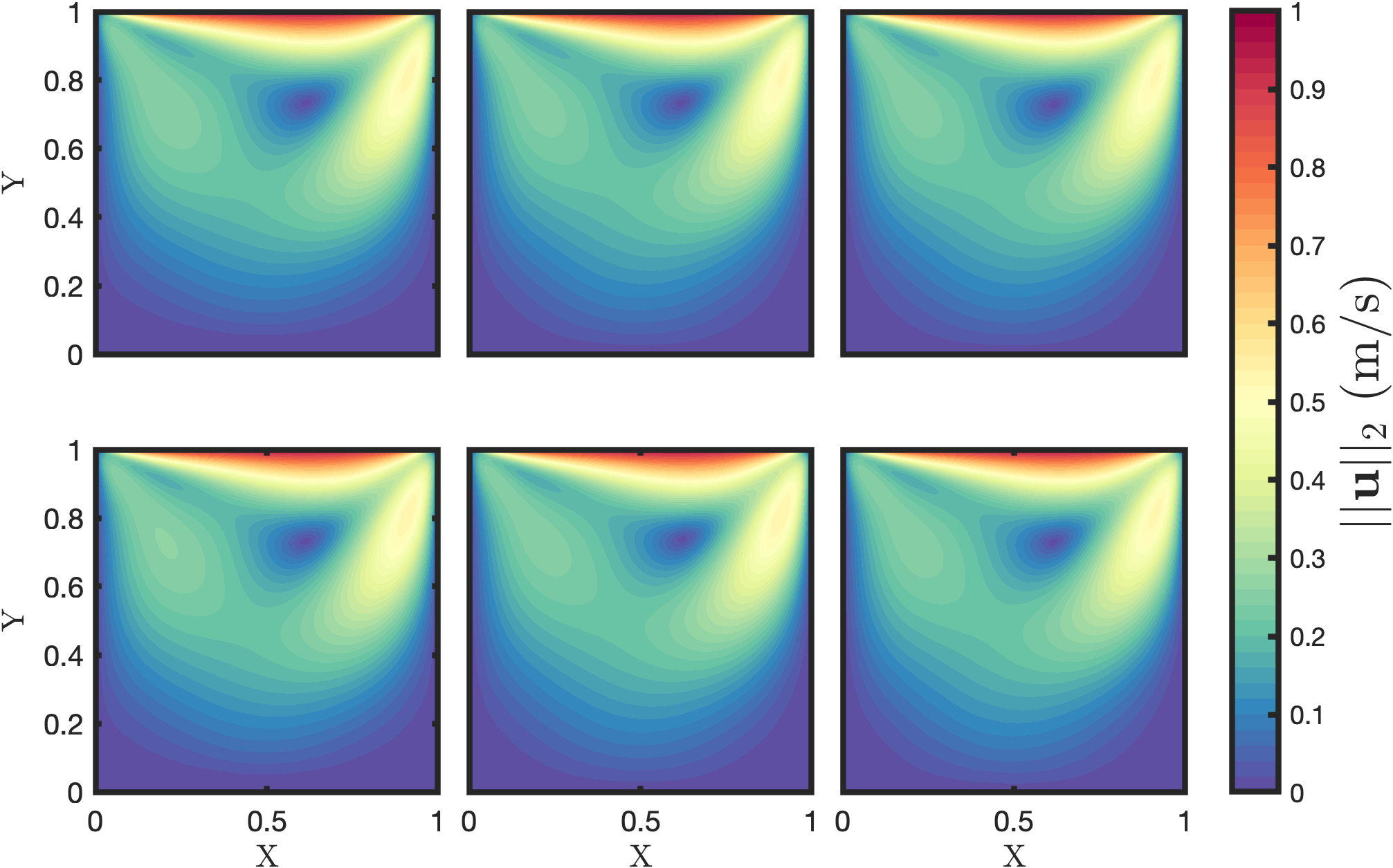}
	\caption{Snapshot at $t=5$ s of the reconstructed velocity field by GPOD (top row) and APG (bottom row). Columns from left to right indicate Case 1 - Case 3 of the PMOR basis selection.} 
	\centering
	\label{fig:LDC_reconstructive_velocity}
\end{figure}

\begin{figure}[t!]
	\centering
	\includegraphics[trim = 0cm 0cm 0 0, scale=0.5]{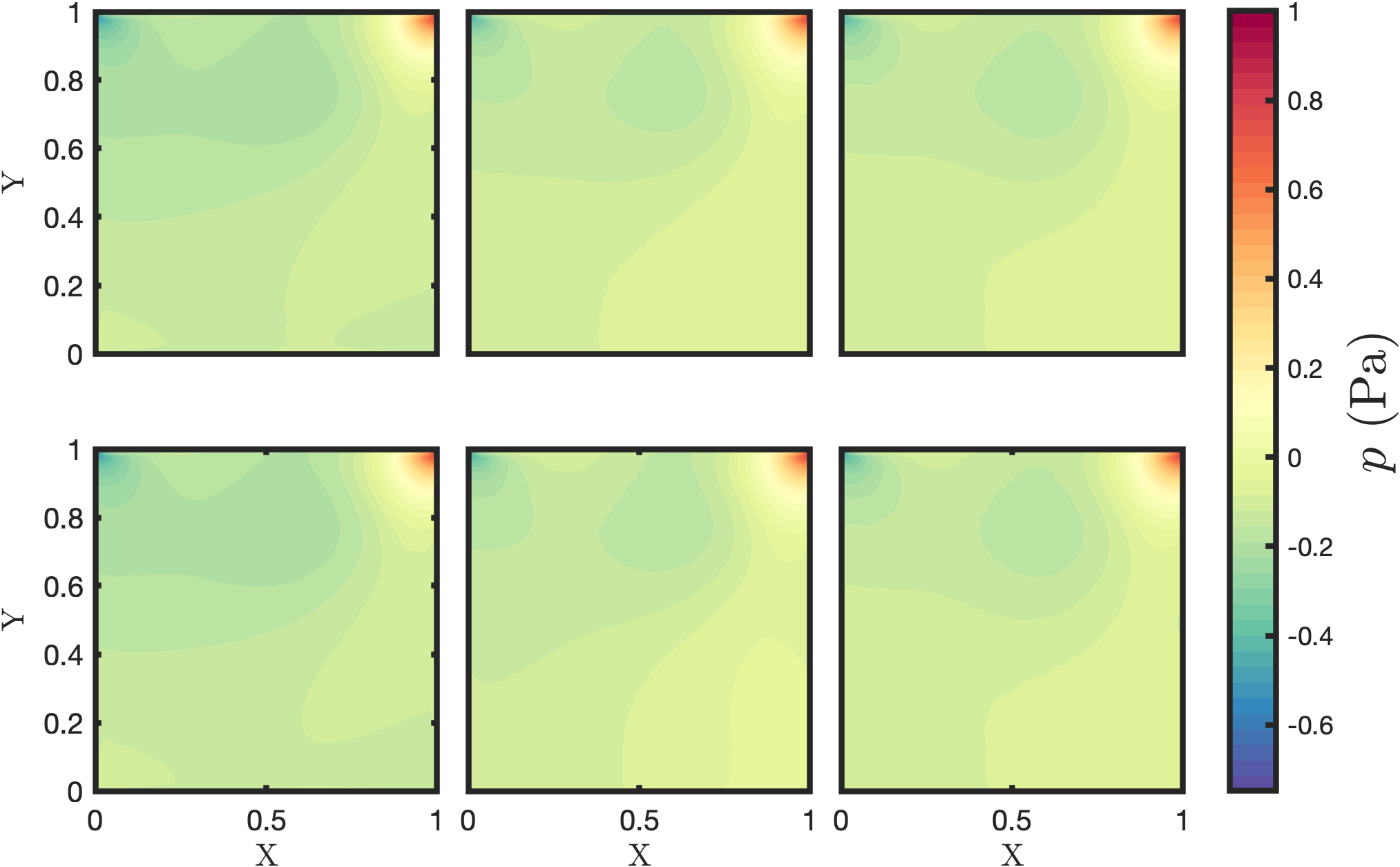}
	\caption{Snapshot at $t=5$ s of the reconstructed pressure field by GPOD (top row) and APG (bottom row). Columns from left to right indicate Case 1, 2 and 3 of the PMOR basis selection.}
	\centering
	\label{fig:LDC_reconstructive_pressure}
\end{figure}

Reconstructive flow field snapshot results for $t=5$ s are shown in Figs.~\ref{fig:LDC_reconstructive_velocity} and \ref{fig:LDC_reconstructive_pressure}. For both GPOD and APG methods, the following cases were tested: POD basis dimension of Case 1) $M=10$; Case 2) $M=20$; Case 3) $M=30$. The selected basis dimensions result in compression factors for GPOD/APG Case 1, 2, and 3 of $CF = 20, 000$, $CF = 10, 000$ and $CF = 6, 667$, respectively. Finally, a memory length of $\tau=10^{-7}$ s was selected for the APG method. A more detailed analysis to guide the selection of the memory length and quantifying its dependence on the weakly-compressible assumption are outside the scope of this work and is a promising area for future research. 

\begin{figure*}[t!]
	\centering
	\begin{subfigure}[t]{0.5\textwidth}
		\centering
		\includegraphics[trim = {0cm 0cm 0 0},  scale=0.325]{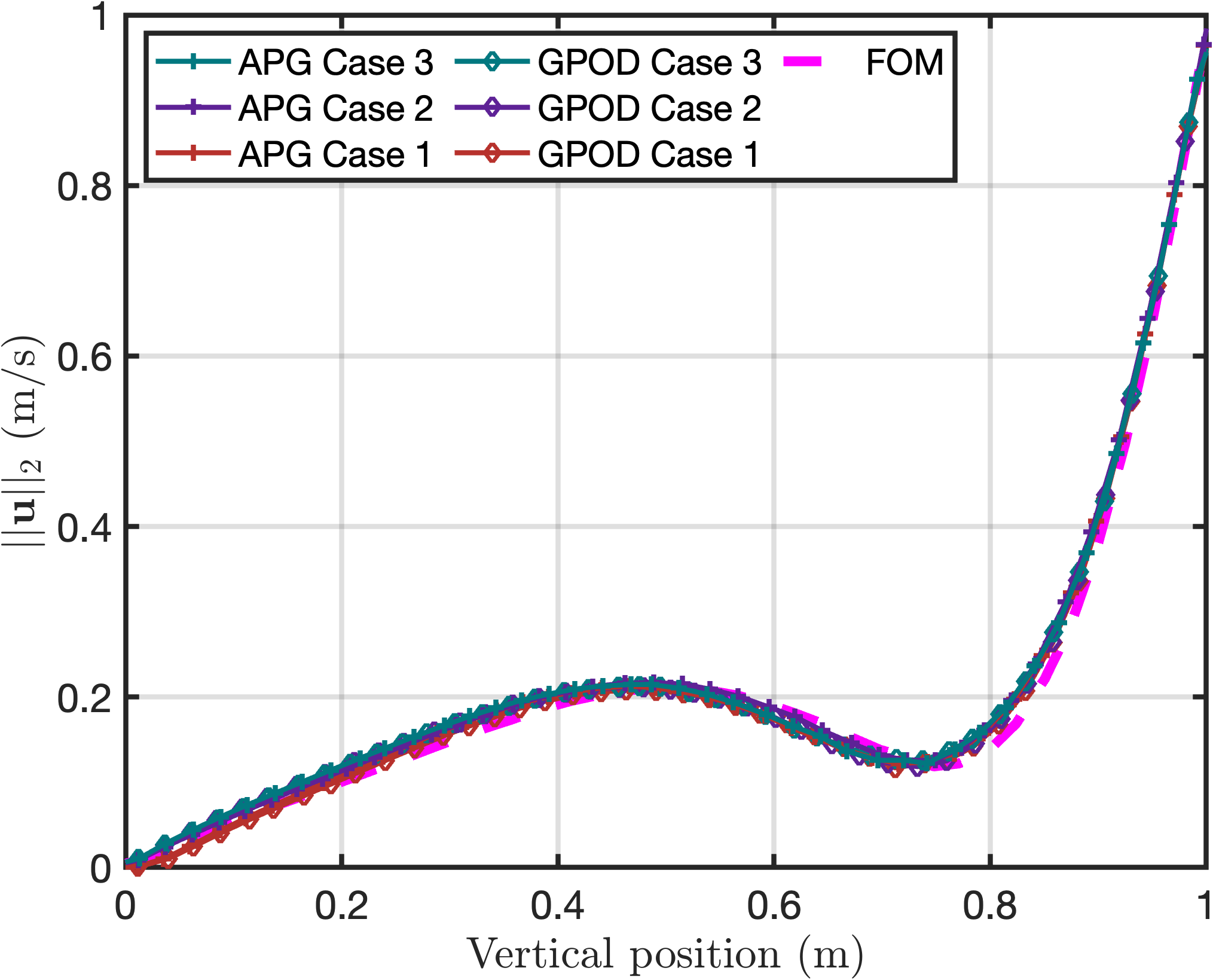}
		\caption{Velocity field at vertical centerline}
     	\label{fig:LDC_centerline_velocity_reconstructive}
	\end{subfigure}%
	\hfill
	\begin{subfigure}[t]{0.5\textwidth}
		\centering
		\includegraphics[ trim={0cm 0cm 0 0}, clip, scale=0.325]{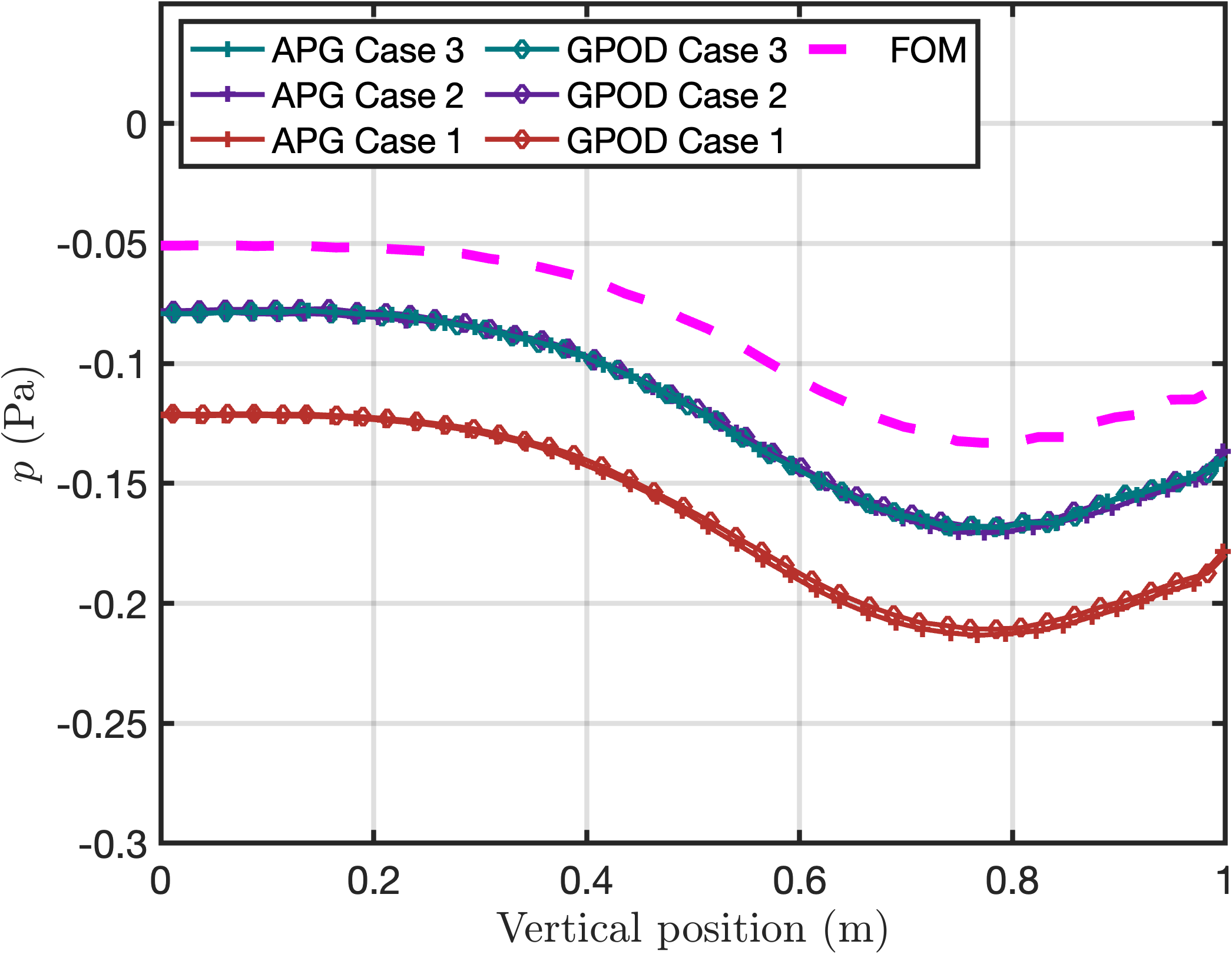}
		\caption{Pressure field at vertical centerline}
		\label{fig:LDC_centerline_pressure_reconstructive}
	\end{subfigure}
	\caption{Vertical centerline snapshot at $t=5$ s.}
\end{figure*}

Qualitative results show good agreement in the velocity and pressure field for all cases from GPOD and APG methods. Closer examination of the vertical centerline plots in Fig.~\ref{fig:LDC_centerline_velocity_reconstructive} shows that the velocity field closely aligns with centerline profile of the FOM for all cases. However, the pressure field centerline in Fig.~\ref{fig:LDC_centerline_pressure_reconstructive} shows that for GPOD and APG Case 1 a larger local discrepancy between profiles is observed, while for Case 2 and 3, a smaller discrepancy is present. It is important to highlight, however, that the centerline pressure fields contribute relatively low energy content into the dynamics, as opposed to the top corners of the domain. Therefore, the centerline plots exhibit a local discrepancy that is small relative to the global dynamics of the pressure field, which aligns well with the FOM, as shown in Fig.~\ref{fig:LDC_reconstructive_pressure}.

\begin{figure*}[t!]
	\centering
	\begin{subfigure}[t]{0.5\textwidth}
		\centering
		\includegraphics[trim = {0cm 0cm 0 0},  scale=0.325]{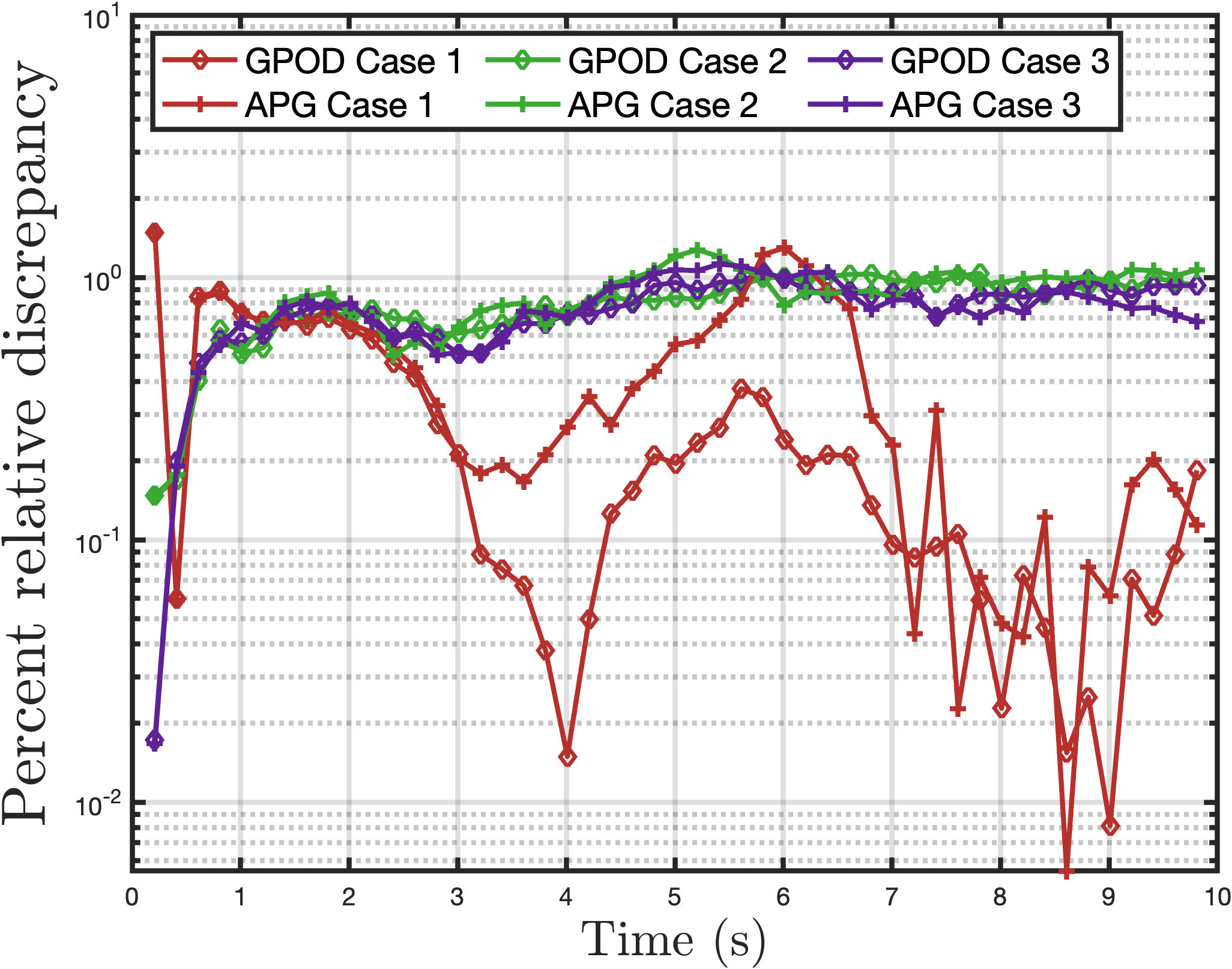}
		\caption{Velocity field relative discrepancy}
		\label{fig:LDC_centerline_velocity_error}
	\end{subfigure}%
	\hfill
	\begin{subfigure}[t]{0.5\textwidth}
		\centering
		\includegraphics[ trim={0cm 0cm 0 0}, clip, scale=0.325]{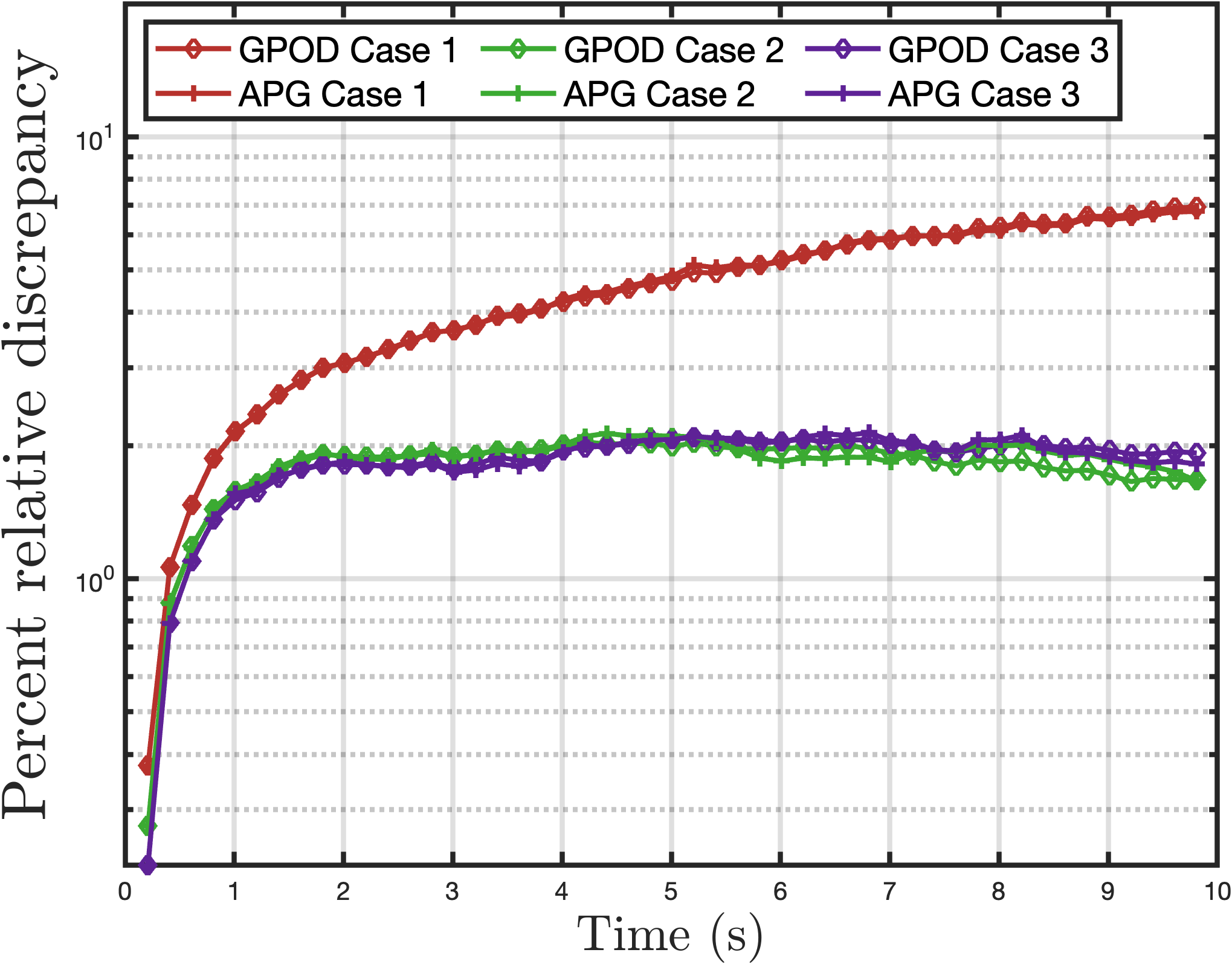}
		\caption{Pressure field relative discrepancy}
		\label{fig:LDC_centerline_pressure_error}
	\end{subfigure}
	\caption{Time histories of centerline relative discrepancy errors.}
		\label{fig:LDC_centerline_errors}
\end{figure*}

Time history relative discrepancy errors for velocity and pressure fields are presented in Fig.~\ref{fig:LDC_centerline_errors}. The velocity field results show good agreement across all cases, where both the GPOD and APG Case 1 show the highest error near 2\% at the initial stages of the simulation, and then outperforms all cases as the steady state is reached. All other GPOD and APG cases exhibit a steady error near 1\% throughout the entire simulation. On the other hand, the pressure field shows peak errors of 7\% for both GPOD and APG case 1. For all other Cases, peak errors lie within 0.8\% and 1.5\%. It is important to highlight that the APG method provided no significant benefit over GPOD in this numerical experiment. This stems from a combination of the choice of a relatively low memory length and the role of the weakly compressible assumption for low-density fluids. Investigating the role of the memory length in the APG method for weakly-compressible SPH would require a thorough standalone  investigation across the density, PMOR dimensionality, and weakly-compressible parametric space, and is therefore left for future studies.

\subsubsection{Parametric results}
For predictive numerical experiments, the parametric space of interest is defined by $Re=[50, 200]$. A local reduced basis approach is adopted \cite{amsallem2012nonlinear}, as it has been shown to provide improved results over a global basis approach for parametric spaces with moving features or features that vary considerably in space. Therefore, since the location of the prominent vortex feature in the lid-driven cavity is Reynolds number dependent, the present work derives three local bases. Namely, separate bases for prediction within a local space of $Re=[50, 100)$, $Re=[100, 150)$, and $Re=[150, 200]$, are employed. Finally, a local basis of dimension $M=15$ was chosen, which corresponds to $99.998\%$ of the cumulative energy, with respect to the Reynolds number, for all local training data. 

\begin{figure}[t!]
	\centering
	\includegraphics[trim = 0cm 0cm 0 0, scale=0.5]{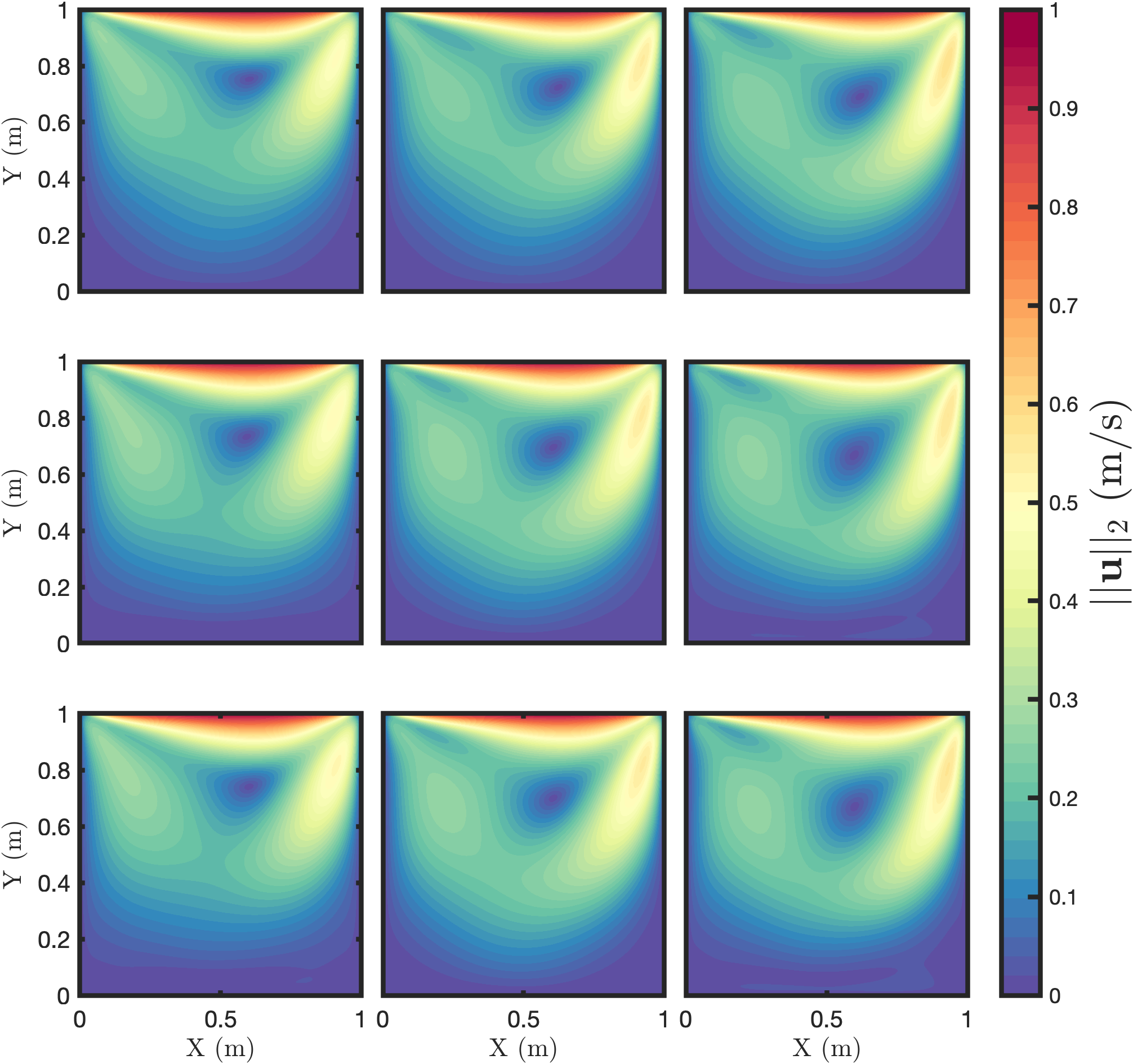}
	\caption{Lid-driven cavity velocity fields at $t=5$ s. Columns from left to right: $Re=75,\: 125,\:175$. First row: FOM; Second row: GPOD; Third row: APG.} 
	\centering
	\label{fig:LDC_parametric_velocity}
\end{figure}

\begin{figure}[t!]
	\centering
	\includegraphics[trim = 0cm 0cm 0 0, scale=0.5]{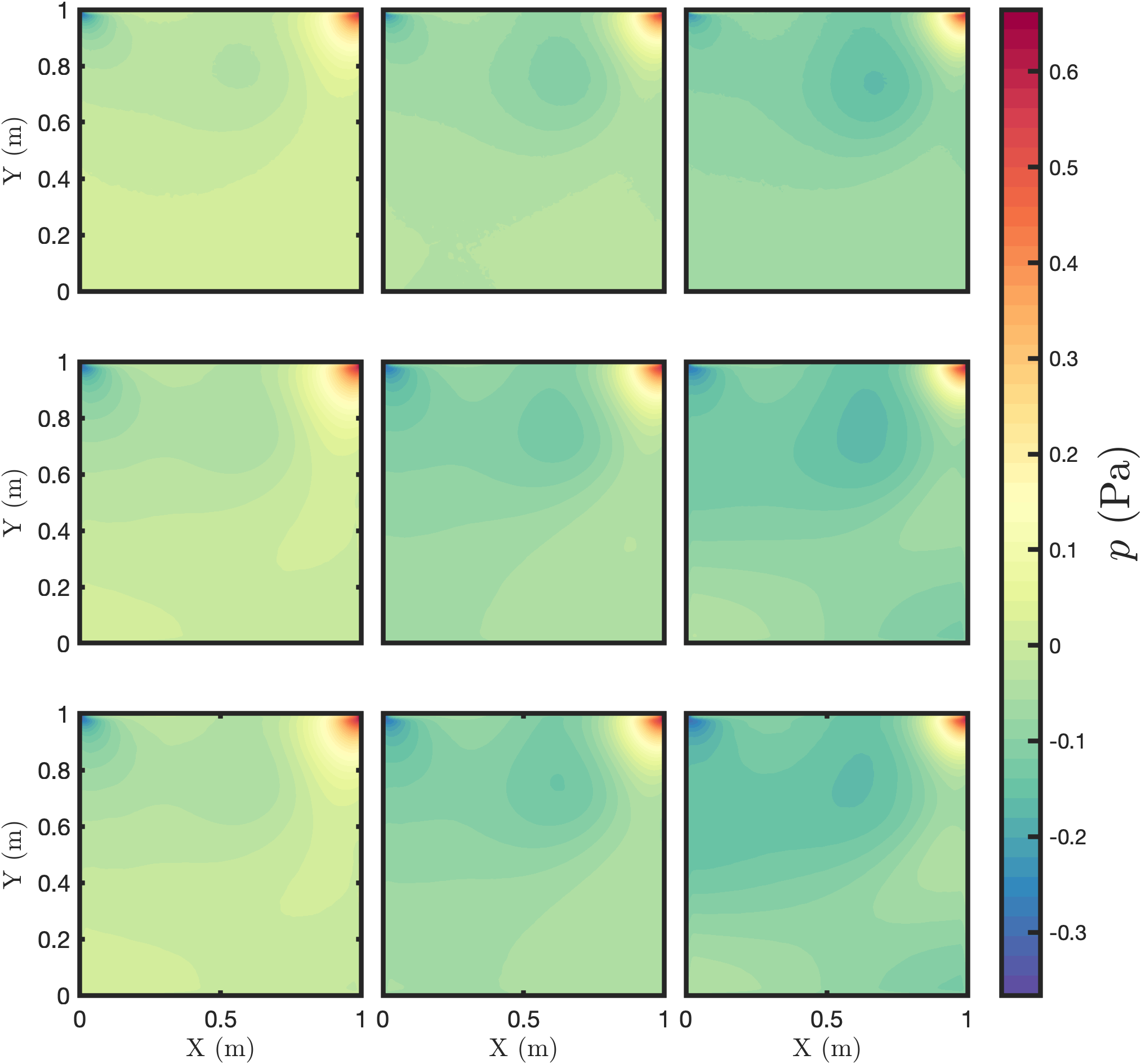}
		\caption{Lid-driven cavity pressure fields at $t=5$ s. Columns from left to right: $Re=75,\: 125,\:175$. First row: FOM; Second row: GPOD; Third row: APG.} 
	\centering
		\label{fig:LDC_parametric_pressure}
\end{figure}

\begin{figure}
	\centering
	\begin{subfigure}{.4\textwidth}
		\centering
		\includegraphics[width=.9\linewidth]{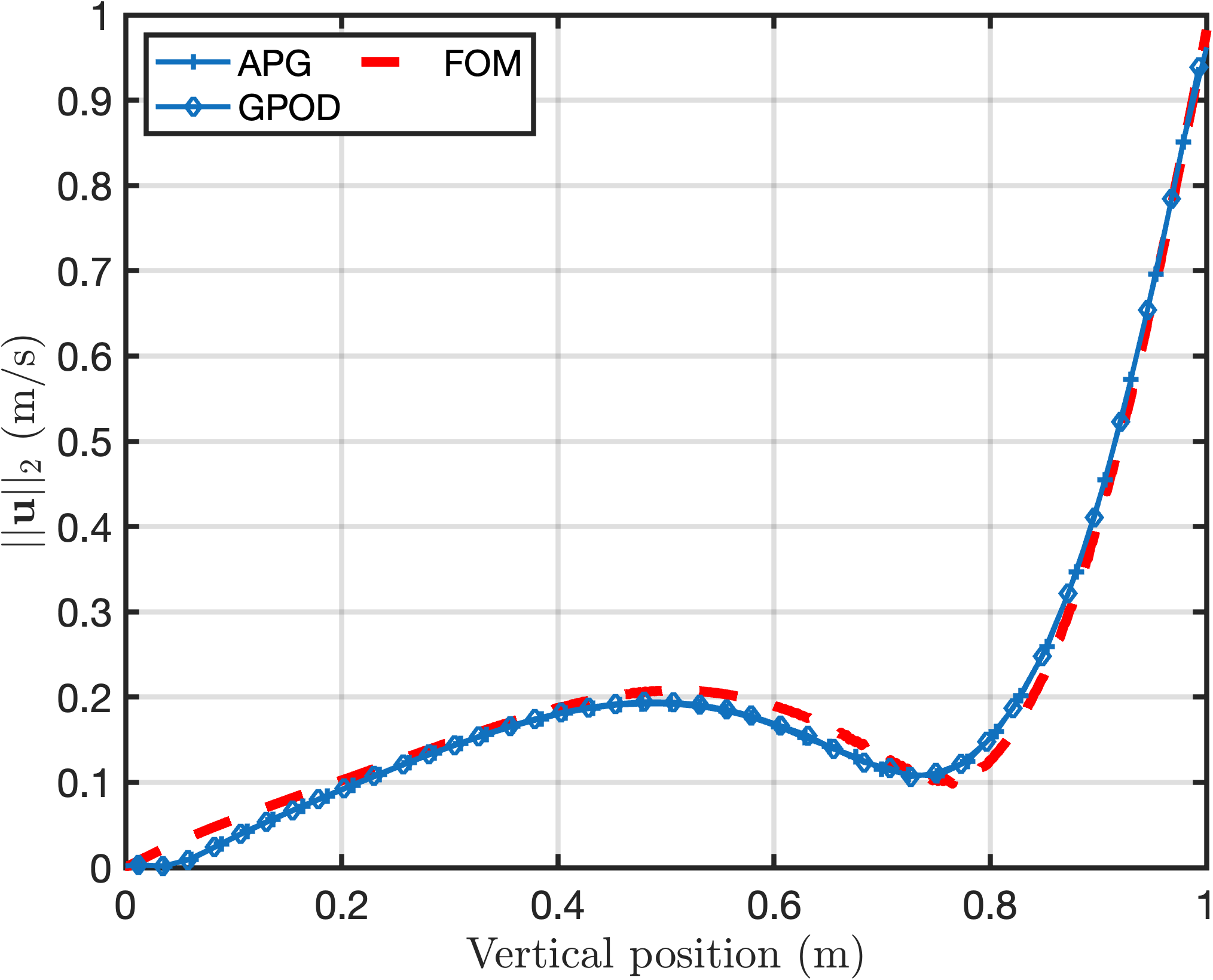}  
		\caption{$Re=75$}
		\label{fig:LDC_centerline_Re75_vel}
	\end{subfigure}
	\hfill
	\begin{subfigure}{.4\textwidth}
		\centering
		\includegraphics[width=.9\linewidth]{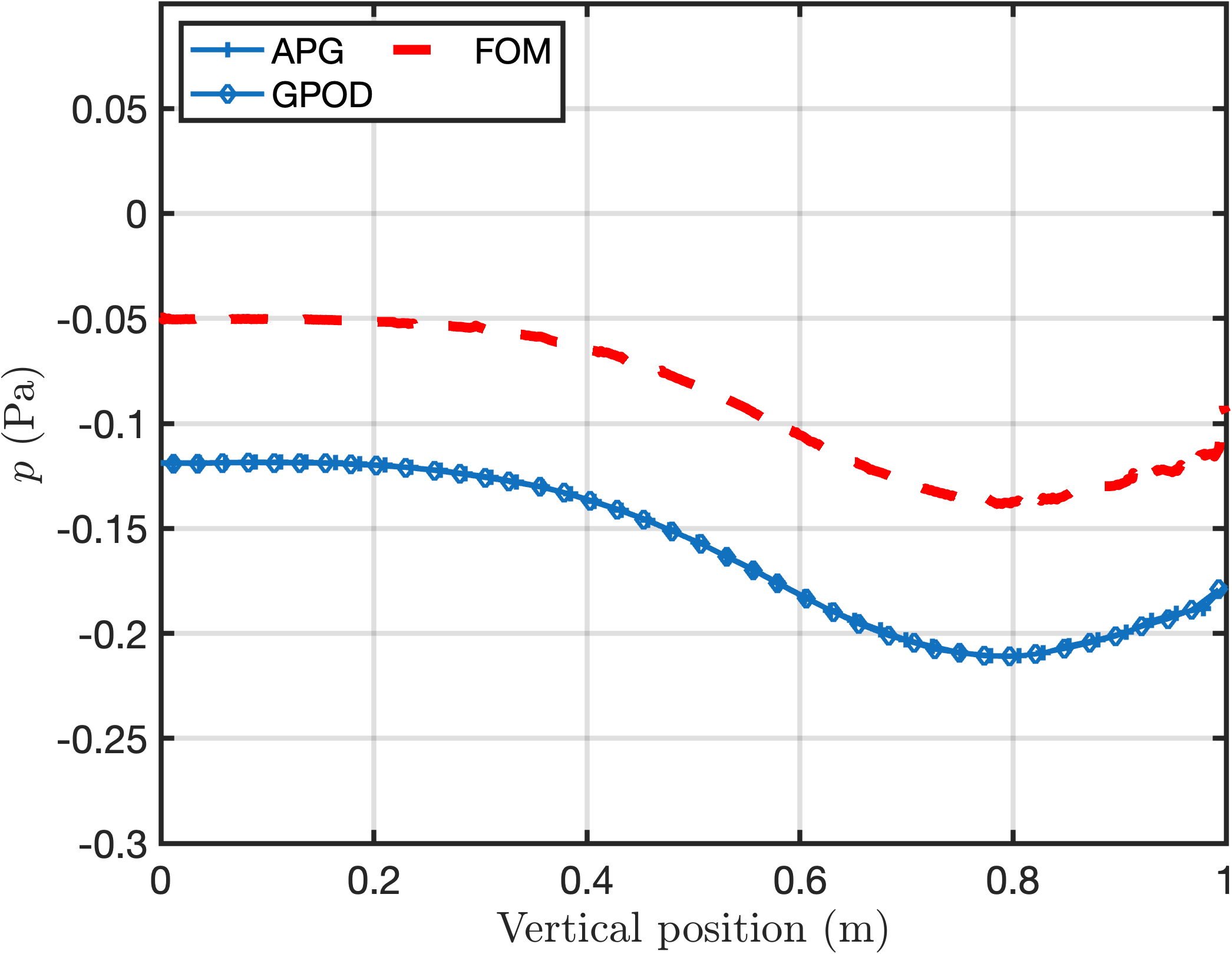}  
		\caption{$Re=75$}
		\label{fig:LDC_centerline_Re75_pressure}
	\end{subfigure}
	\begin{subfigure}{.4\textwidth}
		\centering
		\includegraphics[width=.9\linewidth]{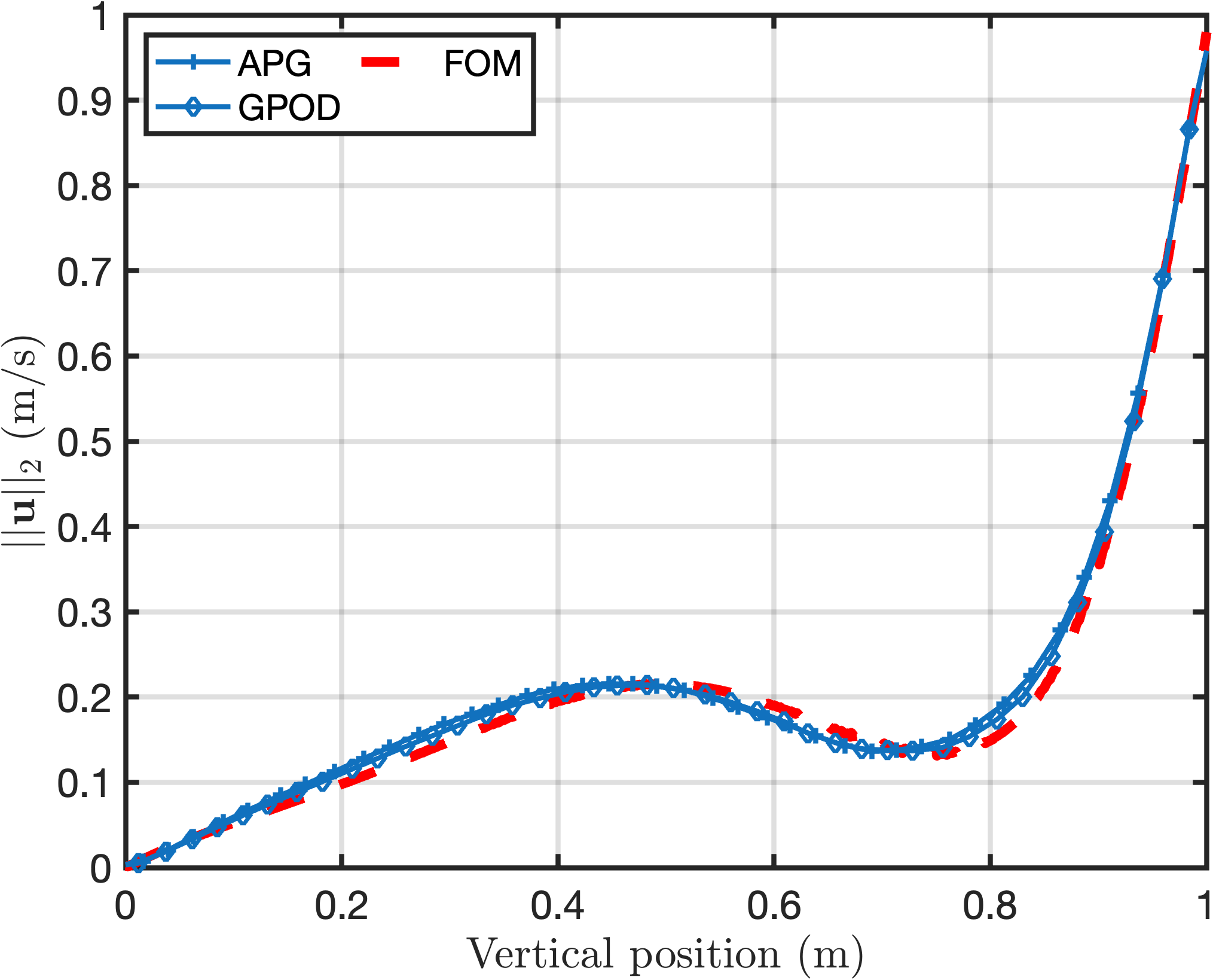}  
		\caption{$Re=125$}
		\label{fig:LDC_centerline_Re125_vel}
	\end{subfigure}
	\hfill
	\begin{subfigure}{.4\textwidth}
		\centering
		\includegraphics[width=.9\linewidth]{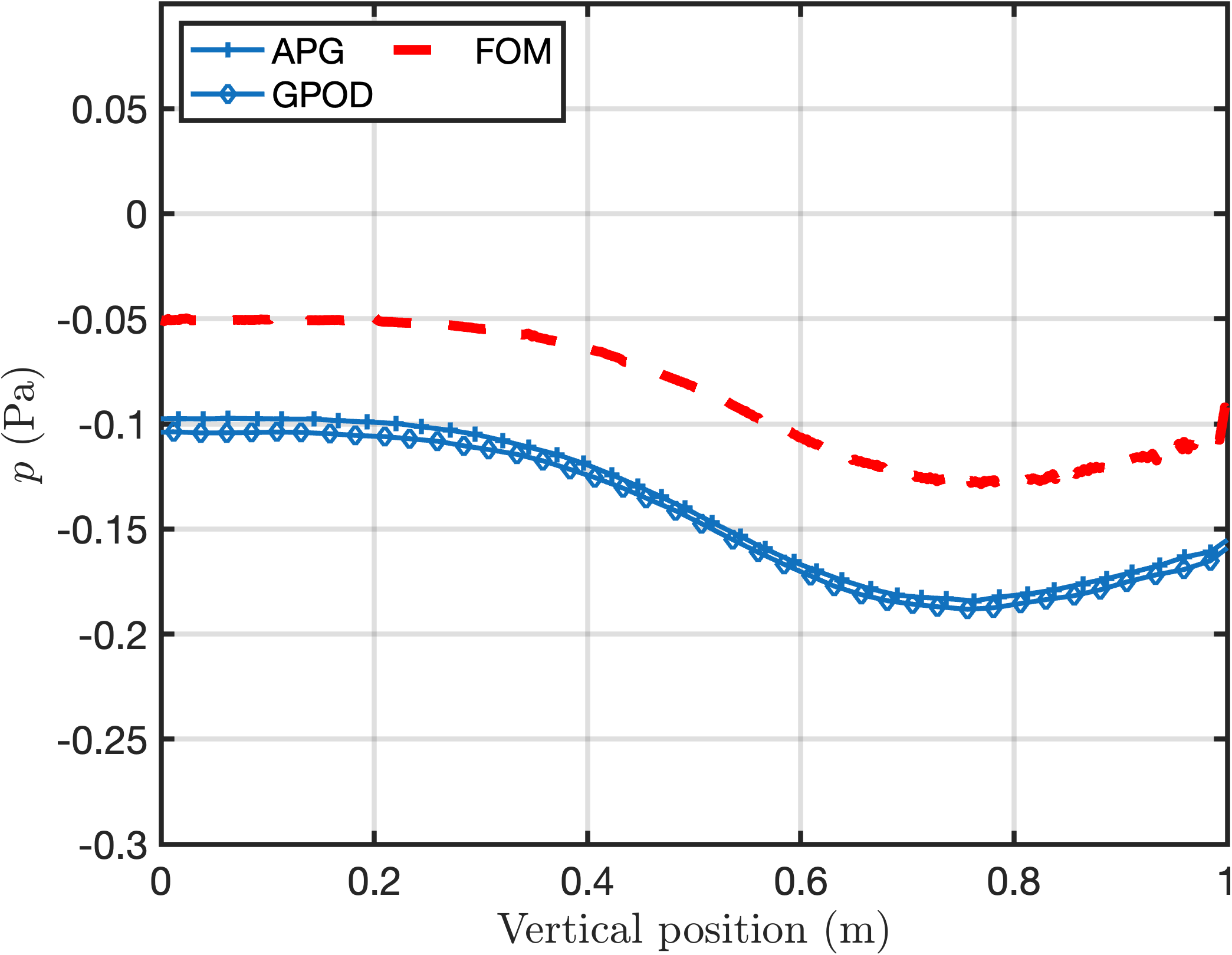}  
		\caption{$Re=125$}
		\label{fig:LDC_centerline_Re125_pressure}
	\end{subfigure}
	\begin{subfigure}{.4\textwidth}
		\centering
		\includegraphics[width=.9\linewidth]{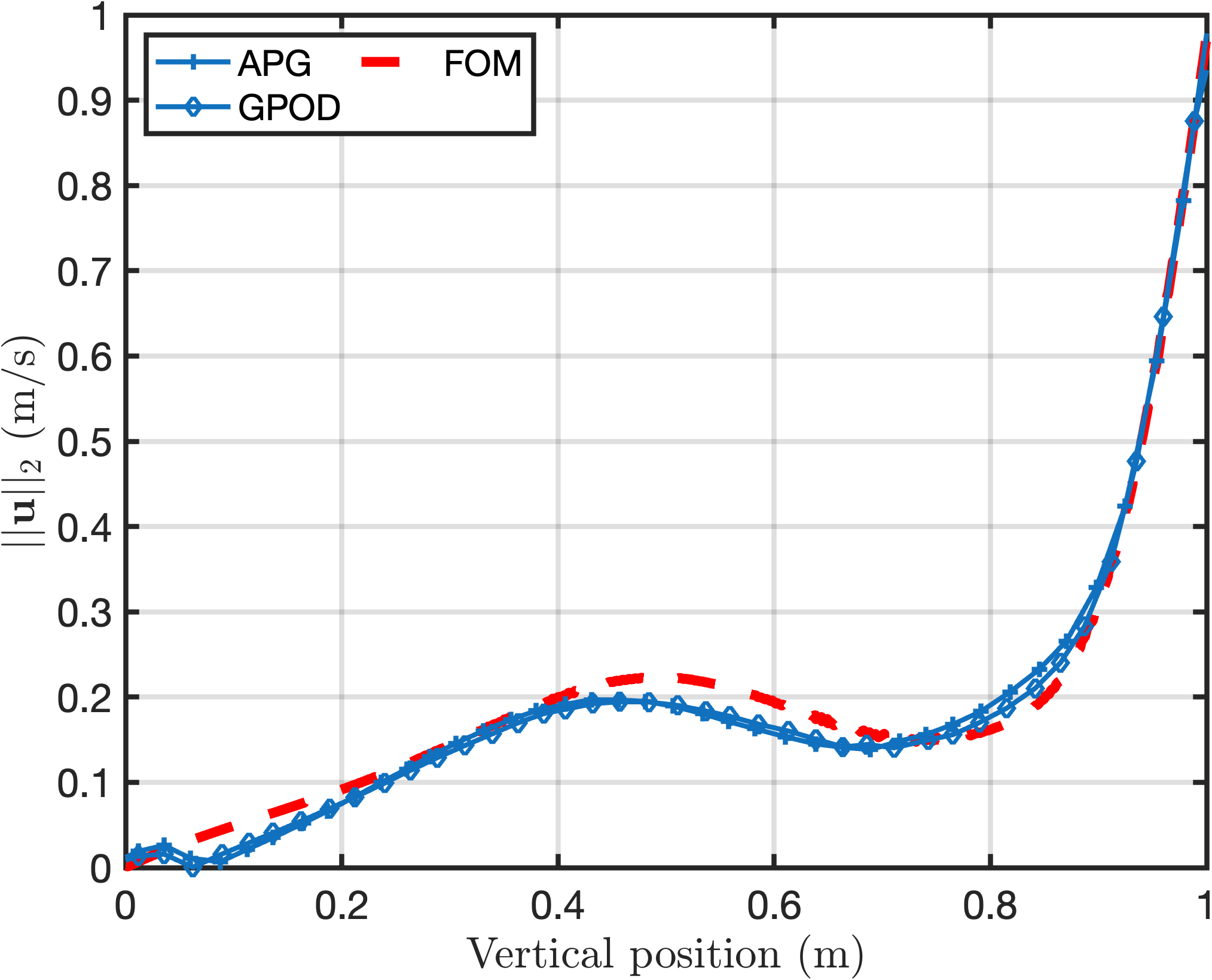}  
		\caption{$Re=175$}
		\label{fig:LDC_centerline_Re175_vel}
	\end{subfigure}
	\hfill
	\begin{subfigure}{.4\textwidth}
		\centering
		\includegraphics[width=.9\linewidth]{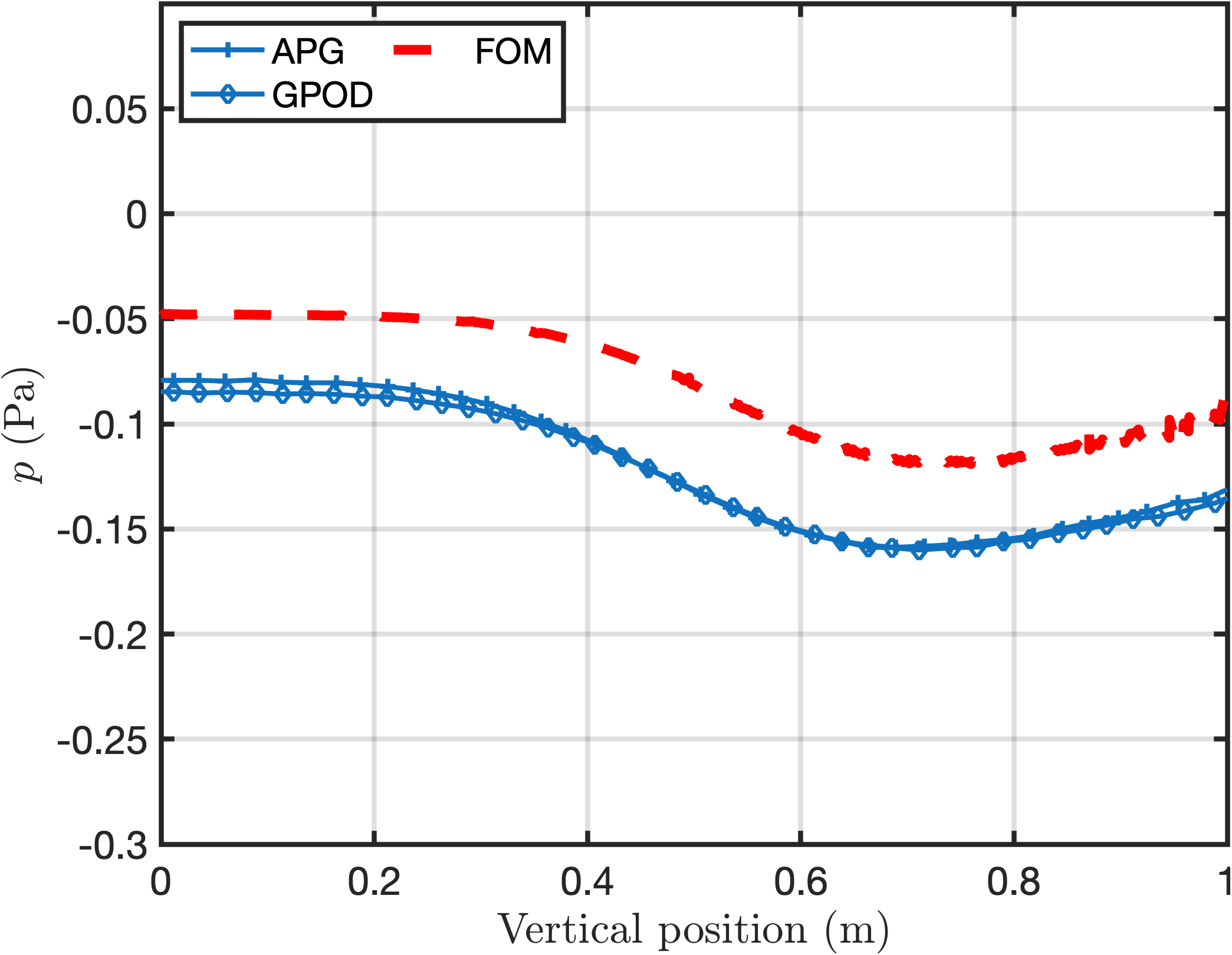}  
		\caption{$Re=175$}
		\label{fig:LDC_centerline_Re175_pressure}
	\end{subfigure}
	\caption{Vertical centerline parametric predictions of the lid-driven cavity at $t=5$ s. Left column: velocity profile. Right column: pressure profile.}
	\label{fig:LDC_centerline_parametric}
\end{figure}

Qualitative results for velocity and pressure fields are shown in Figs.~\ref{fig:LDC_parametric_velocity} and \ref{fig:LDC_parametric_pressure}. Results show good agreement across all Reynolds numbers in both velocity and pressure fields. The location of the vortex roll-up across all Reynolds numbers agrees with the FOM results and is reinforced by the centerline velocity plots presented in Fig.~\ref{fig:LDC_centerline_parametric}. Specifically, the FOM peak and troughs between 0.4 and 0.8 m along the vertical position correspond to the vortex roll-up at the centerline, where both APG and GPOD results show favorable predictive agreement. Centerline plots of the pressure field in Fig.~\ref{fig:LDC_centerline_parametric} show a notable local discrepancy. However, this is mainly an artifact from the centerline dynamics between ROMs and the FOM being close in scale relative to the global dynamics. 

\begin{figure*}[t!]
	\centering
	\begin{subfigure}[t]{0.5\textwidth}
		\centering
		\includegraphics[trim = {0cm 0cm 0 0},  scale=0.325]{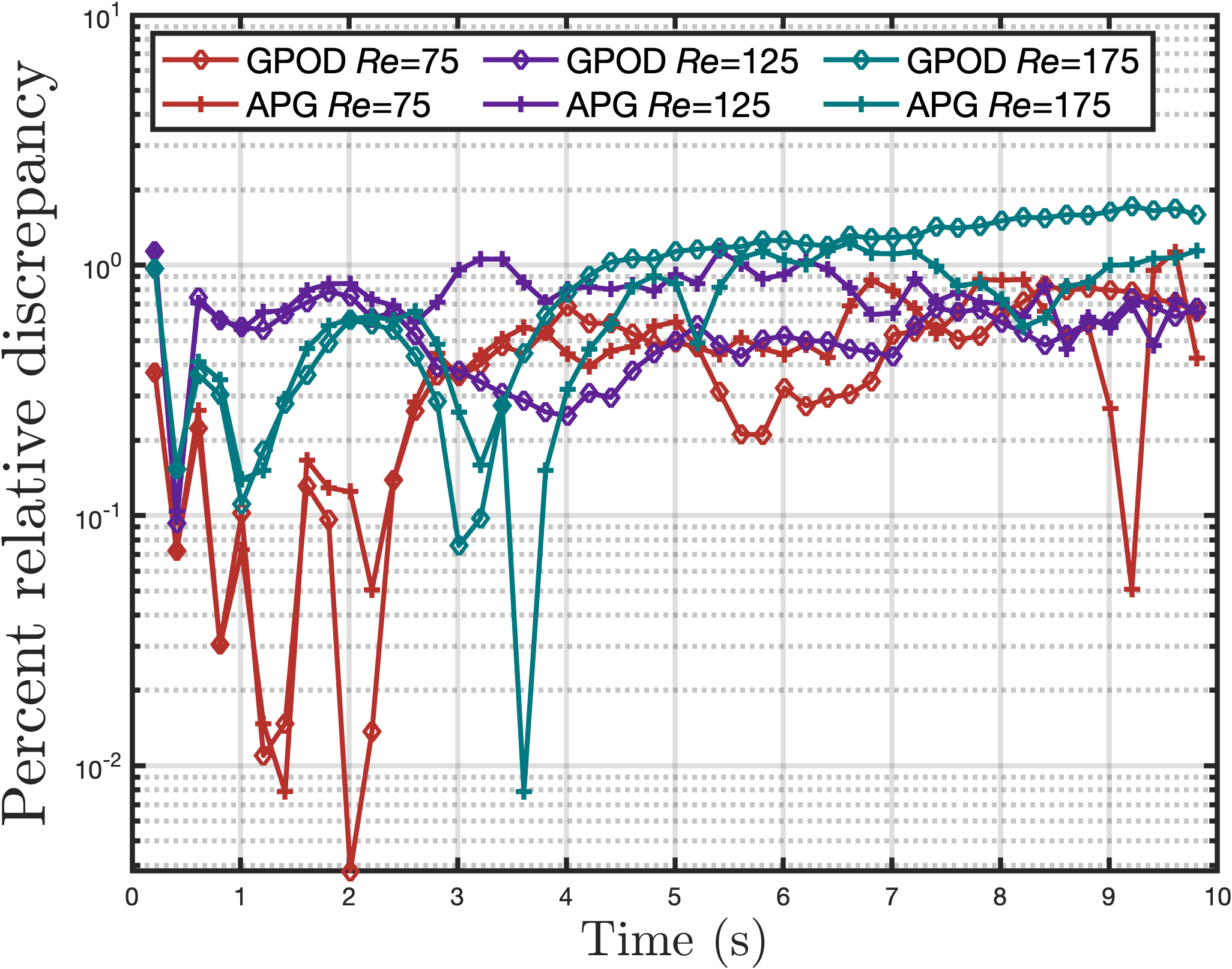}
		\caption{FOM velocity field}
	\end{subfigure}%
	\hfill
	\begin{subfigure}[t]{0.5\textwidth}
		\centering
		\includegraphics[ trim={0cm 0cm 0 0}, clip, scale=0.325]{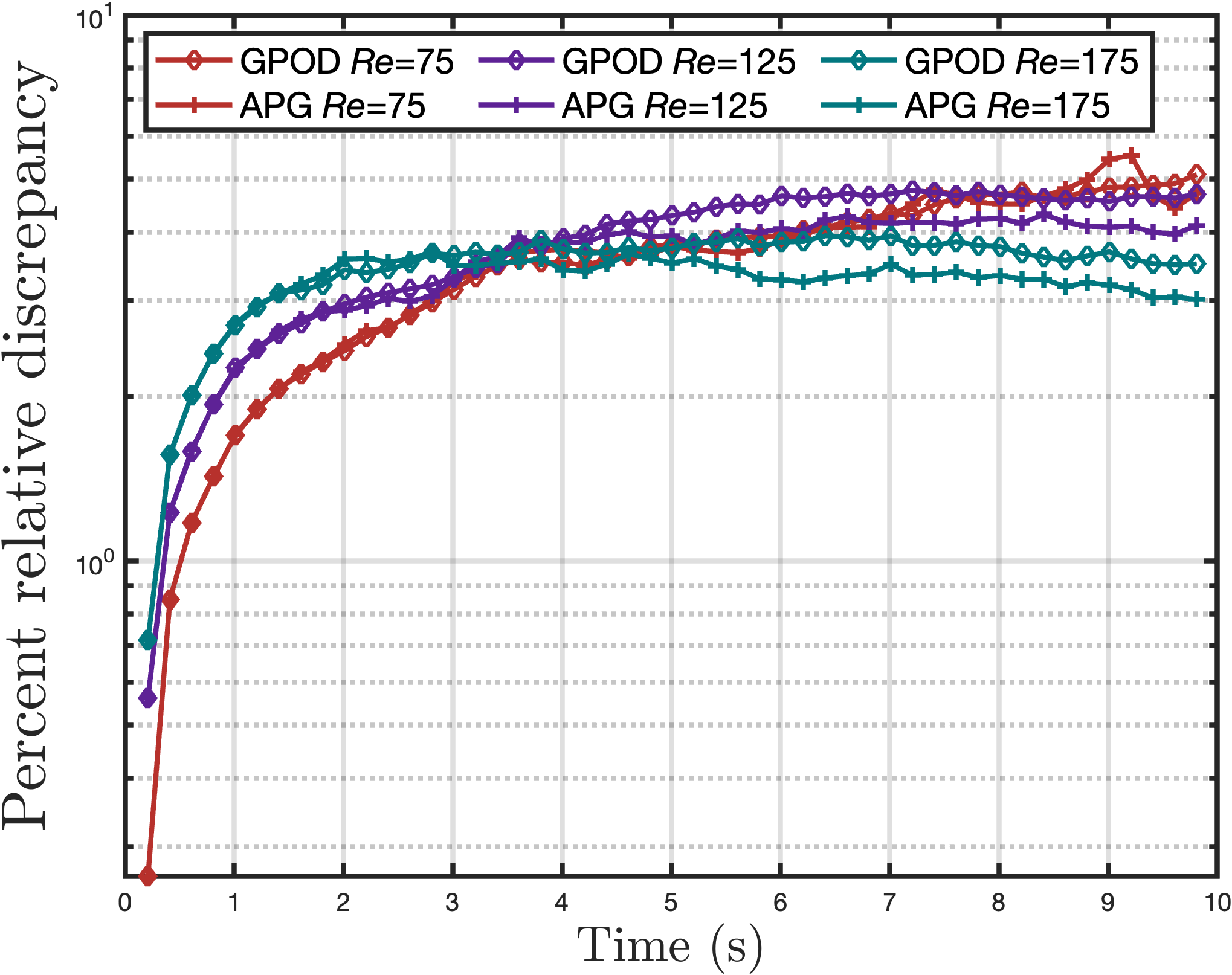}
		\caption{FOM pressure field}
	\end{subfigure}
	\caption{Time histories of vertical centerline relative discrepancy.}
	\label{fig:LDC_parametric_relative_discrepancy}
\end{figure*}

Time history of relative discrepancy errors for velocity and pressure fields are presented in Fig.~\ref{fig:LDC_parametric_relative_discrepancy}. The velocity field results show good agreement across all cases, where both the GPOD and APG at $Re=175$ show the highest error near 2\%. Predictive pressure field results peak at about 5\% in relative discrepancy error. Similar conclusions from previous reconstructive numerical experiment can be drawn about these errors. Overall, the proposed meshless PMOR provides good performance in a predictive and parametric setting for velocity and pressure fields, and further work is needed to more rigorously understand the role weak compressibility has on the spectral content and subspace embeddings.

%
\subsection{Flow over open cavity}
Flow past an open cavity is tested as the final benchmark problem in the current investigation. The problem domain consists of an open channel 0.4 meters long and 0.1 meters in height. The open cavity is centered along the length of the open channel and is 0.2 meters long and 0.1 meters in height. Here, the streamwise flow is implemented as a body force  immediately above the cavity in the $x$-direction, $\bm{b}=\{b_x,0\}$, as was done in \cite{adami2012generalized}. The body force is ramped up to a fixed value using the logistic function in Eq.~\ref{eqn:logistic_function}. The function parameters are set to $k=400$ 1/s, $k_0=0.001$, $f_0=0.01$ s, $f=-\tilde{n}\Delta t$ s, and $b_0=100$ m/s$^2$. Future work will look into more rigorous and robust open flow boundary conditions used in SPH.  A fluid reference density of $\rho_0=1$ kg/m$^3$ is employed and the resulting Reynolds number, $Re\approx2400$, is based on the length of the cavity. The simulations employ a particle resolution of $\Delta x=0.001$ m, a smoothing length of $h=2\Delta x$, three no-slip ghost particles for boundary conditions at the walls, a time-step of $\Delta t=2.5 \times 10^{-5}$ s, and a Mach number of $M_a=0.1$ with a reference speed $U_{\textup{ref}}=2$ m/s. The PMOR training data employed is generated from the final second of the quasi steady-state results, where FOM snapshot data was sampled at intervals of 50 for a total of 800 snapshots.  

\begin{equation}
	b_x=
	b_0 \left(\frac{1}{1+\exp(-k(f+f_0))}+k_0\right).
	\label{eqn:logistic_function}
\end{equation}

\subsubsection{Reconstructive results}
\begin{figure}[t!]
	\centering
\centering
	\begin{subfigure}[t]{0.5\textwidth}
		\centering
		\includegraphics[trim = {0cm 0cm 0 0},  scale=0.325]{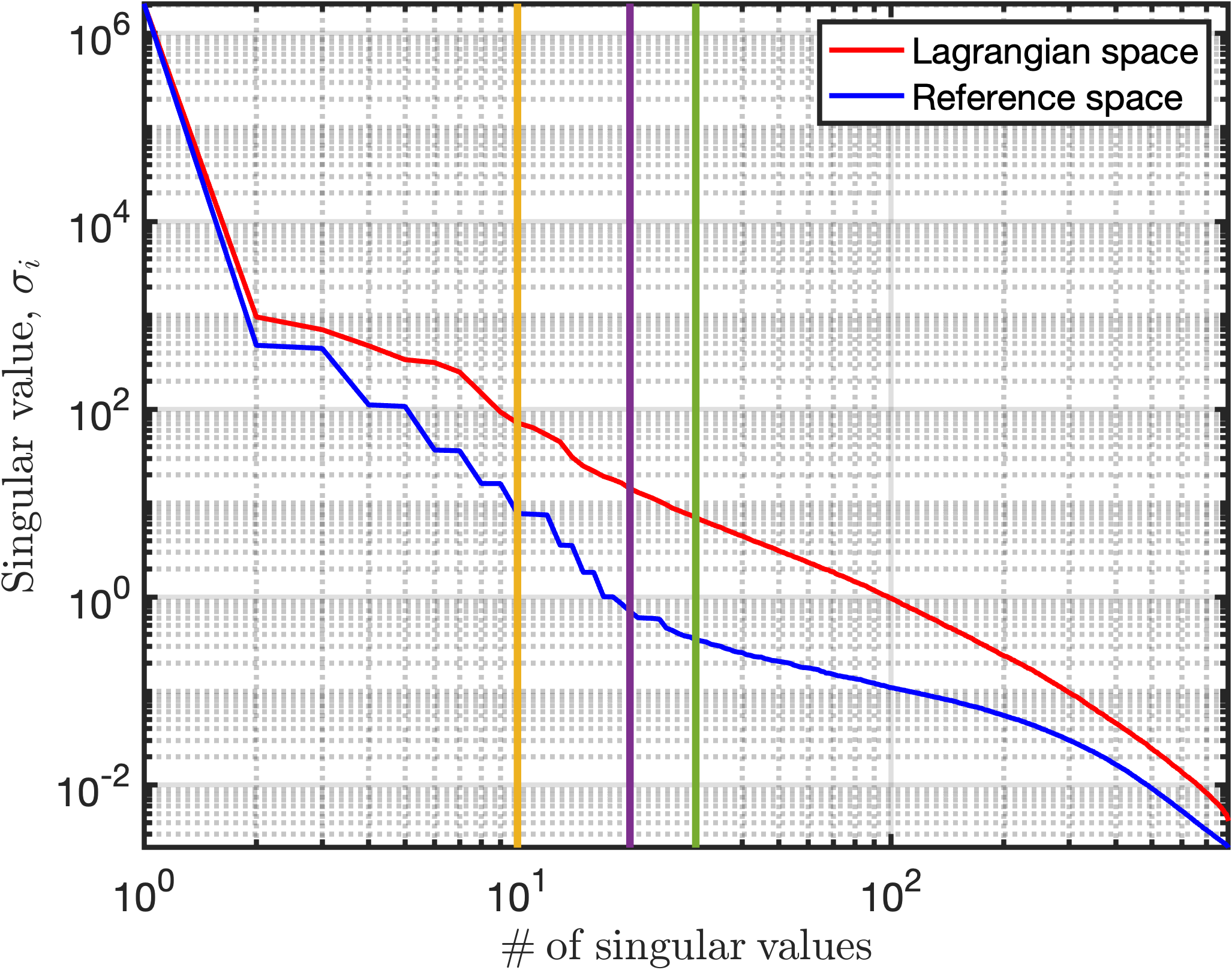}
		\caption{Singular value decay}		
	\end{subfigure}%
	\hfill
	\begin{subfigure}[t]{0.5\textwidth}
		\centering
		\includegraphics[ trim={0cm 0cm 0 0}, clip, scale=0.325]{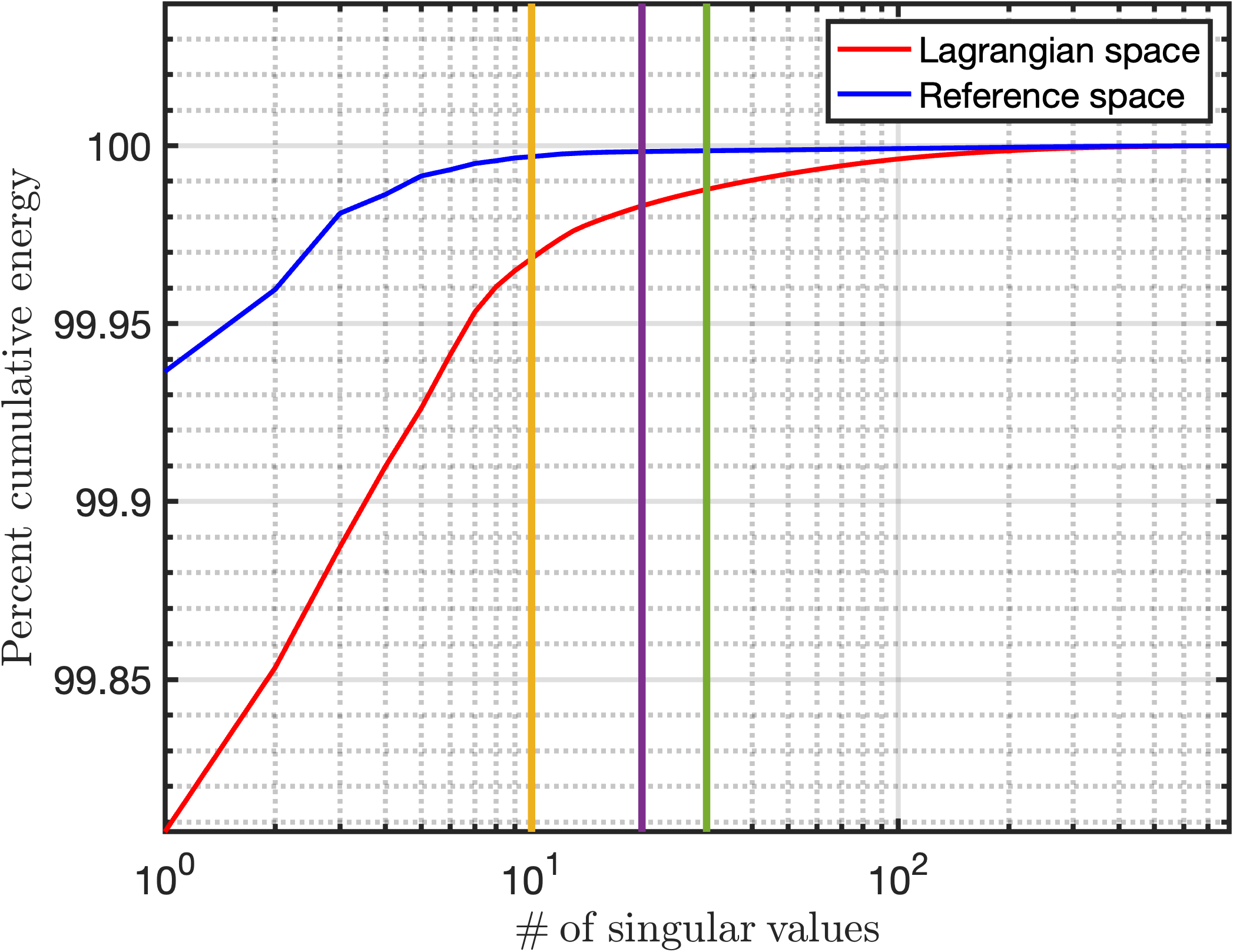}
		\caption{Cumulative energy contribution}
	\end{subfigure}
	\caption{Singular values derived from Lagrangian and reference space snapshot matrices. Vertical colored lines are meant to highlight the differences in singular value decay and cumulative energy between Lagrangian and reference space at $M=10$ (yellow), $M=20$ (purple), and $M=30$ (green).} 	
    \label{fig:FPOC_svdDecay}
	\centering
\end{figure}

\begin{figure}[t!]
	\centering
	\includegraphics[trim = 0cm 0cm 0 0, scale=0.325]{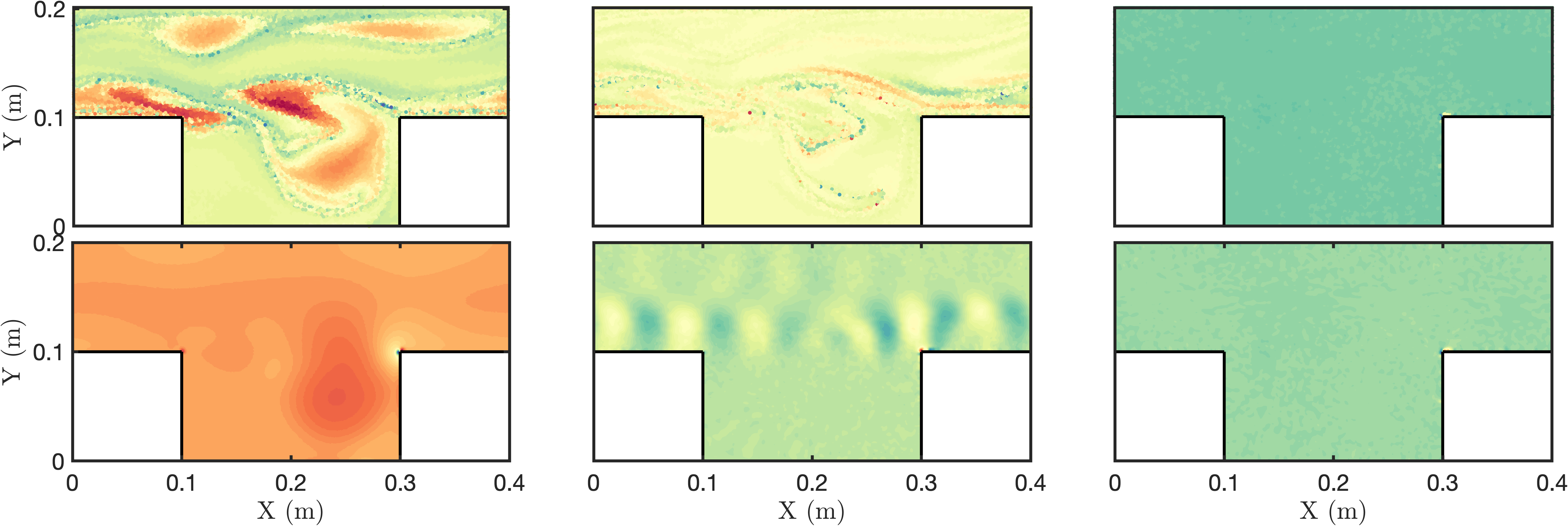}
	\caption{Density field modes, from left to right $M=10, 20, {\text{and}\:} 30$. Top row: Lagrangian space; Bottom row: Reference space.} 
	\centering
		\label{fig:FPOC_rhomodes}
\end{figure}

\begin{figure}[t!]
	\centering
	\includegraphics[trim = 0cm 0cm 0 0, scale=0.325]{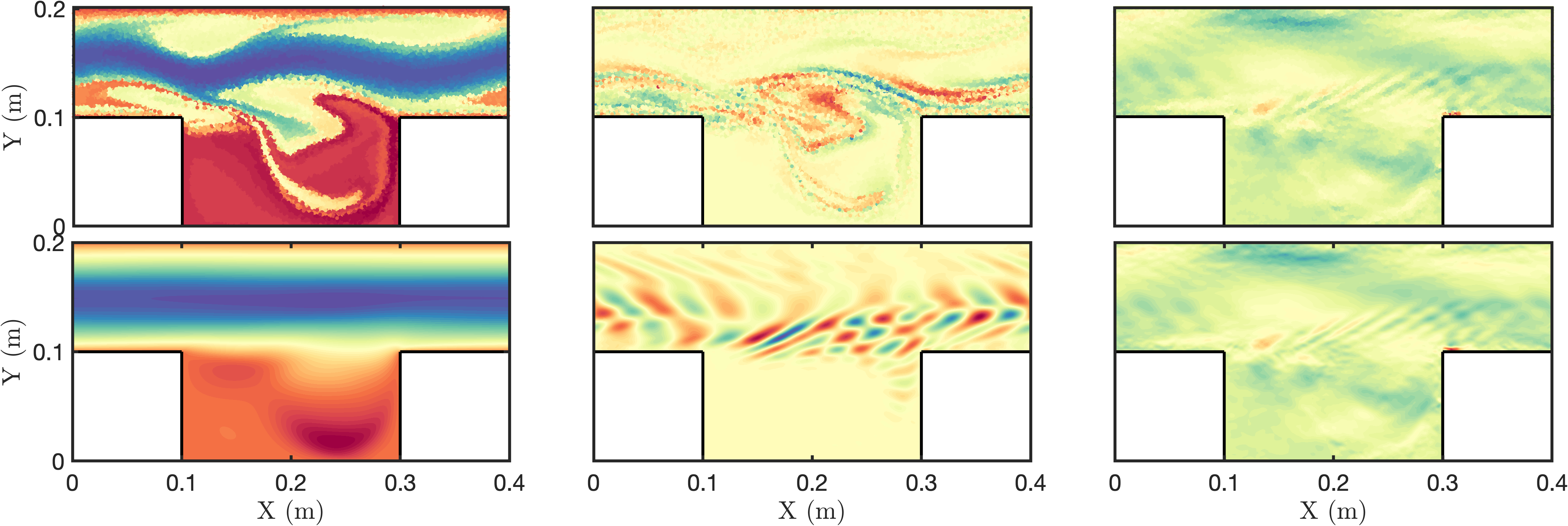}
	\caption{Velocity field $x-$component modes, from left to right $M=10, 20, {\text{and}\:} 30$. Top row: Lagrangian space; Bottom row: Reference space.} 
	\centering
	  \label{fig:FPOC_velxmodes}
\end{figure}

\begin{figure}[t!]
	\centering
	\includegraphics[trim = 0cm 0cm 0 0, scale=0.325]{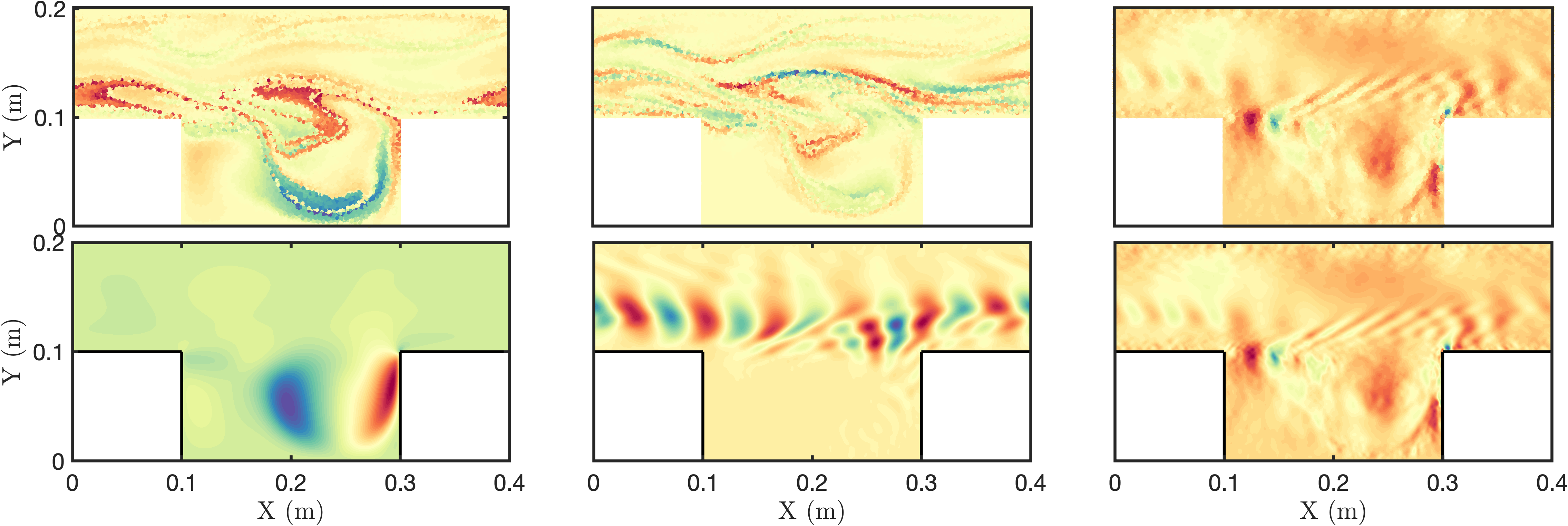}
	\caption{Velocity field $y-$component modes, from left to right $M=10, 20, {\text{and}\:} 30$. Top row: Lagrangian space; Bottom row: Reference space.}
	\centering
		\label{fig:FPOC_velymodes}
\end{figure}

Results from the dimensional compression of the FOM snapshots are shown in Figs.~\ref{fig:FPOC_svdDecay}-\ref{fig:FPOC_velymodes}. The corresponding singular value decay is shown in Fig.~\ref{fig:FPOC_svdDecay}, and a notable difference in decay rates is again observed between reference and Lagrangian space. The current reconstructive experiment employs a $M=20$ dimensional subspace corresponding to 99.9984\% of the cumulative energy ($CF=15,750$), and a memory length of $\tau=10^{-7}$ s was selected for the APG method.

Modes 1, 10, and 20 are shown in Figs.~\ref{fig:FPOC_rhomodes}-\ref{fig:FPOC_velymodes}. It is important to highlight the impact of the mixing numerical topology in this experiment. In the current experiment, the periodic vortex shedding off the corner in the back wall creates chaotic mixing of the numerical topology as particles re-enter the periodic domain. Slight perturbations in the flow create different trajectories for individual particles as time progresses. The mixing numerical topology, therefore, generates mixing modes in Lagrangian space that are clearly seen in Figs.~\ref{fig:FPOC_rhomodes}-\ref{fig:FPOC_velymodes}. On the other hand, reference space modal quantities depict the expected low-dimensional and coherent mode shapes from the current numerical experiment. 

\begin{figure}[t!]
	\centering
		\begin{subfigure}{.45\textwidth}
		\centering
		\includegraphics[width=.9\linewidth]{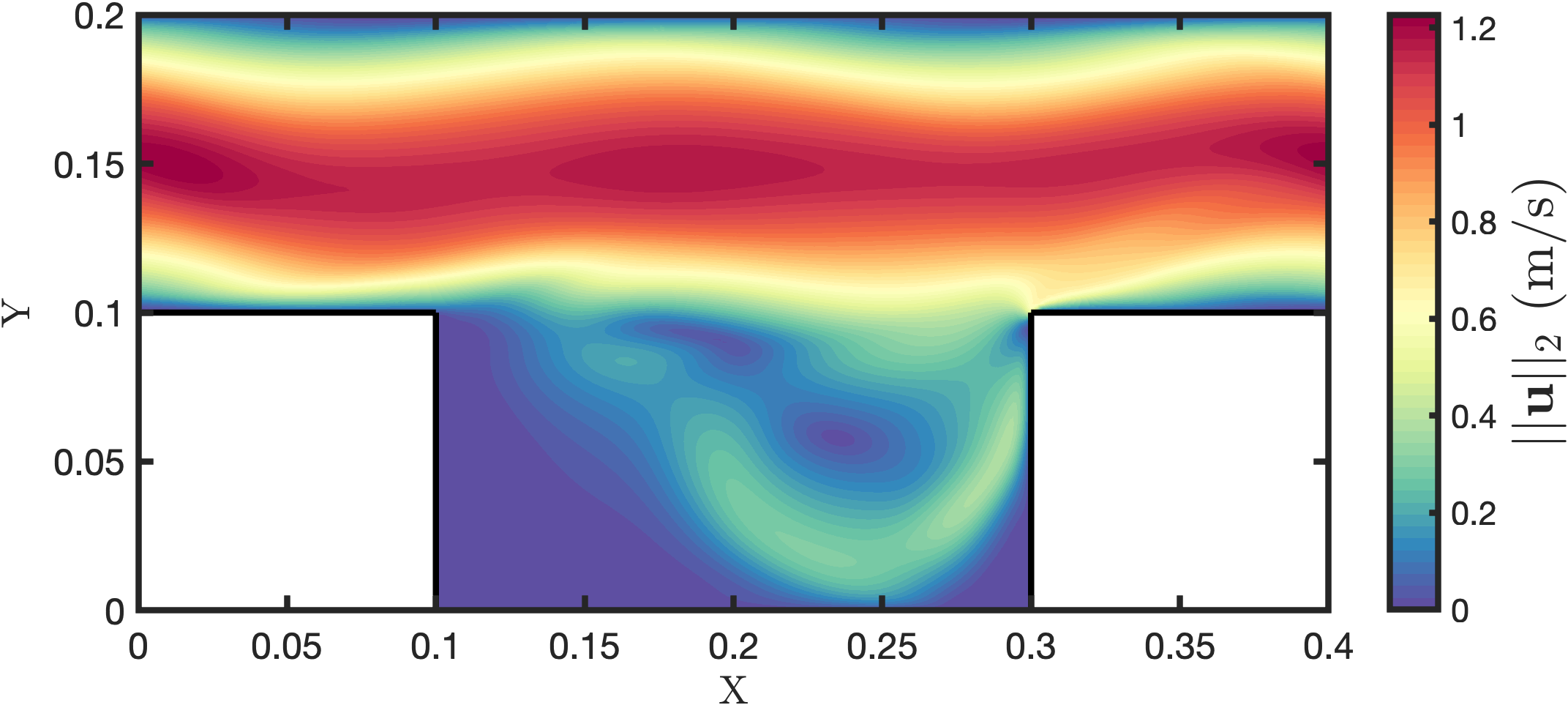}  
		\caption{FOM velocity field}
		\label{fig:FPOC_FOM_velocity}
	\end{subfigure}
	\hfill
	\begin{subfigure}{.45\textwidth}
		\centering
		\includegraphics[width=.9\linewidth]{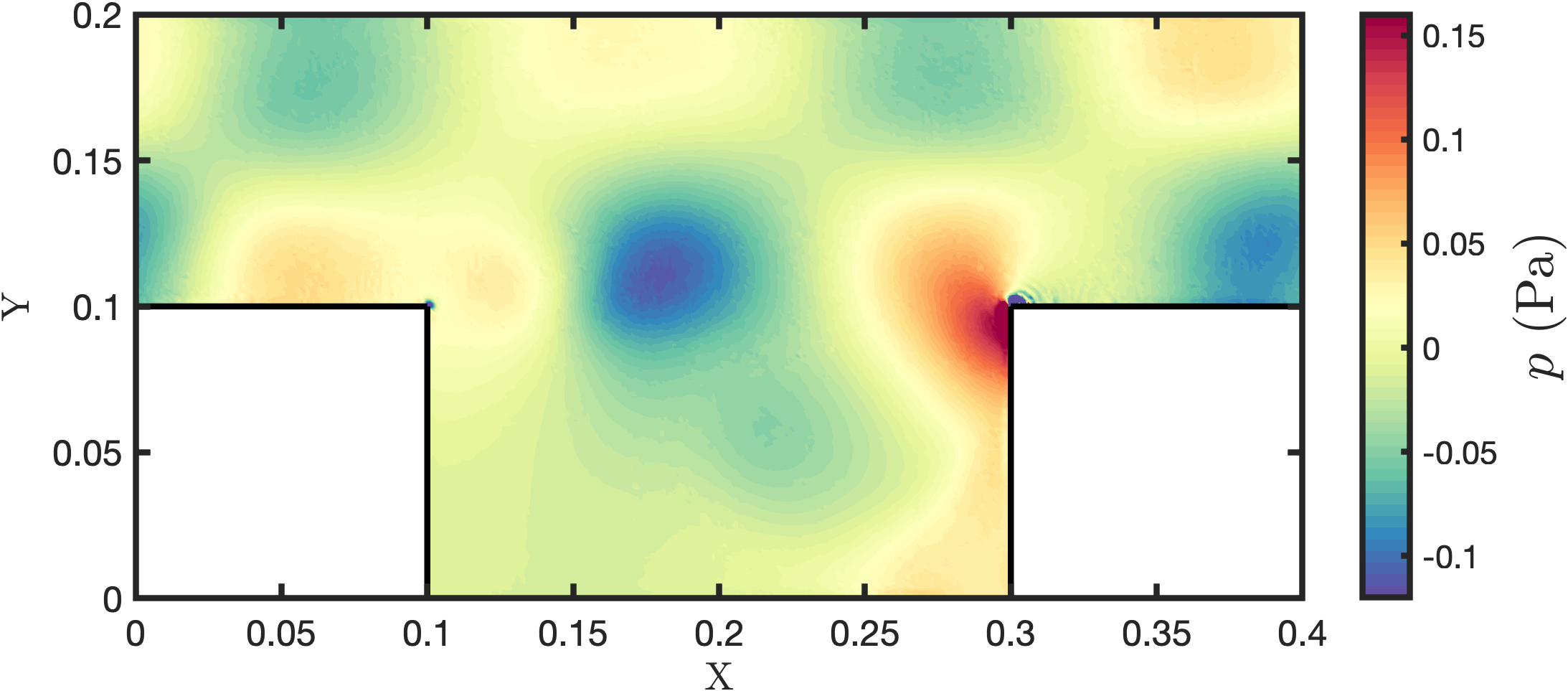}  
		\caption{FOM pressure field}
		\label{fig:FPOC_FOM_pressure}
	\end{subfigure}
	\begin{subfigure}{.45\textwidth}
		\centering
		\includegraphics[width=.9\linewidth]{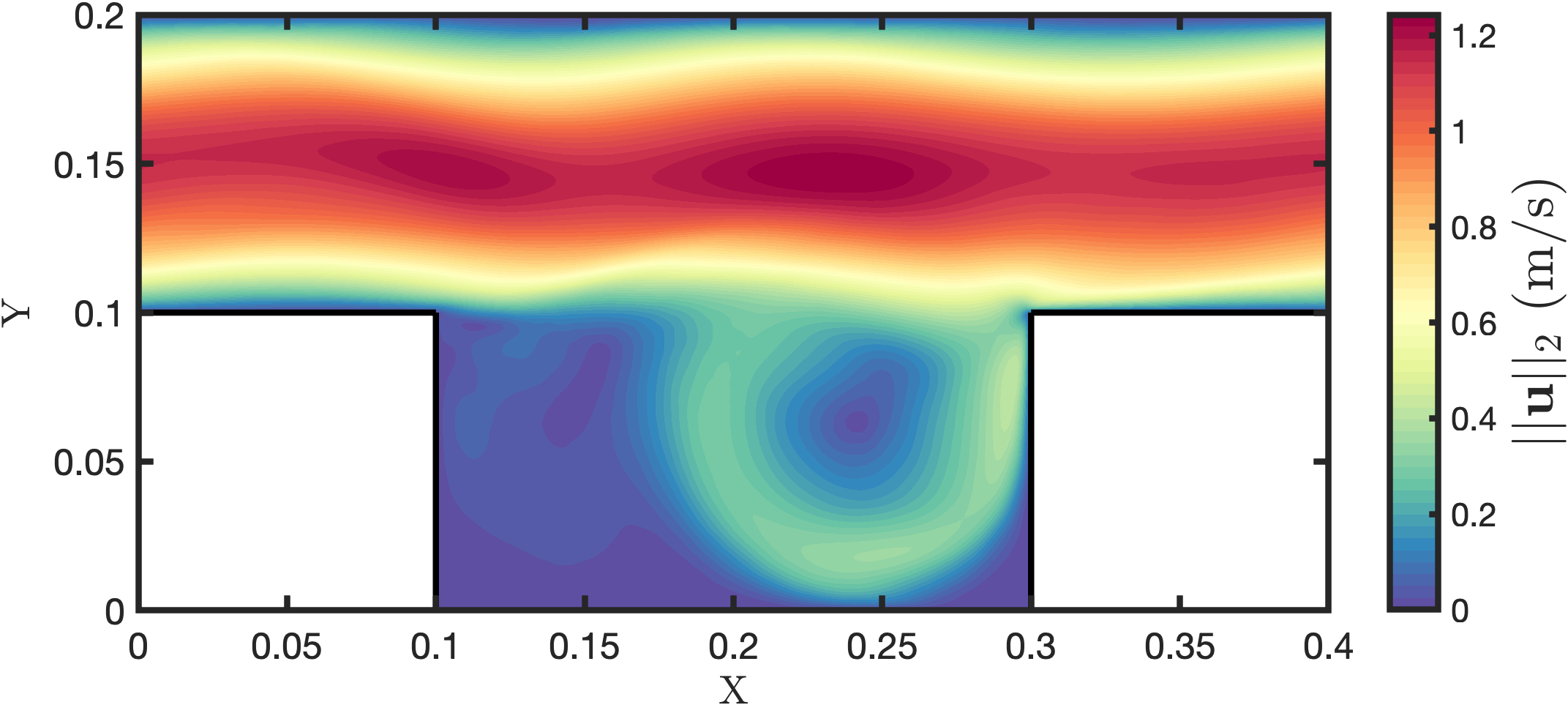}  
		\caption{GPOD velocity field}
		\label{FPOC_GPOD_velocity}
	\end{subfigure}
	\hfill
	\begin{subfigure}{.45\textwidth}
		\centering
		\includegraphics[width=.9\linewidth]{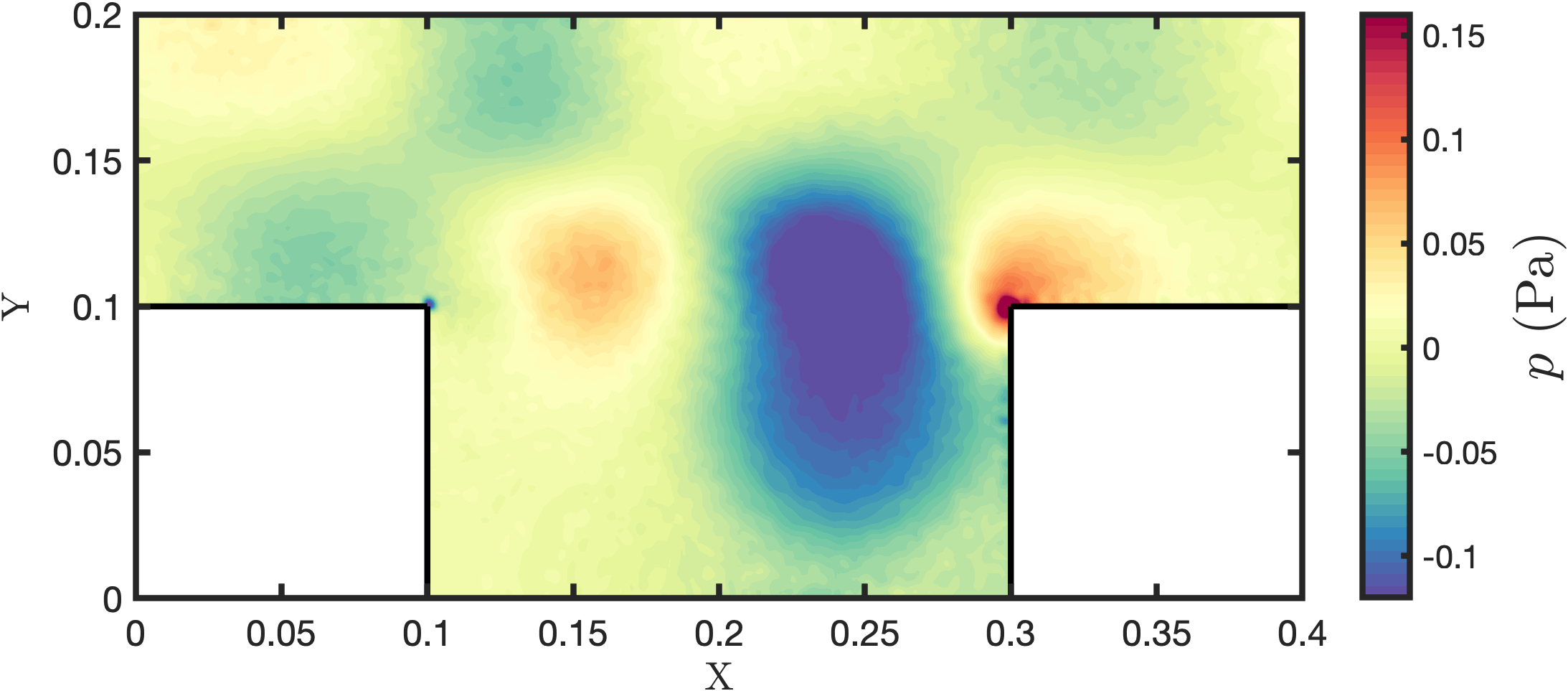}  
		\caption{GPOD pressure field}
		\label{fig:FPOC_GPOD_pressure}
	\end{subfigure}
	\begin{subfigure}{.45\textwidth}
		\centering
		\includegraphics[width=.9\linewidth]{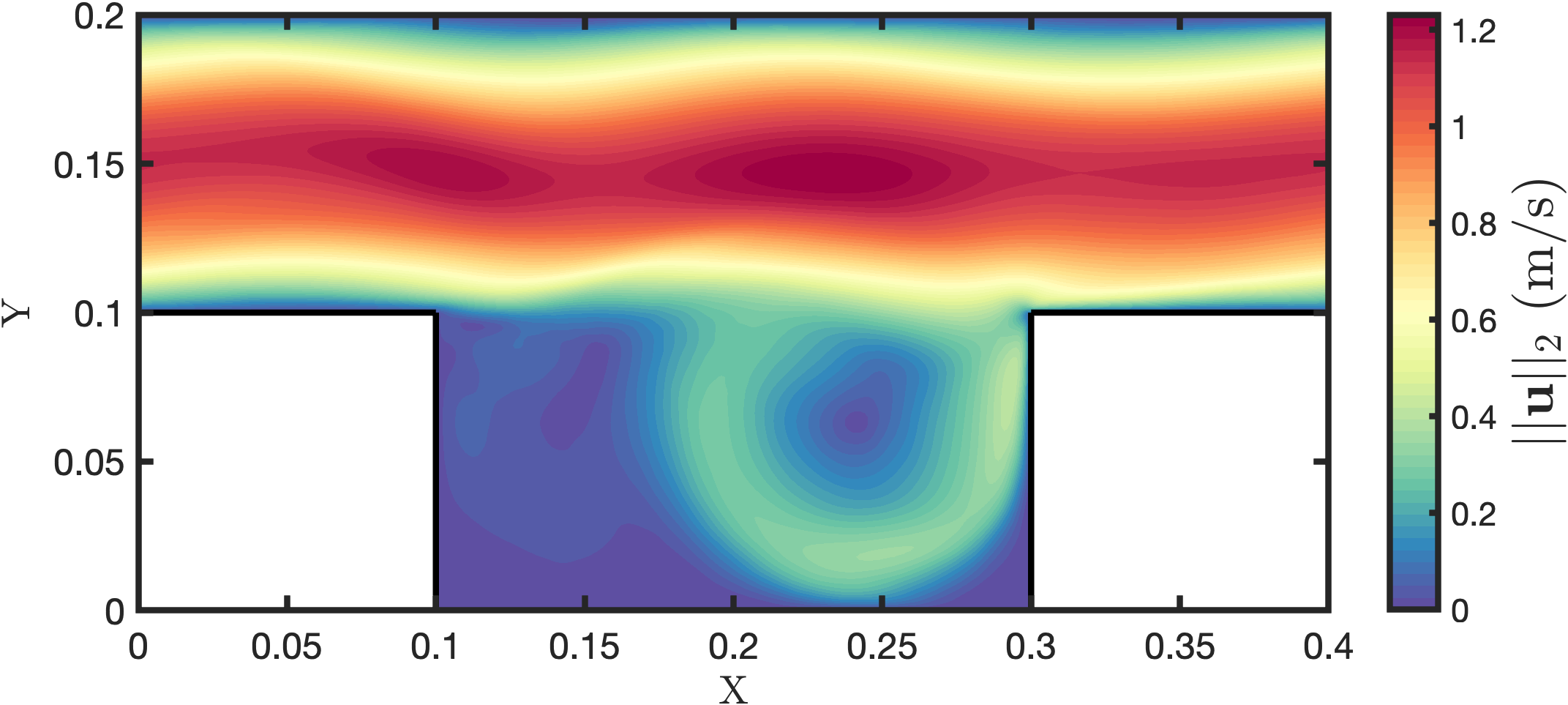}  
		\caption{APG velocity field}
		\label{fig:FPOC_APG_velocity}
	\end{subfigure}
	\hfill
	\begin{subfigure}{.45\textwidth}
		\centering
		\includegraphics[width=.9\linewidth]{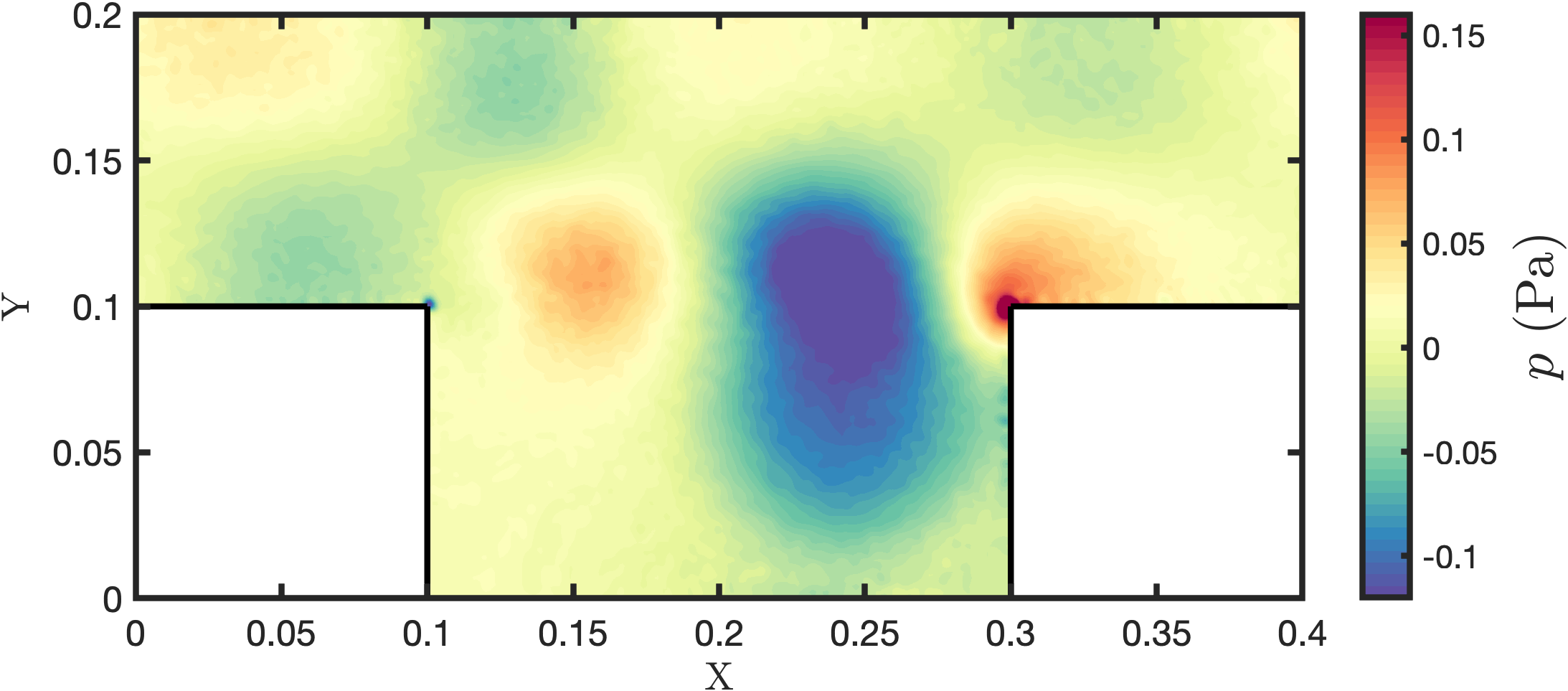}  
		\caption{APG pressure field}
		\label{fig:FPOC_APG_pressure}
	\end{subfigure}
	\caption{Snapshots of the velocity (left column) and pressure (right column) field at $t=0.625$ s. Top row: FOM; Middle row: GPOD; Bottom row: APG.}
			\label{fig:FPOC_ flowfieldresutls}
\end{figure}

Qualitative flow field results are shown in Fig.~\ref{fig:FPOC_ flowfieldresutls} at snapshot $t=0.625$ s. The development of the periodic vortex inside of the cavity between 0.2 and 0.3 meters is  a prominent feature in the velocity field across FOM and PMOR results. However, it is important to note the the vortex shows higher-frequency content not captured by either PMOR methods, while in both the GPOD and APG methods, the vortex exhibits lower frequency and more coherent shapes. Another attribute that is important to highlight is the periodic structure of the flow field through the top of the cavity. As the flow passes over the cavity and the periodic vortex shedding occurs, the oscillation in the fluid flow creates a periodic structure in the channel. What is interesting to note is that the phase of this structure in both PMOR methods does not align with the structure in the FOM. In fact, the structure seems slightly shifted to the right for both GPOD and APG. The discrepancies in the phase of the structure are likely due to the accumulation of errors in the low-dimensional approximation through several cycles of vortex shedding. Future work will consider symplectic time integration and embeddings to appease accumulation error over time. Overall, despite the lack of high-frequency structure, the velocity field reconstruction results qualitatively agree with the FOM. On the other hand, the PMOR pressure fields show vastly different structure than the FOM. However, it is important to highlight that the phase shift previously discussed is also present in the pressure field. Specifically, a high-pressure region is generated in the FOM at $t=0.625$ s  as the vortex roll-up on the left corner of the cavity begins. In both PMORs, the roll-up has already detached from the left corner and has begun approaching the right wall. Furthermore, the FOM vortex roll-up on the left corner is adjacent to a prominent negative pressure region. In both PMOR results, this negative pressure region is higher in amplitude and is closer to the right wall, indicating that the vortex shedding phase is ahead of the FOM. 

\begin{figure*}[t!]
	\centering
	\begin{subfigure}[t]{0.5\textwidth}
		\centering
		\includegraphics[trim = {0cm 0cm 0 0},  clip, scale=0.325]{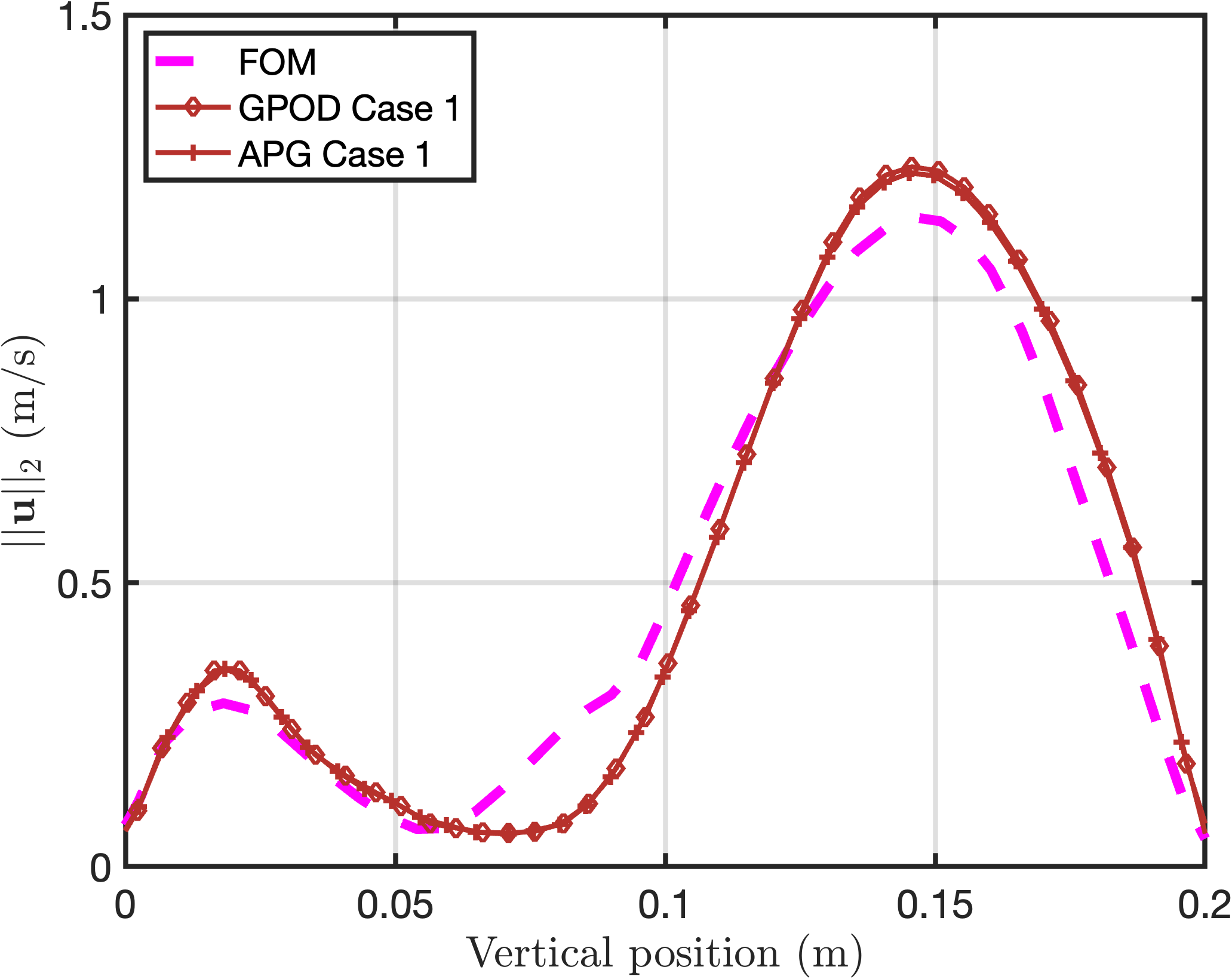}
		\caption{FOM velocity field}
				\label{fig:FPOC_vertical_slices_velocity}
	\end{subfigure}%
	\hfill
	\begin{subfigure}[t]{0.5\textwidth}
		\centering
		\includegraphics[ trim={0cm 0cm 0 0}, clip, scale=0.325]{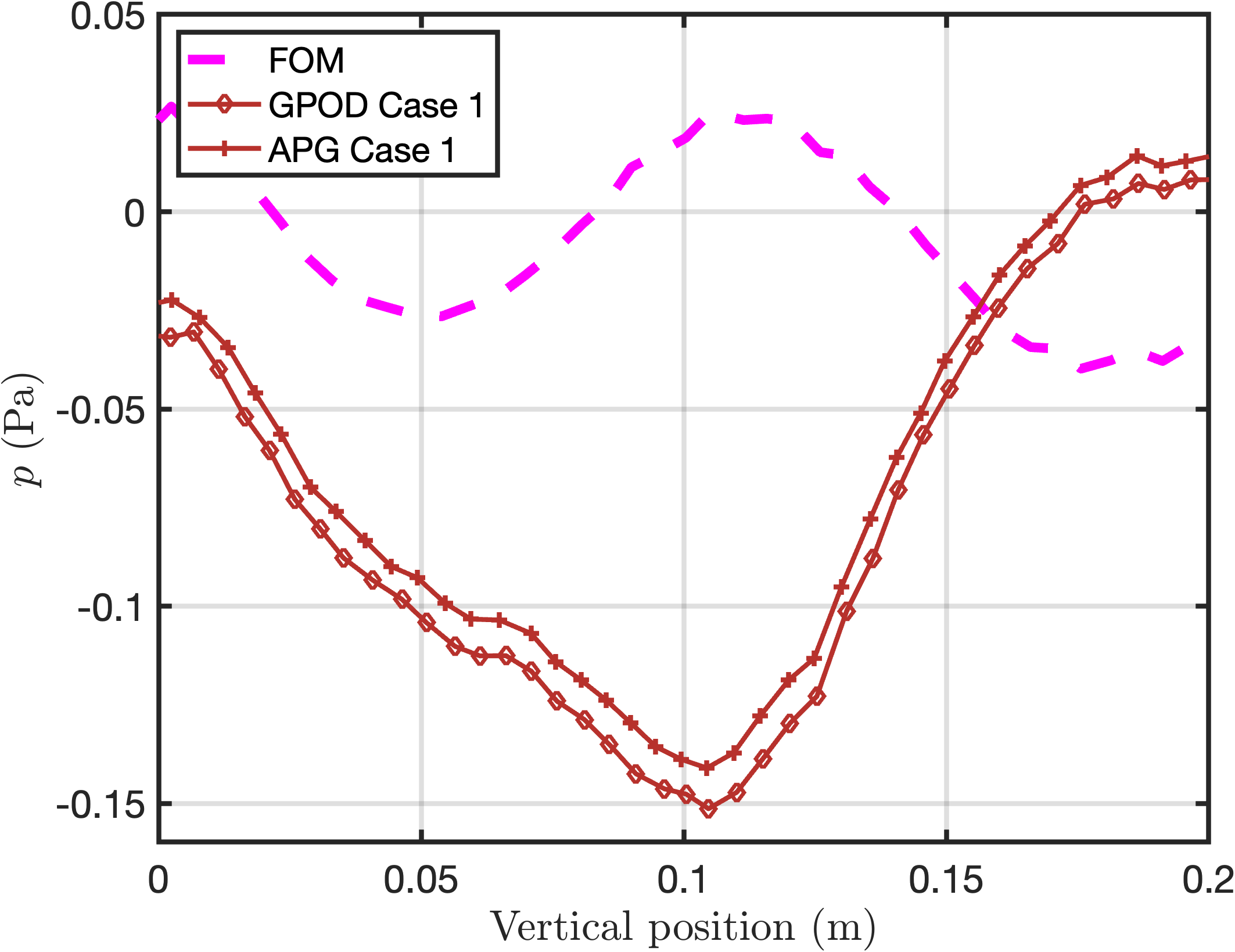}
		\caption{FOM pressure field}
		\label{fig:FPOC_vertical_slices_pressure}
	\end{subfigure}
	\caption{Snapshot at $t=0.625$ s of vertical slices at $X=0.25$ m.}
	\label{fig:FPOC_vertical_slices}
\end{figure*}

Vertical slices at $X=0.25$ m of the velocity and pressure field profiles are shown in Fig.~\ref{fig:FPOC_vertical_slices}. The velocity profiles in Fig.~\ref{fig:FPOC_vertical_slices_velocity} show that both GPOD and APG methods capture the vortex in the cavity and the profile of the open channel. However, as previously indicated, both PMORs are slightly shifted in phase, and the FOM exhibits additional nonlinear behavior across the trough of the cavity vortex, which is not captured by both PMORs. Figure \ref{fig:FPOC_vertical_slices_pressure} highlights significant differences in the pressure field profiles, which is expected due to the phase shift previously observed in the velocity profiles. Here, the APG method has a marginally smaller pressure amplitude than the GPOD method, but shares similar profile shape and spatial characteristics. Overall, qualitative results indicate that both PMORs can capture spatial velocity field characteristics but exhibit phase discrepancies due to the vortex shedding sensitivity and long-time accumulation of errors in subspace approximation errors.

\begin{figure*}[t!]
	\centering
	\begin{subfigure}[t]{0.5\textwidth}
		\centering
		\includegraphics[trim = {0cm 0cm 0 0},  clip, scale=0.325]{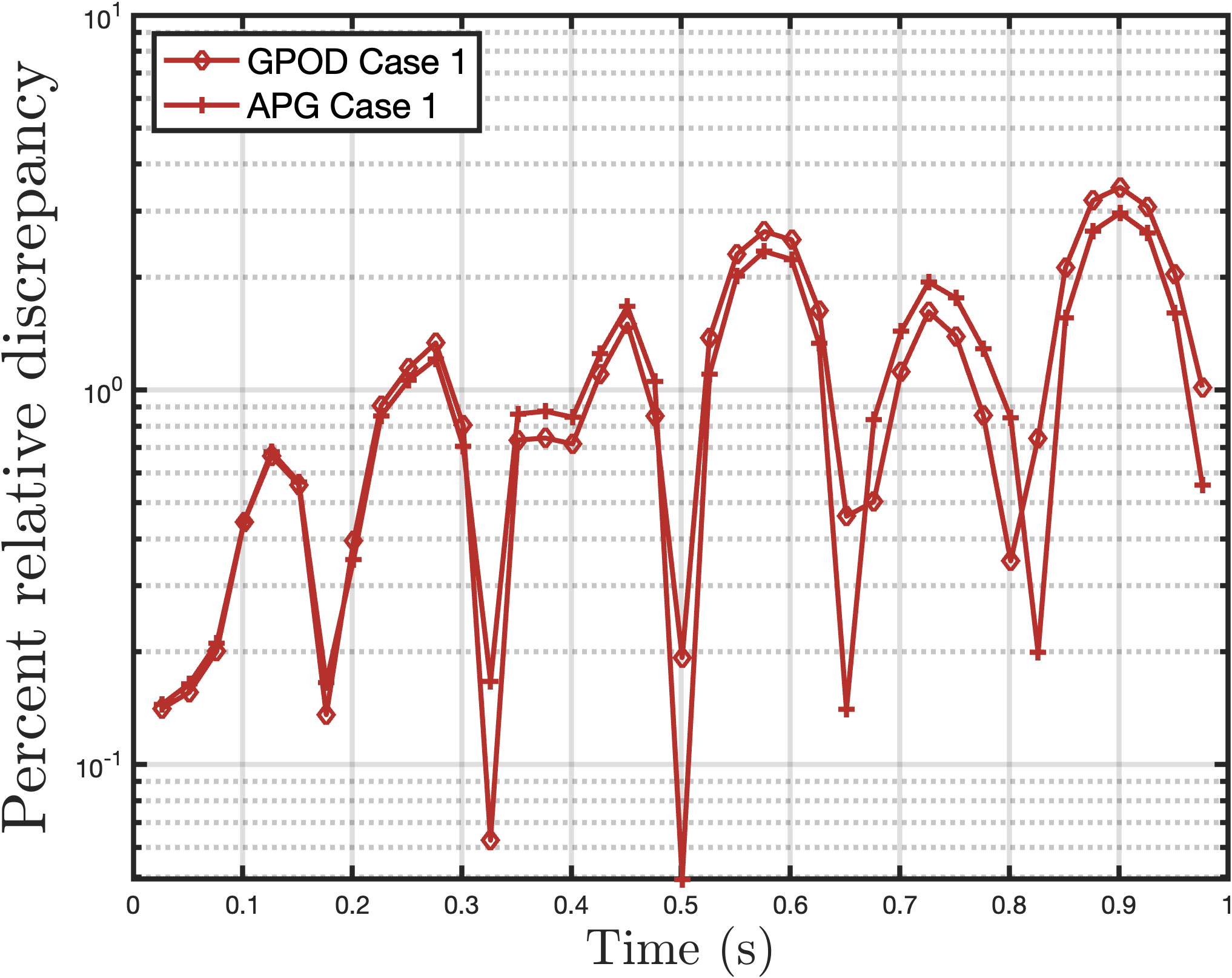}
		\caption{FOM velocity field}
	\end{subfigure}%
	\hfill
	\begin{subfigure}[t]{0.5\textwidth}
		\centering
		\includegraphics[ trim={0cm 0cm 0 0}, clip, scale=0.325]{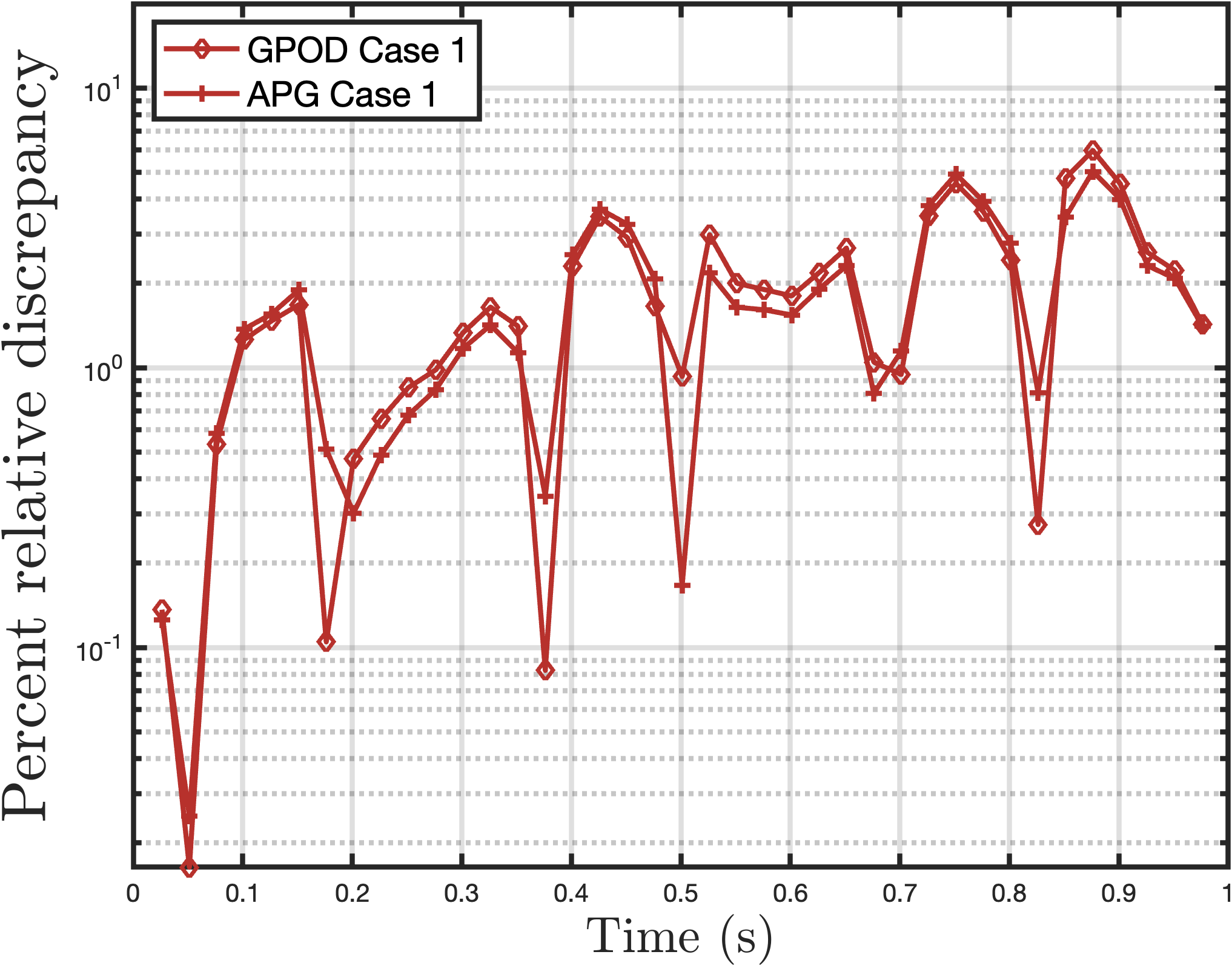}
		\caption{FOM pressure field}
	\end{subfigure}
	\caption{Time histories of relative discrepancy errors at $X=0.25$ m vertical slices.}
	\label{fig:FPOC_relative_discrepancy}
\end{figure*}

Relative discrepancy errors for velocity and pressure fields are shown in Fig.~\ref{fig:FPOC_relative_discrepancy}. Velocity field errors peak near 4\% while pressure fields peak near 6\%. Results further highlight the ability of the proposed PMOR to capture the velocity and pressure fields in the FOM, but additional work is needed to improve fidelity in capturing density variations to more faithfully capture pressure fields derived from the stiff weakly compressible equation of state in the SPH framework employed.

\section{Conclusions and Future Work} \label{Section:Conclusion}

A projection-based model-order reduction framework for meshless numerical methods is introduced in this paper. The proposed approach is built upon a traditional PMOR architecture while maintaining the benefits meshless numerical methods offer. This work chooses the weakly compressible smoothed-particle hydrodynamics numerical method for the demonstration of the model-order reduction framework, due to its wide applications across science and engineering. The proposed PMOR method enables the projections of field quantities derived from mixing numerical topologies onto a low-dimensional subspace, where the meshless framework could evolve forward in time. Two projection based approaches are incorporated into the presented meshless projection framework; the Galerkin POD method, and the Adjoint Petrov--Galerkin method. The GPOD method is selected as a test case since it is ubiquitous and was considered state-of-the-art for many years. The Adjoint Petrov--Galerkin method is a recent nonlinear framework derived from the variational multiscale method and the Mori--Zwanzig formalism. The APG method provides stabilization properties with matrix-free Jacobian operations and does not need to be cast into an implicit time-integration scheme, which is beneficial for SPH frameworks, where their explicit formulation is one of their main advantages.

Numerical experiments with the Galerkin POD and Adjoint Petrov--Galerkin projection methods were presented for 1) the Taylor--Green vortex; 2) the lid-driven cavity and 3) flow past an open cavity. In all examples, it was shown that the proposed method can reconstruct and predict velocity fields to within at most 10\% in both reconstructive and parametric settings for the present examples. However, it was empirically shown that reconstructing and predicting pressure fields is sensitive to the choice of basis dimensionality. For instance, it is conjectured that the weakly compressible assumption employed in the current SPH method introduces numerical acoustic noise into the latent spaces of the density field derived from training data. Given the stiff equation of state in the adopted SPH method, any numerical noise embedded into the density field latent space would result in a sensitive basis selection process and limits the level of approximation available in the PMOR without compromising stability. Future work will look into dimensionality reduction techniques that are capable of robustly constructing smooth subspaces in the presence of such numerical acoustic noise or high-frequency data. Nevertheless, despite a sensitive selection process of basis dimensionality, the proposed PMOR delivered agreeable relative discrepancy errors across all reconstructive and predictive numerical experiments. 

The present work, to the best of the authors' knowledge, provides a first step toward enabling projection-based model-reduction for smoothed-particle hydrodynamics. It is important to reiterate that the present work focused on developing the low-dimensional projection framework, which has not been developed for meshless numerical methods. Therefore, the performance and cost-savings of the proposed PMOR is beyond the scope of the present work. Nevertheless, the development of hyper-reduction methods to enable cost-saving under the proposed meshless PMOR umbrella is underway.  


\section*{Acknowledgements}

S.~N.~Rodriguez and S.~L.~Brunton would like to acknowledge OUSD R\&E for funding through the Laboratory and University Collaborative Initiative (LUCI). L.~K. Magargal acknowledges current support from the Department of Defense (DoD) through the National Defense Science \& Engineering Graduate (NDSEG) Fellowship Program and prior support by the Office of Naval Research NREIP program. S.~N.~Rodriguez, J.~C.~Steuben, N.~A.~Apetre, A. Iliopoulos, and J.~G.~Michopoulos, would like to acknowledge ONR through NRL core funding. J.~W.~ Jaworski ackowledges support from ONR under grant N00014-24-1-2111, monitored by Dr.~Yin Lu (Julie) Young.

\bibliographystyle{siamplain}
\bibliography{references}
\end{document}

%% file: ex_shared.tex
\usepackage{lineno,hyperref}
\usepackage{bm}
\usepackage{amsmath}
\usepackage{amsfonts}
\usepackage{amsmath}
\usepackage{amssymb}
\usepackage{xcolor}
\usepackage{enumitem}
\usepackage{graphicx}
\usepackage{enumitem}
\usepackage{algpseudocode}
\usepackage{setspace}
\usepackage{mathtools}
\usepackage{caption}
\usepackage{subcaption}
\usepackage{ragged2e} 
\usepackage[sort]{cite}
\usepackage{pdflscape}
\usepackage{fancyhdr}
\usepackage{hhline}
\usepackage{multirow}
\usepackage{array}
\usepackage{booktabs}
\usepackage{nccmath}
\usepackage{enumitem}
\usepackage{xcolor}
\usepackage{algorithm}
\usepackage[algo2e,ruled]{algorithm2e} 
\usepackage{xpatch}
\usepackage{setspace}

\SetKwInOut{Input}{Input~}
\SetKwInOut{Output}{Output~}
\SetKwInOut{Note}{Note}
\SetKwFunction{Function}{Function}

\makeatletter
\xpatchcmd\SetKwInOut
{\hangafter=1\parbox[t]}
{\hangafter=1\justify\parbox[t]}
{}{\fail}
\makeatother

\algnewcommand\algorithmicinput{\textbf{Input:}}
\algnewcommand\NewInput{\item[\algorithmicinput]}

\ifpdf
  \DeclareGraphicsExtensions{.eps,.pdf,.png,.jpg}
\else
  \DeclareGraphicsExtensions{.eps}
\fi

\newsiamremark{remark}{Remark}
\newsiamremark{hypothesis}{Hypothesis}
\crefname{hypothesis}{Hypothesis}{Hypotheses}
\newsiamthm{claim}{Claim}

\fancypagestyle{plain}{%
	\fancyhf{} 
	\cfoot{\vspace{0.25cm} \\\smaller{DISTRIBUTION STATEMENT A. Approved for public release; distribution is unlimited.
}\\ }\rfoot{\thepage}
	
}
\title{Meshless projection model-order reduction via reference spaces for smoothed-particle hydrodynamics}

\author{
		Steven N. Rodriguez\thanks{Material Science and Technology Division, U. S. Naval Research Laboratory, Washington, DC.}\and 
		Steven L. Brunton \thanks{Mechanical Engineering Department, University of Washington, Seattle, WA.}\and 
		Liam K. Magargal\thanks{Mechanical Engineering and Mechanics Department, Lehigh University, Bethlehem, PA.}\and
		Parisa Khodabakhshi\footnotemark[3]\and
		Justin W. Jaworski\thanks{Department of Aerospace and Ocean Engineering, Virginia Tech, Blacksburg, VA.}\and
		Nicoleta A. Apetre\footnotemark[1]\and
		John C. Steuben\footnotemark[1]\and
		John G. Michopoulos\footnotemark[1]\and
		Athanasios Iliopoulos\footnotemark[1]
	}

\usepackage{amsopn}

\makeatletter
\newcommand*{\addFileDependency}[1]{
  \typeout{(#1)}
  \@addtofilelist{#1}
  \IfFileExists{#1}{}{\typeout{No file #1.}}
}
\makeatother

\newcommand*{\myexternaldocument}[1]{%
    \externaldocument{#1}%
    \addFileDependency{#1.tex}%
    \addFileDependency{#1.aux}%
}

\algrenewcommand\algorithmicrequire{\textbf{Input:}}
\algrenewcommand\algorithmicensure{\textbf{Output:}}

\headers{Meshless projection model-order reduction via reference spaces}{S. N. Rodriguez ET AL.}